\documentclass[final,5p,times,twocolumn]{elsarticle}

\usepackage{graphicx}

\usepackage{amssymb}
\usepackage{amsthm}
\usepackage{amsmath}
\usepackage{physics}
\usepackage{bbm}
\usepackage{bm}
\usepackage{enumitem}
\usepackage{nicematrix}
\biboptions{numbers,sort&compress}
\usepackage{printlen}
\usepackage{booktabs}
\uselengthunit{in}
\usepackage{lineno}
\usepackage{enumitem} 
\allowdisplaybreaks

\usepackage{acro}
\DeclareAcronym{lbm}{short=\textsc{LBM}, long=Lattice Boltzmann Method}
\DeclareAcronym{bgk}{short=\textsc{BGK}, long=Bhatnagar-Gross-Krook}
\DeclareAcronym{dem}{short=\textsc{DEM}, long=Discrete Element Method}
\DeclareAcronym{cfd}{short=\textsc{CFD}, long=Computational Fluid Dynamics}
\DeclareAcronym{lbe}{short=\textsc{LBE}, long=Lattice Boltzmann Equation}
\DeclareAcronym{trt}{short=\textsc{TRT}, long=Two Relaxation Time}
\DeclareAcronym{stl}{short=\textsc{STL}, long=Standard Template Library}
\DeclareAcronym{aos}{short=\textsc{AoS}, long=array of structures}
\DeclareAcronym{soa}{short=\textsc{SoA}, long=structure of arrays}
\DeclareAcronym{oop}{short=\textsc{OOP}, long=object-oriented programming}
\DeclareAcronym{rbc}{short=\textsc{RBC}, long=red blood cell, long-plural=s}
\acsetup{list/display=all}

\usepackage{tikz}
\usetikzlibrary{shadows,positioning,backgrounds,fit,chains,scopes}
\usetikzlibrary{calc}
\usetikzlibrary{arrows.meta}
\usetikzlibrary {patterns,patterns.meta}
\usetikzlibrary{decorations.pathreplacing,calligraphy}
\usetikzlibrary{decorations.pathmorphing,decorations.markings}
\usetikzlibrary{backgrounds}
\usetikzlibrary{angles,quotes}

\usepackage{subcaption}
\usepackage{layouts}

\usepackage{float}
\usepackage{gensymb}
\usepackage[c]{esvect}

\makeatletter
\newif\ifreview
  \reviewtrue
  \reviewfalse
\makeatother

\newcounter{bla}

\usepackage{hyperref}

\hypersetup{
    colorlinks=true,
    linkcolor=red,
    citecolor=red,
    urlcolor=red
}

\journal{Computer Physics Communications}

\begin{document}

\newcommand{\fin}{f^{in}}
\newcommand{\fout}{f^{out}}
\newcommand{\feq}{f^{eq}}
\newcommand{\finv}{\vv{f^{in}}}
\newcommand{\foutv}{\vv{f^{out}}}
\newcommand{\feqv}{\vv{f^{eq}}}
\newcommand{\Dt}{\Delta t}
\newcommand{\Dx}{\Delta x}
\newcommand{\CC}[1]{\textcolor{blue}{#1}}
\newcommand{\JL}[1]{\textcolor{red}{#1}}
\newcommand{\MR}[1]{\textcolor{orange}{#1}}

\begin{frontmatter}



\title{LEDDS: Portable LBM-DEM simulations on GPUs}

\author[a]{Raphael Maggio-Aprile\fnref{equal}}
\author[a]{Maxime Rambosson\fnref{equal}}
\author[b,a]{Christophe Coreixas}
\author[a]{Jonas Latt\corref{author}}

\cortext[author] {Corresponding author.\\\textit{E-mail address:} jonas.latt@unige.ch}
\fntext[equal]{These authors contributed equally to this work.}

\address[a]{Computer Science Department, University of Geneva, Carouge 1227, Switzerland.}
\address[b]{Institute for Advanced Study, Beijing Normal - Hong Kong Baptist University, Zhuhai 519087, China.}

\begin{abstract}

Algorithmic formulations of GPU programs provide a high-level alternative to device-specific code by expressing computations as compositions of well-defined parallel primitives (e.g., map, sort, reduce), rather than through handcrafted GPU kernels. In this work, we demonstrate that this paradigm can be extended to complex and challenging problems in computational physics: the simulation of granular flows and fluid-particle interactions.

LEDDS, our open-source framework, performs fully coupled Lattice Boltzmann -- Discrete Element Method (LBM-DEM) simulations using only algorithmic primitives, and runs efficiently on single-GPU platforms. The entire workflow, including neighbor search, collision detection, and fluid-particle coupling, is expressed as a sequence of portable primitives. 
While the current implementation illustrates these principles primarily through algorithms from the C++ Standard Library, with selective use of Thrust primitives for performance, the underlying concept is compatible with any HPC environment offering a rich set of parallel algorithms and is therefore applicable across a wide range of modern GPU systems and future accelerators.

LEDDS is validated through benchmarks spanning both DEM and LBM-DEM configurations, including sphere and ellipsoid collisions, wall friction tests, single-particle settling, Jeffery's orbits, and particle-laden shear flows. Despite its high level of abstraction, LEDDS achieves performances comparable to those of hand-tuned CUDA solvers, while maintaining portability and code clarity.
These results show that high-performance LBM-DEM coupling can be achieved without sacrificing generality or readability, establishing LEDDS as a blueprint for portable multiphysics frameworks based on algorithmic primitives.

\end{abstract}


\end{frontmatter}



{\bf PROGRAM SUMMARY}

\begin{small}
\noindent
{\em Program Title:} \textbf{L}BM-\textbf{E}nhanced \textbf{D}evice-independent \textbf{D}EM \textbf{S}olver (LEDDS) \\
{\em Developer's repository link:} \url{https://gitlab.com/unigehpfs/ledds} \\
{\em Licensing provisions:} MIT  \\
{\em Programming language:} {C++}                                  \\
{\em Nature of problem:}
Fully resolved particle-laden flows involving granular materials and fluid–particle interactions. Achieving particle-scale accuracy in contacts and fluid–solid interactions is computationally demanding, and it requires a careful design and implementation strategy to simulate such flows on limited resources. \\
{\em Solution method:}
In LEDDS, the fluid flow evolution is modelled using the Lattice Boltzmann Method (LBM), while particle motion and contacts are handled by a Discrete Element Method (DEM). Fluid–particle coupling is achieved using a diffuse-interface formulation: the partially saturated method. Simulations are performed on uniform Cartesian grids, and support both spherical and ellipsoidal particles. Parallelization relies on algorithmic primitives implemented in modern C++, and selectively in Thrust, which enables portable high-performance execution on multi-core CPUs and single GPUs.\\
{\em Additional comments including restrictions and unusual features:} 
The LBM and DEM modules can be used independently or fully coupled through a consistent data structure. LEDDS is currently limited to single-phase, isothermal flows and to spherical or ellipsoidal particles. MPI-based or other multi-GPU and multi-node CPU parallelism are not supported in this version. The code requires nvc++ version 24.9 or later; backward compatibility is not guaranteed, but upward compatibility is ensured.
   \\

\end{small}

\section{Introduction}
\label{sec:introduction}

The Discrete Element Method (DEM), originally proposed by Cundall and Strack~\cite{CUNDALL_GEO_29_1979}, is a versatile tool for predicting the behavior of particle-based systems. Specifically, in DEM, individual particles are modeled as discrete entities with their own physical and mechanical properties, such as size, shape, mass, Young’s modulus, and Poisson ratio. Particle motion, both translational and rotational, arises from interactions with other particles or with the environment~\cite{CUNDALL_GEO_29_1979}. Interparticle forces, including cohesive, bridging, and contact forces, are described using models such as spring-dashpot, friction, and collision models. The system evolves over time by integrating Newton’s second law of motion while accounting for external effects, including gravity and interstitial fluids.

Since its advent, DEM has found wide application in studying granular materials, powders, soil mechanics, and particle-fluid interactions~\cite{ZHU_CES_62_2007,ZHU_CES_63_2008}. In the latter case, DEM is coupled with a computational fluid dynamics (CFD) solver~\cite{ZHAO_PT_239_2013,ZHONG_PT_302_2016,KIECKHEFEN_ARCBM_11_2020,MA_PT_412_2022}, forming a CFD-DEM framework. These simulations are computationally demanding, which has motivated the use of general-purpose GPU acceleration. Because GPUs excel at data-parallel tasks, GPU implementations can significantly speed up operations like particle tracking, collision detection, and fluid flow calculations. 
As a result, GPU-accelerated DEM and CFD-DEM simulations can significantly reduce runtime compared with modern multi-core CPU implementations, with reported GPU speedups often reaching one to two orders of magnitude depending on the problem and hardware~\cite{GAN_PT_301_2016,LU_PARTICUOLOGY_62_2022,FU_CEJ_465_2023}.

Different coding strategies can be used to efficiently run CFD-DEM simulations on GPUs. These include hardware-specific low-level languages (CUDA~\cite{NICKOLLS_QUEUE_2008}, HIP~\cite{HIP_WEBSITE}), directive-based programming models (OpenACC~\cite{WIENKE_EUROPAR_2012}, OpenMP~\cite{CHANDRA_Book_2001}), and dedicated libraries or frameworks (OpenCL~\cite{MUNSHI_IEEE_2009}, SYCL~\cite{ALPAY_PIWOCL_2020}). In the present work, we rely on the availability of a set of well-defined algorithmic primitives in many of the high-level framework, allowing the design of platform-agnostic approach through the decomposition of complex problems into such primitives. First formalized by Blelloch~\cite{Blelloch1990} for vector and PRAM models, a parallel paradigm driven by a decomposition into primitives such as \emph{map}, \emph{scan}, and \emph{reduce} has gained renewed attention with massively parallel architectures, owing to the possibility to express complex kernels as compositions of a few generic building blocks~\cite{Sengupta2007,ShunBlelloch2014,Dhulipala2018,Brahmakshatriya2020}. Similar ideas have emerged in sparse linear algebra, where graph and matrix operations are unified under common algebraic primitives~\cite{Buluc2010}.
This decomposition-based approach produces self-documenting, portable programs across different hardware, and reduces concurrency-related errors by avoiding manual synchronization. Notably, several mature frameworks provide efficient implementations of algorithmic primitives for heterogeneous platforms, including Thrust~\cite{BELL_2012}, Kokkos~\cite{CARTEREDWRADS_JPDC_74_2014}, and AMD’s HIPSTDPAR/rocThrust~\cite{AMD_HIPSTDPAR_2023}. Modern C++ standards (C++17/C++20) also offer portable parallel algorithms~\cite{LARKIN_GTC_2022,LARKIN_GTC_2024}, which we adopt as the primary framework, falling back to specific algorithms of the Thrust library selectively when limitations of the C++ standard impacts the performance distinctly. This combination allows the code to run efficiently across different hardware with minimal changes, typically requiring only a compiler or flag adjustment.

The authors have previously developed algorithm-based implementations of fluid solvers using the Lattice Boltzmann Method (LBM). In an earlier work, the fundamental operations of the LBM algorithm were reformulated entirely in terms of algorithmic primitives such as \emph{map} and \emph{reduce}~\cite{LATT_PLOSONE_16_2021}. This approach was subsequently extended to multi-GPU systems, where a large-scale LBM framework was ported using standard parallel algorithms combined with domain decomposition~\cite{LATT_CPC_323_2026}, as well as to LBM simulations on non-uniform static or dynamic grids~\cite{COREIXAS_CPC_317_2025,COREIXAS_3AF_2026,COREIXAS_AIAA_2026}, and even high-order finite-difference schemes~\cite{COREIXAS_HIFILED_LBFDGPU_2022}. It has also been extended to multi-physics configurations, including conjugate heat transfer in porous media, where two LBM solvers are coupled to resolve both the fluid flow and heat conduction within the solid phase~\cite{SCHWENDENER_JFM_1026_2026}. In the present work, we go one step further by addressing the complex case of LBM-DEM coupling within the same algorithmic paradigm, while restricting ourselves to uniform grids only. 

In particle-resolved simulations, coupling a particulate phase to a fluid solver requires accurately representing the fluid-solid interface without resorting to costly remeshing. Two popular diffuse-interface strategies in the context of LBM-DEM coupling are (i) immersed boundary methods (IBM), which enforce the no-slip condition through localized forcing terms that spread particle surface forces over nearby lattice nodes, and (ii) partially saturated cell (PSC/PSM) approaches, where each lattice cell stores a fractional solid occupancy used to modulate the collision operator. Both techniques avoid dynamic mesh generation and naturally fit the Eulerian structure of LBM. IBM-based LBM-DEM coupling has been widely explored, including applications to dense suspensions and rigid spherical particles~\cite{RettingerRude2017Comparative,RettingerRude2018CFDDEM,thorimbert_lbm_ibm_2020}, whereas PSM-like penalization methods have become a standard choice for large particulate systems due to their simplicity and robustness~\cite{noble_lattice-Boltzmann_1998,RETTINGER_JCP_453_2022}. In the present work, we adopt the PSM approach for its high accuracy and algorithmic compatibility with parallel primitives.

This coupling strategy introduces challenges arising not only from the individual solvers but, more importantly, from their interaction. Fluid-particle interactions in densely packed systems are particularly demanding, as the computation of forces exchanged between solid and fluid phases can dominate simulation time. We describe in detail how each component of the coupling algorithm --the identification of fluid-particle interface cells, the computation of per-cell forces, and the subsequent force reduction-- is implemented as a sequence of parallel primitives, primarily \emph{map}, \emph{sort}, and \emph{prefix-sum} operations.

\begin{figure}[hbt]
\centering
\begin{tikzpicture}[scale=0.7]
    \tikzstyle{fluidcell}=[draw=black!50, thin]
    \tikzstyle{coarsegrid}=[draw=black, very thick]
    \tikzstyle{particle}=[draw=black, thick, fill=blue!30, fill opacity=0.5]

    \def\nx{9}
    \def\ny{9}
    \def\dx{1}
    \def\margin{0.2}

    \foreach \i in {0,...,\nx}{
        \draw[fluidcell] (\i*\dx, -\margin) -- (\i*\dx, \ny*\dx+\margin);
    }
    \foreach \j in {0,...,\ny}{
        \draw[fluidcell] (-\margin, \j*\dx) -- (\nx*\dx+\margin, \j*\dx);
    }

    \foreach \i in {0,3,6,9}{
        \draw[coarsegrid] (\i*\dx, -\margin) -- (\i*\dx, \ny*\dx+\margin);
    }
    \foreach \j in {0,3,6,9}{
        \draw[coarsegrid] (-\margin, \j*\dx) -- (\nx*\dx+\margin, \j*\dx);
    }

    \draw[particle] (1.5,1.5) circle (1.2);
    \draw[particle] (3.8,1.8) circle (0.8);
    \draw[particle] (6.5,2.2) circle (1.0);
    \draw[particle] (4.8,6.2) circle (1.3);
    \draw[particle] (1.2,7.0) circle (0.9);
    \draw[particle] (7.0,4.5) circle (1.0);
    \draw[particle] (2.8,3.5) circle (0.9);

    \pgfmathsetmacro{\centerX}{\nx*\dx/2-1}
    \pgfmathsetmacro{\ylabels}{\ny+1}
    \pgfmathsetmacro{\spacing}{3} 

    \draw[coarsegrid] (\centerX-\spacing-0.5, \ylabels) -- ++(0.8,0);
    \node[anchor=west, scale=0.8] at (\centerX-\spacing+0.5, \ylabels) {DEM grid};

    \draw[fluidcell] (\centerX, \ylabels) -- ++(0.8,0);
    \node[anchor=west, scale=0.8] at (\centerX+1, \ylabels) {Fluid grid};

    \draw[particle] (\centerX+\spacing+0.7, \ylabels) circle (0.15);
    \node[anchor=west, scale=0.8] at (\centerX+\spacing+1.2, \ylabels) {Particles};

\end{tikzpicture}

\caption{2D schematic of the coupled particle-fluid simulation. Bold lines indicate the DEM (cell-linked list) grid used for collision detection, thin lines show the LBM fluid grid, and semi-transparent circles show the particles. The relative 1:3 spacing between the grids is for illustration; in practice, the ratio depends on the largest particle size. The algorithm requires the grid spacing to be larger than the particle diameter to guarantee an overlap of a particle with at most 8 cells (in 3D).}
\label{fig:grids}
\end{figure}

Hereafter, we present the new open-source, algorithm-based framework \underline{LEDDS} (\underline{L}BM-\underline{E}nhanced \underline{D}evice-independent \underline{D}EM \underline{S}olver), which integrates an LBM solver for the fluid phase, a DEM solver for the particle phase, and a coupling algorithm based on the Partially Saturated Method (PSM) which is a penalization-like approach~\cite{noble_lattice-Boltzmann_1998,kruger_lattice_2017}.
While the LBM part of LEDDS heavily relies on template codes provided in the STLBM framework, its DEM part is conceptually inspired by the open-source DEM framework LIGGGHTS~\cite{KLOSS_SP_2011}.
This was done to make sure the DEM component of LEDDS is consistent with well-established numerical best practices in the field.
However, LEDDS does not reuse LIGGGHTS code, and instead, its DEM module has been developed from scratch as a fully independent, algorithm-based module.
To achieve this, we isolated each relevant feature of LIGGGHTS, reformulated it in terms of algorithmic primitives, and verified it against the original code. This ensured that all methodological details necessary for a complete DEM implementation were incorporated in LEDDS, e.g., tracking lateral displacements during collisions to compute tangential forces accurately.

On top of this, LEDDS stands out by the uniform implementation of its DEM and LBM solvers as both are developed using modern C++. Hence, they benefit from a common software design and are parallelized using C++ parallel algorithms. In case of GPU execution, both the DEM and the LBM run on the GPU and benefit from an efficient in-memory data exchange. 
To further emphasize the tight interaction between these two aspects of the code, the DEM solver relies on uniform Cartesian background mesh for the detection of particle collision pairs, which naturally relates to the Cartesian grid of the LBM solver (see Figure~\ref{fig:grids}). 
Finally, after thorough performance evaluation, we replaced selected STL algorithm calls with their Thrust counterparts, for example in particle force computations. This decision was driven by the availability of specialized implementations in Thrust, such as radix sort, which outperformed the STL’s comparison-based sorting in our use case.
In both approaches, whether using the C++ Standard Library or Thrust, the implementation relies on unified memory, eliminating the need for explicit data transfers or device-specific programming constructs.

Unlike large production frameworks, LEDDS is designed as a flexible platform to explore whether a complete LBM-DEM solver can be expressed using high-level algorithmic primitives while maintaining competitive performance on heterogeneous hardware. This is investigated in the remainder of the paper that is organized as follows.
Section~\ref{sec:theoretical-approach} reviews the theoretical background of DEM and LBM, including their coupling via the LB-specific penalization-like scheme (PSM). Section~\ref{sec:portable_implementation} presents the implementation strategy,  
highlighting how algorithmic primitives, primarily from the C++ Standard Library with selective use of Thrust, are used to build an efficient 
device-independent CFD-DEM solver. Section~\ref{sec:validation} validates the accuracy and robustness of LEDDS through particle-only and fluid-particle benchmark tests, followed by an in-depth performance analysis in Section~\ref{sec:perfo}. Finally, discussions and conclusions are provided in Section~\ref{sec:summary_and_outlook}.
Appendices summarize DEM-material parameter relations~(\ref{app:material_properties}) and cover single-precision arithmetic~(\ref{app:single_precision}), parameter sensitivity~(\ref{app:param_study}), and GPU memory usage~(\ref{app:gpu_memory_usage}).

\section{Theoretical approach}
\label{sec:theoretical-approach}

\subsection{Discrete Element Method}
\label{subsec:dem}
The \ac{dem} models the interactions between suspended particles inside the fluid.
In LEDDS, particles are represented either as spheres or ellipsoids, and their evolution is based on contact mechanics.
When particles collide, they interact through localized contact regions, where forces are exchanged.
The motion of a particle $i$ thus only depends on two quantities, which are the forces and torques it experiences during contact with other particles:
\begin{align}
    \vv{F_i} = \sum_{ij, i \neq j} \vv{F_{ij}}, \qquad
    \vv{M_i} = \sum_{ij, i \neq j} \vv{M_{ij}}.\label{eq:dem-forces}
\end{align}
There is an interaction between two particles $i$ and $j$ only if they are in contact/overlap. With spheres, this happens when the distance $d$ between the centers of the spheres is smaller than the sum of their radii, denoted as $R_i$ and $R_j$:
\begin{equation}
d < R_i + R_j.
\end{equation}

For ellipsoids, the same collision procedure is applied by locally approximating the surface near the contact point with an equivalent sphere. The associated radius of curvature is computed from an approximation of the mean curvature:
\begin{equation}
    \label{eq:local-curvature}
    R = \frac{b_i c_i x_i^2}{a_i^3} + \frac{a_i c_i y_i^2}{b_i^3} + \frac{a_i b_i z_i^2}{c_i^3}
\end{equation}
with $(a_i, b_i, c_i)$ the radii of the ellipsoid in local frame. This approximation is preferred over the Gaussian curvature radius, as it avoids high-order operations and is therefore \emph{better suited for single-precision computations}.
Using this local spherical approximation, the contact detection consists in iteratively computing a pair of closest points (one on each ellipsoid) such that the line connecting them defines a common normal direction~\cite{LAMBERT_NMF_2020}. Starting from an initial guess on each surface, the points are updated by intersecting the ellipsoids with the line defined by the centers of the associated sphere approximations. The iterations proceed until convergence to a common normal configuration is achieved, subject to a prescribed tolerance and a maximum number of iterations.

Once the contact points and the associated normal direction are determined, the interaction force between two particles can be evaluated. The force exerted by a particle $j$ on a particle $i$ is decomposed into two components: a component $\vv{F_n}$ normal to the contact surface and a component $\vv{F_t}$ parallel to it, where, for brevity, the notation $\vv{F}$ is used in place of $\vv{F_{ij}}$.
These forces will be calculated using a spring-dashpot model \cite{mindlin_compliance_2021}. In this model, the normal and tangential forces can be represented as a system consisting of a spring and a damper. The formula of the forces for this model is: 
\begin{align}
    \vv{F} = \underbrace{\left(- k_n \; \vv{\delta_{n_{ij}}} \; - \; \gamma_n \; \vv{v_{n_{ij}}} \right)}_{\vv{F_n}} 
    +
    \underbrace{\left( - k_t \; \vv{\delta_{t_{ij}}} \; - \; \gamma_t \; \vv{v_{t_{ij}}} \right)}_{\vv{F_t}}
    \label{eq:spring_dashpot_easy},
\end{align}
where $k_n$ and $k_t$ are the normal and tangential elastic coefficients, $\gamma_n$ and $\gamma_t$ the corresponding damping coefficients, $\vv{\delta_{n_{ij}}}$ and $\vv{\delta_{t_{ij}}}$ the normal and tangential overlap vectors, and $\vv{v_{n_{ij}}}$ and $\vv{v_{t_{ij}}}$ the normal and tangential relative velocity vectors (see Figure~\ref{fig:illustration-overlap}.

\begin{figure}
    \centering

\begin{tikzpicture}[scale=0.7,>=stealth]
\coordinate (Cj) at (0,0);        
\coordinate (Ci) at (3.2,0.1);    
\shade[ball color=cyan!40] (Cj) circle (2);
\coordinate (P3) at (0.05, 1.90);     
\coordinate (P4) at (-0.35,-2.90);    
\coordinate (d_far) at (1.7, 0.50);
\def\k{2.0}
\coordinate (P2) at ($(P3) + \k*(d_far)$);
\coordinate (P1) at ($(P4) + \k*(d_far)$);
\filldraw[
    fill=gray!20,
    fill opacity=0.7,
    draw=black
] (P1) -- (P2) -- (P3) -- (P4) -- cycle;
\shade[ball color=purple!50] (Ci) circle (2);
\node[font=\bfseries] at (Ci) {i};
\node[font=\bfseries] at (Cj) {j};
\coordinate (O) at (1.5, -0.1);
\begin{scope}[shift={(O)},rotate=-10]
  \draw[->,thick] (0,0) -- (1.2,0)
    node[midway,below] {$\mathbf{e}_{n_{ij}}$};
  \draw[->,thick] (0,0) -- (0,1.2)
    node[midway,left,xshift=-4pt] {$\mathbf{e}_{t_{ij}}$};
\end{scope}
\node[font=\Large] at (1.55,-1.55) {$H$};
\end{tikzpicture}
    \caption{Intersection plane which characterizes a collision between two spheres. The tangential force is integrated in time along the direction $\mathbf{e}_{t_{ij}}$.}
    \label{fig:illustration-overlap}
\end{figure}
The elastic and damping coefficients ($k_n, k_t, \gamma_n, \gamma_t$) are obtained from the material properties of the solids and the details are given in \ref{app:material_properties}.

As shown in Eq.~\eqref{eq:spring_dashpot_easy}, the normal and tangential forces, $\vv{F_n}$ and $\vv{F_t}$, each contain two contributions: an elastic term opposing displacement and a damping term opposing relative velocity. Depending on the collision phase (particles approaching or separating), these contributions may act in different directions.
Additionally, the tangential overlap $\vv{\delta_{t_{ij}}}$ is limited by a Coulomb friction condition:
\begin{align}
    \norm{\vv{F_t}} \leq \mu \norm{\vv{F_n}}  \label{eq:coulombCoeff}.
\end{align}

Let us detail how the normal force, $\vv{F_n}$, is computed. First, we define the unit vector $\vv{e_{n_{ij}}}$ normal to the contact plane (it goes from j to i direction), while the scalar $\delta_{n_{ij}}$ represents the value of the overlap. The normal overlap is given by the difference between the sum of the radii of the spheres and the distance between their centers:
\begin{align}
    \delta_{n_{ij}} = R_i + R_j - \norm{\vv{r_i} - \vv{r_j}} \label{eq:distanceSpheres}.
\end{align}
If $\delta_{n_{ij}} > 0$, the spheres are in contact.
The unit vector for the normal direction is given by the normalized vector from $j$ center to $i$ center:
\begin{align}
\vv{e_{n_{ij}}} = \frac{\vv{r_i} - \vv{r}_j}{\norm{\vv{r_i} - \vv{r_j}}}.
\end{align}
Then, $\vv{\delta_{n_{ij}}}$ is defined as:
\begin{align}
    \vv{\delta_{n_{ij}}} = -  \delta_{n_{ij}} \vv{e_{n_{ij}}}.
\end{align}
The normal relative velocity is given by the following projection:
\begin{align}
    \vv{v_{n_{ij}}} = \left[ (\vv{v_j} - \vv{v_i}) \cdot \vv{e_{n_{ij}}} \right] \vv{e_{n_{ij}}},
\end{align}
where ``$\cdot$'' is the scalar product, and $\vv{v_i}$ (resp. $\vv{v_j}$) is the velocity of the center of mass of $i$ (resp. $j$).
Therefore, the normal force $\vv{F_n}$ can be rewritten as:
\begin{align}
\vv{F_n} = \left[ k_n \; \delta_{n_{ij}} \; + \; \gamma_n \; (\vv{v_j} - \vv{v_i}) \cdot \vv{e_{n_{ij}}} \right] \vv{e_{n_{ij}}} \label{eq:fnScal}.
\end{align}

When it comes to the tangential force, $\vv{F_t}$, we first compute the relative tangential velocity $\vv{v_{t_{ij}}}$.
\noindent To do so, we need to compute the relative velocity between the surfaces of the spheres:
\begin{align}
    \vv{v_{rel_{ij}}} = (\vv{v_i} - \vv{v_j}) - (R_i \vv{\omega_i} + R_j \vv{\omega_j}) \cross \vv{e_{n_{ij}}} \label{eq:relative_velocity},
\end{align}
where $\vv{\omega}_i$ and $\vv{\omega}_j$ are the the angular velocities of the spheres $i$ and $j$, and ``$\cross$'' is the cross product.
To get the tangential component of the relative velocity, we then remove the normal component:
\begin{align}
    \vv{v_{t_{ij}}} & = \vv{v_{rel_{ij}}} - \vv{v_{n_{ij}}}
\end{align}

After this, the tangential relative displacement, $\vv{\delta_{t_{ij}}}$, needs to be computed. Since it depends on the relative movement of the surfaces over time, we need to integrate the tangential relative velocity $v_{t_{ij}}$ over the duration of the collision:
\begin{align}
    \vv{\delta_{t_{ij}}}(t) = \int_{t_0}^{t} \vv{v_{t_{ij}}}(u) \; du,\label{eq:trangential-history}
\end{align}
where $t_0$ is the time at which the collision starts. Consequently, the ``displacement history'' must be tracked across iterations, as a single collision may span multiple DEM steps within one fluid time step~\cite{MAGGIOAPRILE_Master_2023}.

Eventually, the torque $\vv{M}_{ij}$ exerted by $j$ on $i$ is obtained from the force $\vv{F}$ through:
\begin{align}
    \vv{M_{ij}} = \vv{rc_i} \cross \vv{F_{ij}}.\label{eq:torque}
\end{align}
where $\vv{rc_i}$ denotes the vector from the center of the particle to the contact point.
In the spherical case, this formula can be simplified to:
\begin{align}
    \vv{M_{ij}} = - R_i \cdot \vv{e_{n_{ij}}} \cross \vv{F_{t_{ij}}}.
\end{align}
It is then convenient to store the torque normalized by the radius, $\vv{M_{ij}}/R_i$, since this quantity is identical for both particles in contact and avoids redundant computations.
However, this simplification does not hold for an ellipsoid, as the normal at the contact point is not guaranteed to pass by the solid's center.

For the interaction of particles with walls, walls are treated like a static particle of a given shape (for example an infinitely extending plane) and a given surface velocity. Further details for the handling of walls are provided in~\cite{MAGGIOAPRILE_Master_2023}.

\subsection{The integration scheme}

The motion of a particle is governed by Newton’s equations of translation and rotation: 
\begin{align}
    & \vv{F} + \vv{F}_{\mathrm{ext}} = m \vv{a} 
    = m \frac{d\vv{v}}{dt} 
    = m \frac{d^2 \vv{r}}{dt^2}, \\
    & \vv{M} = \mathbf{I} \, \vv{\alpha} 
    = \mathbf{I} \frac{d\vv{\omega}}{dt} 
    = \mathbf{I} \frac{d^2 \vv{d}}{dt^2},
\end{align}
where $m$ is the particle mass, $\vv{a}$ the linear acceleration, $\mathbf{I}$ the inertia tensor, $\vv{\alpha}$ the angular acceleration, and $\vv{d}$ a body-fixed vector defining the particle’s orientation. The term $\vv{F}_{\mathrm{ext}}$ collects all external forces, and in this work, it reduces to gravity only: $\vv{F}_{\mathrm{ext}} = m \, \vv{g}$.

These equations are integrated in time using the second-order accurate, time-reversible Verlet algorithm~\cite{VERLET_PR_159_1967,SKEEL_1999}. The translational degrees of freedom are updated as
\begin{subequations}
\begin{align}
    & \vv{r}(t + \delta t) = \vv{r}(t) + \delta t \,\vv{v}(t) + \frac{\delta t^2}{2} \, \vv{a}(t), \label{eq:advance-r}\\
    & \vv{a}(t + \delta t) = \tfrac{1}{m}\vv{F}(t + \delta t) + \vv{g}, \\
    & \vv{v}(t + \delta t) = \vv{v}(t) + \tfrac{\delta t}{2} \left[\vv{a}(t) + \vv{a}(t + \delta t)\right],
\end{align}
\end{subequations}
The implicit nature of the scheme, due to the anticipated acceleration term $\vv{a}(t + \delta t)$, is resolved through a single predictor-corrector step. The rotational degrees of freedom evolve according to
\begin{subequations}
\begin{align}
    & \vv{d_i}^{*}(t + \delta t) 
        = \vv{d_i}(t) + \delta t \,\big(\vv{\omega}(t) \cross \vv{d_i}(t)\big), \\
    & \vv{d_i}(t + \delta t) 
        = \vv{d_i}(t) + \delta t \,\Big\{\vv{\omega}(t) \cross \tfrac{1}{2}\big[\vv{d_i}(t) + \vv{d_i}^{*}(t + \delta t)\big]\Big\}, \\
    & \vv{\alpha}(t + \delta t)
         \begin{aligned}[t]
             &=\mathbf{R}^{-1}(t)\,\mathbf{I}^{-1}
             \Big\{
                 \mathbf{R}(t)\,\vv{M}(t + \delta t)
             \\
             &-\,\mathbf{R}(t)\,\vv{\omega}(t)
                     \times \big[\mathbf{I}\,\mathbf{R}\,\vv{\omega}(t)\big]
             \Big\},
         \end{aligned}
         \\ 
    & \vv{\omega}(t + \delta t) 
        = \vv{\omega}(t) + \tfrac{\delta t}{2} \left[\vv{\alpha}(t) + \vv{\alpha}(t + \delta t)\right].\label{eq:advance-omega}
\end{align}
\end{subequations}
Here, $\mathbf{R}(t)$ is the rotation matrix that maps vectors from the body-fixed frame at time $t$ to the reference frame aligned with the coordinate axes at $t=0$. 
Since the particle’s orientation is fully described by the orthonormal triad $\{\vv{d_1}(t), \vv{d_2}(t), \vv{d_3}(t)\}$, the $i$-th row of $\mathbf{R}(t)$ is given by the components of $\vv{d_i}(t)$.

Assuming a homogeneous particle of mass $m$, the inertia tensor is defined as
\begin{align}
    \mathbf{I} = \int_V \rho \left( |\vv{r}|^2 \mathbf{I}_3 - \vv{r}\vv{r}^\top \right) \, dV,
\end{align}
where $\rho = m/V$ is the constant mass density, $\vv{r}$ is the position vector relative to the particle’s center of mass, and $\mathbf{I}_3$ is the $3 \times 3$ identity matrix.
For the particle types implemented here, and assuming axis alignment, the inertia tensors reduce to
\begin{subequations}
\begin{align}
    & \mathbf{I}_{\text{sphere}} = \frac{2 m}{5} 
    \begin{pmatrix}
        r^2 & 0 & 0 \\
        0 & r^2 & 0 \\
        0 & 0 & r^2 \\
    \end{pmatrix}, \\[1em]
    & \mathbf{I}_{\text{ellipsoid}} = \frac{m}{5}
    \begin{pmatrix}
        r_y^2 + r_z^2 & 0 & 0 \\
        0 & r_x^2 + r_z^2 & 0 \\
        0 & 0 & r_x^2 + r_y^2 \\
    \end{pmatrix},
\end{align}
\end{subequations}
where $r$ is the radius of the sphere, and $(r_x, r_y, r_z)$ are the semi-axes of the ellipsoid.

\subsection{The lattice Boltzmann method}
\label{subsec:lbm}

Contrary to standard CFD solvers which explicitly work with macroscopic quantities, LBM focuses on the evolution of $V$ velocity distribution functions $f_i(\vv{r},t)$ with $i\in\{1,...,V\}$. This mesoscopic quantity represents the number of \emph{fictitious} particles at a point $(\vv{r},t)$ with a given velocity $\vv{c_i}$. In that context, macroscopic quantities are byproducts of $f_i(\vv{r},t)$, and they can be computed as its statistical moments. As an example, density $\rho$ and momentum $\rho\vv{u}$ are computed as
\begin{subequations}
\begin{align}
    \rho(\vv{r}, t) &= \sum_i f_i(\vv{r}, t), \label{eq:rho} \\
    \rho(\vv{r}, t)\, \vv{u}(\vv{r}, t) &= \sum_i f_i(\vv{r}, t) \, \vv{c_i}. \label{eq:momentum}
\end{align}
\end{subequations}

The choice of discrete velocities $\vv{c_i}$ is particularly important for the LBM as it dictates its macroscopic behavior~\cite{SHAN_JFM_550_2006}. To simulate the physics of \emph{isothermal and weakly compressible flows} with an LBM, it is sufficient to consider the D3Q19 lattice which is a 3D model composed of 19 discrete velocities~\cite{kruger_lattice_2017}. 
D3Q19-LBMs then solve a coupled set of 19 lattice Boltzmann equations: $\forall i \in \{1,\ldots,19\},$
\begin{equation}\label{eq:collAndStream}
f_i(\vv{r}+\vv{c_i}\Delta t, t+\Delta t)=\big[f_i + \Omega_i\big](\vv{r}, t),
\end{equation}
In practice, solving each equation involves two successive steps~\cite{DELLAR_CMA_65_2013}:
\begin{align}
\mathrm{Collision:}\quad &f_i^*(\vv{r}, t)=\big[f_i + \Omega_i\big](\vv{r}, t),\\ 
\mathrm{Streaming:}\quad &f_i\left(\vv{r}+\vv{c_i} \Delta t, t+\Delta t \right)=f_i^*(\vv{r}, t),
\end{align}
where $f^*_i$ represents the post-collision populations, and $\Omega_i$ is the collision model in its general form. These steps collectively form the ``Collide and Stream'' numerical scheme, well-recognized for its efficiency~\cite{SCHORNBAUM_SIAM_38_2016,LATT_PLOSONE_16_2021} and accuracy~\cite{MARIE_JCP_228_2009,SUSS_CF_257_2023}.

The most common collision model considered in the LB literature is the BGK operator, which was named after its authors Bhatnagar, Gross, and Krook~\cite{bhatnagar_model_1954}. It corresponds to a simple relaxation process that makes $f_i(\vv{r}, t)$ tend towards an equilibrium state $f^{eq}_{i}(\rho, \vv{u})$. It reads
\begin{equation}\label{eq:BGK}
    \Omega^{\mathrm{BGK}}_i = -\frac{\Delta t}{\tau} \big[f_i(\vv{r}, t) - f^{eq}_{i}(\rho, \vv{u})\big],
\end{equation}
where the relaxation time is
\begin{equation}
\tau/\Delta t = \nu / c_s^2 + 1/2\label{eq:tau}
\end{equation}
$\nu$ is the kinematic viscosity of the fluid, and $c_s/(\Delta x / \Delta t)= 1/3$ is the speed of sound of the D3Q19 lattice~\cite{kruger_lattice_2017}. The equilibrium state $f^{eq}_{i}(\rho, \vv{u})$ is an Hermite polynomial expansion of the Maxwellian, that is commonly truncated to second-order for isothermal and weakly compressible LBM~\cite{SHAN_JFM_550_2006}:
\begin{equation}\label{eq:equilibrium}
f^{eq}_{i}(\rho, \vv{u}) = \rho t_i\left[1 + \dfrac{\vv{u}\cdot\vv{c_i}}{c^2_s} + \dfrac{(\vv{u}\cdot\vv{c_i})^2}{2c^4_s}-\dfrac{u^2}{2c^2_s}\right].
\end{equation}
$t_i$ are the weights of the D3Q19 lattice that can be found in any textbook~\cite{kruger_lattice_2017}.

Since the advent of LBM, a large number of collision terms $\Omega_i$ have been proposed~\cite{COREIXAS_PRE_100_2019}. Some of them target an improved stability in the low viscosity regime~\cite{COREIXAS_RSTA_378_2020}, while others are dedicated to increasing the accuracy of LBMs for under-resolved low-Reynolds simulations~\cite{GINZBURG_CCP_3_2008a,DHUMIERE_CMA_58_2009}. The latter approach, commonly referred to as two-relaxation-time (TRT) LBM, introduces a set of two parameters to independently control the kinematic viscosity and the accuracy of the model. In practice, this is achieved by considering two temporary distributions 
$(f^+_i,f^-_i)$ where
\begin{equation}\label{eq:trt_populations}
    f^+_i = 0.5 (f_i + f_{\bar{i}})\quad \text{and}\quad  f^-_i = 0.5 (f_i - f_{\bar{i}}).
\end{equation}
$\bar{i}$ is defined as the opposite direction of $i$ so that $\vv{c_{\bar{i}}}=-\vv{c_i}$. A temporary pair of equilibrium states $(f^{eq+}_i,f^{eq-}_i)$ following the same rules~(\ref{eq:trt_populations}) is also introduced. One eventually ends up with the TRT collision model:
\begin{equation}\label{eq:TRT}
\begin{split}
    \Omega^{\mathrm{TRT}}_i = 
    & -\frac{\Delta t}{\tau^+} \big[f^+_i(\vv{r}, t) - f^{eq+}_{i}(\rho, \vv{u})\big] \\
    & -\frac{\Delta t}{\tau^-} \big[f^-_i(\vv{r}, t) - f^{eq-}_{i}(\rho, \vv{u})\big].
\end{split}
\end{equation}
In this work, $\tau^+=\tau$ controls the kinematic viscosity whereas $\tau^-$ is fixed to 
\begin{equation}
\frac{\Delta t}{\tau^-} = \frac{8 (2 - \Delta t/\tau^+)}{8 - \Delta t/\tau^+}    
\end{equation}
for an improved accuracy in under-resolved and low Reynolds conditions~\cite{PAN_CF_35_2006}.

In LEDDS, both BGK and TRT collision models have been implemented, but additional models can be easily included through code copy/paste from the STLBM framework. For the purposes of this study, we focus exclusively on the TRT model, as it is particularly well suited for low-Reynolds-number simulations of fluid-particle interactions.

Eventually, Dirichlet boundary conditions are enforced using the standard and velocity-modified bounce-back schemes~\cite{LADD_JFM_271_1994a}. The initial macroscopic fields are prescribed by setting the distribution functions to their corresponding equilibrium values~(\ref{eq:equilibrium}).

In most simulations, the physical setup is specified through a characteristic length in physical unit $L$, the Reynold number $Re$, and a reference velocity in physical unit $u$. The numerical parameters are fixed by the choice of the characteristic length in lattice unit $L_{lb}$ and the velocity in lattice unit $u_{lb}$, from which LEDDS computes the space discretization $\Delta x = \frac{L}{L_{lb}}$ and discrete timestep $\Delta t = \Delta x\,\frac{u_{lb}}{u}$. The relaxation time $\tau$ is computed from the kinematic viscosity $\nu = u \frac{L}{Re}$ according to Eq.~(\ref{eq:tau}). 

In practice, the time step of the fluid $\Delta_t$ is chosen to match stability and accuracy constraints, through the choice of $u_{lb}$. The DEM solver can run with the same time, which sometimes however leads to numerical instability. In this case, we resort to substepping, where the particle time step $\Delta_{tp}$ is smaller than $\Delta_t$, and the ratio $\Delta_t/\Delta_{tp}$ is integer.

\subsection{Fluid-solid coupling through partially saturated cells}
\label{subsec:coupling}

To model fluid-solid interactions, we adopt a \emph{two-way coupling} approach, where the fluid influences the solids and the solids in turn influence the fluid. Our study focuses on the (fully) resolved case, in which particle diameters are several times larger than the fluid cell size. 

In this work, no explicit tagging of cells as fluid or solid is performed. Instead, for cells intersected by a particle, we compute the fraction of solid volume contained within each fluid cell. This quantity provides a continuous measure of the local fluid–solid overlap and allows for a smooth representation of the interface on the Cartesian grid.
This approach avoids a sharp classification of cells and instead relies on a local solid fraction to account for partially occupied cells, offering a good compromise between accuracy, computational efficiency, and ease of implementation. This method, known in the LB community as the \emph{partially saturated method} (PSM), is closely related to the \emph{volume penalization method} widely used in CFD~\cite{NGUYEN2021225}.

In the PSM, each lattice cell may be partially occupied by one or more solids, and the local collision operator is adjusted according to the cell’s solid fraction. This effectively creates a penalization-like blending between fluid and solid behavior, by enforcing the no-slip condition through a smooth transition across the interface~\cite{noble_lattice-Boltzmann_1998}. The inclusion of partially saturated cells modifies the standard LBM collision step as follows
\begin{equation}
f_i^*(\vv{r}, t)=\left[f_i + \Omega^{\mathrm{PSM}}_i\right](\vv{r}, t),\label{eq:psm1}
\end{equation}
where
\begin{equation}
    \Omega^{\mathrm{PSM}}_i = \left( 1 - B_{tot} \right) \Omega_i^{\mathrm{TRT}} + \sum_k B_k \Omega^s_{i,k}\label{eq:lbePS}
\end{equation}
for a fluid collision operator based on the TRT approach. Here, $\Omega^s_{i,k}$ is the \emph{solid collision operator} that accounts for the influence of solid particle $k$ on the local fluid cell. The weighting factor $B_k$ depends on the volume fraction of the cell occupied by solid $k$, while $B_{\mathrm{tot}} = \sum_k B_k$ represents the cumulative contribution of all solids present in the cell. 

The solid collision operator is expressed as~\cite{noble_lattice-Boltzmann_1998, tsigginos_coupled_2022}:
\begin{align}
    \Omega_{i,k}^s = \left[ f_{\Bar{i}}(\vv{r}, t) - f^{eq}_{\Bar{i}}(\rho, \vv{u}) \right] - \left[ f_{i}(\vv{r}, t) - f^{eq}_{i}(\rho, \vv{u_k}) \right],
\end{align}
where $\vv{u_k}$ is the surface velocity of the solid $k$ \cite{noble_lattice-Boltzmann_1998, tsigginos_coupled_2022} and is given by:
\begin{align}
    \vv{u_k} = \vv{v_k} - \vv{\omega_k} \cross (\vv{r_j} - \vv{r_k}).
\end{align}
$\vv{r_k}$ is the center of mass of the solid $k$, while $\vv{\omega_k}$ is the angular velocity of the sphere, and $\vv{r_j}$ is the position of the fluid cell $j$.
The coefficient $B_k$ is given by:
\begin{align}
B_k = \frac{\beta_k (\tau - \frac{1}{2})}{(1 - \beta_k) + (\tau - \frac{1}{2})},\label{eq:bk}
\end{align}
where $\beta_k$ is the the so-called solid fraction (which is the proportion of the cell occupied by the solid $k$). 

When multiple solids collide, they may overlap within the same fluid cell, potentially resulting in a total value of $B_{tot}$ greater than one. In our framework, we address this challenge by employing a straightforward solution, which is truncating this sum to one:
\begin{align}
    & B_{tot} = \min \left( \sum_k B_k, 1 \right).\label{eq:btot}
\end{align}
Interestingly, when a fluid cell contains no solid, $\beta_k$ equals zero, and the PSM formulation reverts to the standard ``Collide and Stream'' algorithm.

Ensuring a smooth evolution of the solid fraction is crucial to avoid accuracy and stability issues. In practice, there are different ways to compute that quantity (see~Ref.\cite{jones_fast_2017} for more details). In LEDDS, two of these methods are implemented. The first method, called the \emph{subdivision method}, involves picking a certain number of points spread evenly inside the cell. We then calculate the ratio between the total number of points in the cell and the number of points in the solid:
\begin{align}
    & \beta_k = \frac{1}{N} \sum_{i = 1}^N \mathbbm{1}_{\norm{\vv{r_j} - \vv{s_i}} < R_k},
\end{align}
where $s_i$ are points taken at known locations and uniformly distributed inside a fluid cell $j$, $\vv{r_k}$ is the position of the sphere $k$, $N$ is the total number of points, and $R_k$ is the radius of the solid sphere. This method ensures a smooth evolution of the solid fraction as the solid moves from one fluid cell to another. However, it can also be computationally intensive if the number of points inside a fluid cell is too high. Imposing 10 points per direction ($N=1000$) was found to be a good tradeoff, but the user can adjust this parameter if needed.
The second approach, referred to as the \emph{distance-based method}, relies solely on a geometric distance computation~\cite{KEMMLER_IJHPCA_39_2025}. The solid fraction contribution of particle $k$ to cell $j$ is estimated as
\begin{align}
    \label{eq:naiveSolidFrac}
    & \beta_k = \begin{cases}
        0 \quad & \text{ if } \quad \delta_{sc} < 0\\
        1 \quad & \text{ if } \quad \delta_{sc} > 2 R_c\\
        \frac{\delta_{sc}}{2 R_c} \quad & \text{ else }
    \end{cases},\\
    & \delta_{sc} = R_k + R_c - \norm{\vv{r_k} - \vv{r_j}}, \qquad R_c = \Dx \sqrt[3]{\frac{3}{4 \pi}},
\end{align}
\noindent where $\vv{r_j}$ is the centered position of the fluid cell $j$. The calculation of $\delta_{sc}$ is actually just the overlap distance between the solid sphere $k$ and a sphere of volume $\Dx^3$ centered at the cell's center. 
Alternative distance-based formulations exist. For example, Jones et al.~\cite{jones_fast_2017} proposed a fast approximation based on a purely linear solid-fraction model using precomputed coefficients that depend only on the particle radius, demonstrating good accuracy for spherical particles. However, such an approach is not considered here because extending it to ellipsoidal particles requires significant modifications of the original, as the distance relation must account for particle orientation. In this work, the subdivision method is therefore used for physical validation (Section~\ref{sec:validation}), while the distance-based approach serves primarily for performance comparison with the state-of-the-art solver waLBerla, which relies on the same strategy~\cite{KEMMLER_IJHPCA_39_2025} (see Section~\ref{sec:perfo}).

It is interesting to note that other approaches for the distance computation exist, such as Ref.~\cite{jones_fast_2017}, which showcases good results by approximating the solid fraction for spheres purely linearly, relying on a precomputed term dependent on the radius of the spheres. This methodology is however not explored in the present work, as it would be less straightforward to extend to ellipsoidal particles, where the distance also depends on the orientation of the particles. Instead, this work showcase the subdivision method for physical validation, whereas the distance-based approach is used for performance comparison with state-of-the-art solver waLBerla which also relies on this approach.

Finally, the forces and torques exerted by the fluid on a solid $k$ can also be linked to the coefficient $B_k$~\cite{noble_lattice-Boltzmann_1998, seil_lbdemcoupling_2016}:
\begin{align}
    & \vv{F_{sf}} = -\frac{\Dx^3}{\Dt} \sum_j \left[ B_k(\vv{r_j}) \sum_i \Omega_{k,i}^{s}(\vv{r_j}) \vv{c_i} \right] \label{eq:forceSF},\\
    & \vv{M_{sf}} = -\frac{\Dx^3}{\Dt} \sum_j \left[ B_k(\vv{r_j}) \cdot \left( (\vv{r_j} - \vv{r_k}) \times \sum_i \Omega_{k,i}^{s}(\vv{r_j}) \vv{c_i} \right) \right] \label{eq:torqueSF},
\end{align}
where $j$ is the index of the cells (partially) contained inside the solid, and $\vv{c_i}$ are the velocities of the lattice. $\Omega_{k,i}^{s}(\vv{r_j})$ is the $i$th component of the solid collision term corresponding to the $k$th solid inside the cell $j$.
\section{Portable implementation strategy\label{sec:portable_implementation}}

\begin{table*}[t]
\centering
\renewcommand{\arraystretch}{0.9}
\setlength{\tabcolsep}{5pt}
\begin{tabular*}{\textwidth}{@{\extracolsep{\fill}} l l l c l @{}}
\toprule
\textbf{Task / Step} & \textbf{Primitive}
& \textbf{Size} & \textbf{Further details} \\
\midrule

{\textbf{Particle advancement}}&&&Section~\ref{subsubsec:grid} \\
Integrate (Velocity-Verlet)            & Map               
& $N_\text{part}$  & Eq.~(\ref{eq:advance-r}) to~(\ref{eq:advance-omega}) \\
Update cell-linked list                      & Map               
& $N_\text{grid}$  &  \\
Compute number of neighbors      & Map               
& $N_\text{part}$  &  \\[3pt]
\arrayrulecolor{lightgray}\midrule\arrayrulecolor{black}

{\textbf{Collision list construction}}&&&Section~\ref{subsubsec:grid} \\
Compute offset of collision pairs & Prefix Sum        
& $N_\text{grid}$  &  \\
Create list of collision pairs             & Map               
& $N_\text{grid}$  & See Fig.~\ref{fig:collision-scheme}(b) \\
Make collisions adjacent          & Sort              
& $N_\text{grid}$  & See Fig.~\ref{fig:collision-scheme}(c) \\
Remove duplicates                 & Unique            
& $N_\text{grid}$  & \\[3pt]
\arrayrulecolor{lightgray}\midrule\arrayrulecolor{black}

{\textbf{Compute DEM forces}}&&&Section~\ref{subsubsec:DEMforces} \\
Compute force and torque          & Map               
& $N_\text{coll}$  & Eqs.~(\ref{eq:spring_dashpot_easy}) and~(\ref{eq:torque}) \\
Update tangential displacement    & Map               
& $N_\text{coll}$  & Eq.~(\ref{eq:trangential-history}) \\
Compute particle range            & Map               
& $N_\text{coll}$  & See Fig.~\ref{fig:forces-sum}(left) \\[3pt]
\arrayrulecolor{lightgray}\midrule\arrayrulecolor{black}

{\textbf{Reduce DEM forces}}&&&Section~\ref{subsubsec:DEMforces} \\
Left pair-member: reduce force    & Map               
& $N_\text{part}$  & See Fig.~\ref{fig:forces-sum}(left) \\
Right pair-member: sort           & Sort              
& $N_\text{coll}$  & See Fig.~\ref{fig:forces-sum}(right) \\
Right pair-member: compute range  & Map               
& $N_\text{coll}$  & See Fig.~\ref{fig:forces-sum}(right)\\
Right pair-member: reduce force   & Map               
& $N_\text{part}$  & See Fig.~\ref{fig:forces-sum}(right)\\[3pt]
\arrayrulecolor{lightgray}\midrule\arrayrulecolor{black}

{\textbf{PSM fluid-particle coupling}}&&&Section~\ref{subsubsec:coupling} \\
Compute solid fractions           & Map               
& $N_\text{fluid}$ & Eq.~(\ref{eq:btot}) \\
Compute hydrodynamic forces       & Map               
& $N_\text{fluid}$ & Eq.~(\ref{eq:forceSF}) \\
LBM collision-streaming           & Map               
& $N_\text{fluid}$ & As in Ref.~\cite{LATT_PLOSONE_16_2021}, with Eqs.~(\ref{eq:psm1}) and~(\ref{eq:lbePS}) \\[3pt]
\arrayrulecolor{lightgray}\midrule\arrayrulecolor{black}

{\textbf{Force reduction}}&&&Section~\ref{subsubsec:coupling} \\
Sort forces per particle          & Sort              
& $N_\text{fluid}$ &  \\
Reduce forces                     & Reduce\_by\_key   
& $N_\text{fluid}$ &  \\

\bottomrule
\end{tabular*}
\caption{
Computational workflow of LEDDS, expressed entirely in terms of algorithmic primitives. 
In practice, LEDDS calls either C++ Standard Library (STL) algorithms or Thrust algorithms for any of these primitives. 
This unified representation covers both LBM and DEM stages without requiring any explicit synchronization or device-specific code.
The table also reports the relevant problem size for each step, denoted by $N_\text{part}$ (number of particles), $N_\text{grid}$ (number of grid cells for the cell-linked list), $N_\text{coll}$ (number of inter-particle collisions), and $N_\text{fluid}$ (number of fluid cells).
}
\label{tab:workflow}
\end{table*}

The main methodological novelty of this work lies in expressing all stages of a coupled LBM–DEM simulation (neighbor search, contact detection, DEM force evaluation, etc) exclusively as compositions of high-level algorithmic primitives. 
This formulation enables GPU acceleration without relying on device-specific kernels, while remaining applicable to non-trivial particle geometries such as ellipsoids, where contact detection and interaction forces depend on particle orientation and gap size. Below, we describe the general strategy behind LEDDS, while emphasizing both portability and performance. 
We first present the high-level design principles, including the use of algorithmic primitives and the structure-of-arrays memory layout, which together ensure clear, maintainable, and fully data-parallel code. 
We then describe the implementation details that realize these principles in practice for DEM and LBM-DEM coupling. 

\subsection{General principles}

\subsubsection{Algorithmic primitives}

The design of LEDDS is based on a simple programming philosophy: all parallelism is expressed exclusively through high-level algorithmic primitives provided by the programming framework. Hence, the user code clearly expresses the intent of the computation, while the compiler and runtime determine how these operations are executed in parallel on the available hardware. This approach keeps the code simple, and maintainable, while avoiding manual handling of concurrency and synchronization. As a result, undefined behavior due to race conditions can be excluded by construction, as explained below.

The concept of \emph{algorithmic primitives} originates from Blelloch~\cite{Blelloch1990}, who demonstrated that a small set of data-parallel operations, such as \emph{map}, \emph{reduce}, and \emph{scan}, forms a complete basis for expressing parallel algorithms. 
Originally developed for vector and PRAM architectures, this abstraction has regained importance with the rise of GPUs and heterogeneous platforms, where explicit data parallelism maps naturally to hardware. A central goal of modern high-performance computing is therefore to express algorithms entirely in terms of such primitives, enabling performance portability across current GPUs and future accelerators. 
According to Blelloch’s taxonomy, three fundamental categories of primitives can be defined:  
(1) \emph{elementwise operations} such as \texttt{map}, \texttt{gather}, and \texttt{scatter};  
(2) \emph{reductions and scans} such as \texttt{reduce}, \texttt{scan}, and \texttt{segmented\_scan}; and  
(3) \emph{permutation and filtering} operations such as \texttt{sort} and \texttt{partition}.  
For simplicity, we also treat \texttt{unique} as a primitive, even though it can be expressed as a combination of \texttt{map}, \texttt{scan}, and \texttt{scatter}.

In LEDDS, the LBM-DEM coupling algorithm is fully formulated in terms of four such primitives: \texttt{map}, \texttt{reduce\_by\_key}, \texttt{unique}, and \texttt{sort}. Table~\ref{tab:workflow} summarizes the entire workflow, showing how these primitives are applied across the main computational steps—from particle advancement and collision detection to DEM force computation and fluid-solid coupling. Most operations rely on standard C++ parallel algorithms, while a few performance-critical ones make use of the Thrust library, which provides optimized GPU implementations. This mixed use of C++ and Thrust strikes a balance between portability and efficiency: C++ primitives ensure maintainable and hardware-agnostic code, while Thrust provides the necessary high-performance support for segmented reductions and sorting.

In this context, each primitive serves a specific role: 
\begin{itemize}[leftmargin=*, itemsep=1pt, topsep=1pt]
    \item \textbf{Map} (C++ STL) \\
    Applies a user-defined operation independently to each element of an array, producing one output element per input element. Implemented using \texttt{std::transform} from the C++ Standard Template Library (STL).

    \item \textbf{Reduce-by-key} (Thrust) \\
    Combines consecutive values sharing the same key using an associative operator, producing one output per unique key. Standard C++ parallel algorithms do not yet support segmented reduction, so LEDDS uses \texttt{thrust::reduce\_by\_key} with logarithmic parallel span.

    \item \textbf{Unique} (C++ STL) \\
    Removes consecutive duplicate elements from a sequence, leaving only the first occurrence of each distinct element. Although it can be constructed from lower-level primitives (prefix-sum \& scatter), it is treated as a standalone operation for clarity. Implemented using \texttt{std::unique}.

    \item \textbf{Sort} (Thrust) \\
    Reorders elements by key so that elements with equal keys are contiguous. LEDDS uses a radix-based GPU implementation (\texttt{thrust::sort\_by\_key}) for linear work and logarithmic span on fixed-width integer or floating-point keys, which is more efficient than the comparison-based \texttt{std::sort} with higher asymptotic cost.
\end{itemize}

To exclude the possibility for race conditions during the application of the \texttt{map} primitive, two common cases must be considered. First, non-local overlapping reads and writes must be avoided; this is typically ensured by separating input and output arrays, as in the LBM two-population collision-streaming kernel. Second, when reductions or data reordering occur at the write location, a PULL-style parallel processing approach is employed to ensure thread safety, for example when constructing particle and collision lists (Figure~\ref{fig:collision-scheme} a-b)

From a performance standpoint, primitives with logarithmic parallel span, such as \texttt{reduce\_by\_key}, \texttt{unique}, and \texttt{sort}, are effectively limited by memory bandwidth rather than computation. On GPUs, global memory access dominates runtime, so the theoretical $\mathcal{O}(\log N)$ span does not introduce significant latency. This is particularly true for radix-based sorting, whose observed scaling is essentially linear for data size of typical LEDDS simulation runs; a comparison-based implementation would make the logarithmic factor more apparent. Consequently, all primitives in LEDDS are fully data-parallel and warrant portability of performance across platforms.

\subsubsection{Data structure}

LEDDS primarily employs a structure-of-arrays (SoA) data layout, where each field is stored in a separate array. 
The only exception concerns vector quantities (e.g., position, velocity, and force) whose three components are kept contiguous in memory. 
The SoA layout promotes coalesced memory access and often yields significantly higher performance, for instance when applied to the memory alignment of LBM populations for efficient collision-streaming execution~\cite{LATT_PLOSONE_16_2021}.
It also integrates naturally with the use of algorithmic primitives, since operations such as sorting or reduction can easily exchange one key field for another without restructuring the associated value arrays.

The persistent arrays representing the state of the system at each time step are:

\begin{itemize}[leftmargin=*, itemsep=1pt, topsep=1pt]
    \item \textbf{LB populations:} Two 19-element fields in SoA layout representing the D3Q19 ``two-population'' model for the fluid (used by LBM primitives in Table~\ref{tab:workflow}).
    \item \textbf{Particles:} Two three-component arrays storing particle positions and velocities.
    \item \textbf{Collision pairs:} Each pair identifies a potential contact between two particles. Stored explicitly to index per-contact quantities such as tangential displacements.
    \item \textbf{Lateral displacements:} Track tangential offsets over the lifetime of each contact and are used to compute frictional forces in the DEM model.
\end{itemize}

\noindent In addition to these persistent arrays, LEDDS relies on temporary buffers during specific algorithmic steps:

\begin{itemize}[leftmargin=*, itemsep=1pt, topsep=1pt]
    \item \textbf{Grid for cell-linked list:} The simulation domain is discretized into a uniform grid. Each cell stores a fixed-size list accommodating the maximum number of particles overlapping the cell.
    \item \textbf{DEM forces computation:} Temporary arrays store pairwise contact forces and torques before reduction to per-particle values using \texttt{map} and \texttt{reduce\_by\_key} primitives.
    \item \textbf{Fluid forces computation:} Hydrodynamic forces are evaluated on fluid-solid interface cells and subsequently reduced to per-particle values using the \texttt{reduce\_by\_key} primitive.
\end{itemize}

\noindent Throughout the workflow (Table~\ref{tab:workflow}), the following symbols denote the relevant problem sizes:
\begin{itemize}[leftmargin=*, itemsep=0pt, topsep=1pt]
\item \textbf{$N_\text{fluid}$} Number of cells in the fluid grid.
\item \textbf{$N_\text{grid}$} Number of cells in the cell-linked-list grid.
\item \textbf{$N_\text{part}$} Instantaneous number of particles.
\item \textbf{$N_\text{coll}$} Instantaneous number of inter-particle collisions.
\end{itemize}
Although per-cell operations over the fluid and grid cells share the same asymptotic complexity, 
$\mathcal{O}(N_\text{fluid}) = \mathcal{O}(N_\text{grid})$, 
the number of grid cells $N_\text{grid}$ is typically one to two orders of magnitude smaller than $N_\text{fluid}$. 
Consequently, operations scaling with $N_\text{fluid}$, such as those on LBM populations, tend to dominate the overall computational cost.

\subsection{Implementation of DEM and LBM-DEM coupling}

Hereafter, we focus on the different steps required by LEDDS to update the position of particles, compute the DEM forces, and the fluid-particle interactions. The implementation of the fluid update is detailed in Ref.~\cite{LATT_PLOSONE_16_2021} and is not repeated here. As shown in Figure~\ref{fig:flowchart}, the algorithm for particle progression and interaction requires three main stages: (1) Update of the particles and of the uniform grid / cell-linked list which is used to identify potential collisions; (2) The computation of particle-particle interactions and update of corresponding tangential displacements of particles; (3) The LB update of fluid cells through the PSM, and the computation of particle-fluid interactions. In case of pure DEM simulations, the algorithm reduces to steps (1) and (2). While in principle some stages could be executed asynchronously for fluid cells far from any particle, in practice the strict data dependencies between particle updates, collision detection, and fluid-particle coupling require a serialized timestep-level execution to ensure good GPU performance. The different algorithms composing these steps are summarized in Table~\ref{tab:workflow} and are explained below.
For additional details regarding the memory overhead associated with the coupling, the reader is referred to~\ref{app:gpu_memory_usage}, where a detailed analysis is provided.

\begin{figure}[hbt]
\centering
\resizebox{0.45\textwidth}{!}{
\begin{tikzpicture}
  [
    start chain=p going below,
    every on chain/.append style={etape},
    every join/.append style={line},
    node distance=0.5 and -.25,
    materia/.style={draw, fill=white, text width=6.0em, text centered, minimum height=1.5em, font=\tiny},
    etape/.style={materia, text width=10em, minimum width=5em, minimum height=1em, rounded corners, font=\tiny},
    linepart/.style={draw, thick, color=black!50, -LaTeX, dashed},
    line/.style={draw, thin, color=black!80, -latex},
    ur/.style={draw, text centered, minimum height=0.01em},
    back group/.style={rounded corners, draw=black!50, dashed, inner xsep=5pt, inner ysep=5pt}
  ]
  {
    \node [on chain, join] {Update the particles positions};
    
    \node [on chain, join] {Move the particles in the uniform grid};
    \node [on chain, join] {Determine the list of potential collisions};
    
    \node [on chain, join] {Compute the forces for each pair and update the tangential displacement};
    \node [on chain, join] {Reduce the forces on each particle};
    
    \node [on chain, join] {Update fluid cells following the PSM};
    \node [on chain, join] {Reduce the fluid forces on each solid};
    
    \node [on chain, join] {Update the particles velocities};
  }

  \begin{scope}[on background layer]
    \node (sph) [back group, fill=blue!20] [fit=(p-2) (p-3)] {};
    \node (dem) [back group, fill=red!20] [fit=(p-4) (p-5)] {};
    \node (coupling) [back group, fill=green!20] [fit=(p-6) (p-7)] {};
  \end{scope}
  \node (sphName) [right, rotate=-90, anchor=center, yshift=7pt, font=\tiny] at (sph.east) {{\parbox{3cm}{\centering Particles / Grid}}};
  \node (demName) [right, rotate=-90, anchor=center, yshift=7pt, font=\tiny] at (dem.east) {{\parbox{3cm}{\centering DEM forces computation}}};
  \node (couplingName) [right, rotate=-90, anchor=center, yshift=7pt, font=\tiny] at (coupling.east) {\parbox{3cm}{\centering Fluid-particle coupling}};
\end{tikzpicture}
}
\caption{
Overview of the LEDDS workflow from the particle perspective. 
Colored groups highlight the three main stages: particle/grid updates (blue), DEM force computation (red), and fluid-particle coupling (green), corresponding to Sections~\ref{subsubsec:grid}, \ref{subsubsec:DEMforces}, and \ref{subsubsec:coupling}, respectively. In case of pure DEM, the third stage disappears from the workflow.
}
\label{fig:flowchart}
\end{figure}

\subsubsection{Uniform grid and cell-linked list \label{subsubsec:grid}}
During the first stage, particles are advanced in time using a second-order, time-reversible, explicit Velocity-Verlet scheme, following Eqs.~(\ref{eq:advance-r}) trough~(\ref{eq:advance-omega}).
To avoid a $\mathcal{O}(N^2)$ neighbor search, a uniform grid is set up as a spatial acceleration structure for collision detection, commonly referred to as a cell-linked list, in which each grid cell stores the identifiers of nearby particles. 
In our implementation, each cell uses a fixed-size array rather than a dynamic list, since the number of particles per cell is small and bounded. 
Particles are associated with a grid cell when their axis-aligned bounding box (AABB) intersects the corresponding spatial region (see Figure~\ref{fig:uniform_grid}).

\begin{figure}[h]
\centering
\begin{tikzpicture}[scale=0.5]
    [
        box/.style={rectangle,draw=black,thick, minimum size=1cm},
    ]
\coordinate (c1) at (3.8, 5);
\coordinate (c2) at (7.1, 6.9);
\draw[fill=green, opacity=.2, thick] (0, 2) rectangle +(6,6);
\draw[fill=blue, opacity=.2, thick] (4, 4) rectangle +(6,6);
\draw[style=help lines, thin] (-0.5,-0.5) grid[step=2] (10.5,10.5);

\foreach \xtick in {1,3,...,9} {
    \pgfmathsetmacro\result{\xtick * .5 -.5} 
    \node[font=\scriptsize] at (\xtick,-1) {\pgfmathprintnumber{\result}};
}
\foreach \ytick in {1,3,...,9} {
    \pgfmathsetmacro\result{\ytick * .5 -.5} 
    \node[font=\scriptsize] at (-1,\ytick) {\pgfmathprintnumber{\result}};
}

\draw (c1) node[scale=1] {$\boldsymbol{i}$};
\draw (c2) node[scale=1] {$\boldsymbol{j}$};
\draw (-2,5) node[scale=1] {$\boldsymbol{y}$};
\draw (5,-2) node[scale=1] {$\boldsymbol{x}$};
\draw[thick, opacity=.5] (c1) circle (2);
\draw[thick, opacity=.5] (c2) circle (2.5);
\end{tikzpicture}
\caption{Illustration of the 2D uniform grid. Spheres $i$ and $j$ overlap the cells $(2,2)$ and $(2,3)$. Sphere $i$ is considered present in cells $(0,1)$ and $(0,3)$ due to the use of the AABB.}
\label{fig:uniform_grid}
\end{figure}

 The update of the cell-linked list is simplified by the assumption that particles move no more than one grid-cell width per time step. In this case, the updated particles assigned to a grid cell are computed by parsing all particles currently linked to  neighboring grid cells. This allows updating the cell-linked list through a unique traversal of all grid cells. At each cell, data is gathered from neighboring cell and the local neighbor information can be updated  through a thread-safe PULL-operation. During the same memory traversal, the number of collision pairs for each grid cell is computed to prepare the data structures for the subsequent collision-detection stage. 
These steps are summarized in the following extract of Table~\ref{tab:workflow}:
\begin{center}
\begin{tabular*}{\linewidth}{@{\extracolsep{\fill}} l l l @{}}
\toprule
\textbf{Particle Advancement Steps} & \textbf{Primitive} & \textbf{Size} \\
\midrule
Integrate (Velocity-Verlet) & Map & $N_\text{part}$ \\
Update grid                  & Map & $N_\text{grid}$ \\
Compute number of neighbors  & Map & $N_\text{part}$ \\
\bottomrule
\end{tabular*}
\end{center}

After the grid update, potential collision pairs are identified through a sequence of parallel algorithm calls. Through a parallel map over the grid, a list of particle pairs is first generated at every cell by scanning, as before, the neighboring cells in a PULL-operation. To avoid redundant entries, only pairs $(i, j)$ satisfying $i < j$ are considered. The resulting collision list is dynamically allocated on the fly, providing the flexibility to handle highly dynamic simulations without additional parametrization. The storage offsets of the collision pairs are precomputed as the result of a prefix-sum algorithm call applied to the precomputed number of neighbors, which ensures contiguous and efficient memory allocation. Since the same pair may be detected from multiple neighboring cells, the resulting list can contain duplicates. A global lexicographic \verb|sort| followed by a \verb|unique| operation removes these duplicates, yielding a compact and ordered list of unique collision pairs. These steps are summarized in the following extract of Table~\ref{tab:workflow}:
\begin{center}
\begin{tabular*}{\linewidth}{@{\extracolsep{\fill}} l l l @{}}
\toprule
\textbf{Collision List Construction Steps} & \textbf{Primitive} & \textbf{Size} \\
\midrule
Compute offset of collision pairs & Prefix Sum & $N_\text{grid}$\\
Create list of collision pairs & Map & $N_\text{grid}$ \\
Make collisions adjacent & Sort & $N_\text{grid}$ \\
Make collisions unique & Unique & $N_\text{grid}$ \\
\bottomrule
\end{tabular*}
\end{center}

\begin{figure*}
\centering
\footnotesize
\renewcommand{\arraystretch}{1.2}
\setlength{\arrayrulewidth}{1pt}
\begin{subfigure}{.3\textwidth}
\centering

\begin{tikzpicture}
  \def\cw{0.8cm}   
  \def\ch{0.4cm}   

  \begin{scope}[x=\cw, y=-\ch]
    \tikzset{
      tcell/.style={
        draw,
        line width=0.8pt,
        minimum width=\cw,
        minimum height=\ch,
        inner sep=1pt,
        align=center
      },
      hcell/.style={
        font=\bfseries,
        align=center
      },
      gcell/.style={
        draw,
        fill=gray!25,
        line width=0.5pt,
        minimum width=\cw,
        minimum height=\ch,
        inner sep=1pt
      }
    }

    \node[hcell] at (0,0) {$c_0$};
    \node[hcell] at (1,0) {$c_1$};
    \node[hcell] at (2,0) {$c_2$};
    \node[hcell] at (3,0) {$c_3$};
    \node[hcell] at (4,0) {$c_4$};

    \node[tcell] at (0,1) {A};
    \node[tcell] at (1,1) {B};
    \node[tcell] at (2,1) {A};
    \node[tcell] at (3,1) {G};
    \node[tcell] at (4,1) {A};

    \node[tcell] at (0,2) {B};
    \node[tcell] at (1,2) {D};
    \node[tcell] at (2,2) {E};
    \node[gcell] at (3,2) {};
    \node[tcell] at (4,2) {B};

    \node[tcell] at (0,3) {C};
    \node[gcell] at (1,3) {};
    \node[tcell] at (2,3) {F};
    \node[gcell] at (3,3) {};
    \node[gcell] at (4,3) {};

    \node[gcell] at (0,4) {};
    \node[gcell] at (1,4) {};
    \node[tcell] at (2,4) {G};
    \node[gcell] at (3,4) {};
    \node[gcell] at (4,4) {};

    \node[gcell] at (0,5) {};
    \node[gcell] at (1,5) {};
    \node[gcell] at (2,5) {};
    \node[gcell] at (3,5) {};
    \node[gcell] at (4,5) {};
    \node[gcell] at (0,6) {};
    \node[gcell] at (1,6) {};
    \node[gcell] at (2,6) {};
    \node[gcell] at (3,6) {};
    \node[gcell] at (4,6) {};
  \end{scope}
\end{tikzpicture}

\caption{\parbox{0.85\linewidth}{\centering
List of solids (capital letters) possibly overlapping on a grid cell (column $c_j$).}}
\end{subfigure}%
\begin{subfigure}{.3\textwidth}
\centering

\begin{tikzpicture}
  \def\cw{0.8cm}   
  \def\ch{0.4cm}   

  \begin{scope}[x=\cw, y=-\ch]
    \tikzset{
      tcell/.style={
        draw,
        line width=0.8pt,
        minimum width=\cw,
        minimum height=\ch,
        inner sep=1pt,
        align=center
      },
      hcell/.style={
        font=\bfseries,
        align=center
      }
    }

    \node[hcell] at (0,0) {$c_0$};
    \node[hcell] at (1,0) {$c_1$};
    \node[hcell] at (2,0) {$c_2$};
    \node[hcell] at (3,0) {$c_3$};
    \node[hcell] at (4,0) {$c_4$};

    \node[tcell] at (0,1) {(A,B)};
    \node[tcell] at (1,1) {(B,D)};
    \node[tcell] at (2,1) {(A,E)};
    \node[tcell] at (4,1) {(A,B)};

    \draw[line width=0.8pt]
      ([xshift=-0.5*\cw,yshift=-0.5*\ch]3,0) --
      ([xshift= 0.5*\cw,yshift=-0.5*\ch]3,0);

    \node[tcell] at (0,2) {(A,C)};
    \node[tcell] at (2,2) {(A,F)};

    \node[tcell] at (0,3) {(B,C)};
    \node[tcell] at (2,3) {(A,G)};

    \node[tcell] at (2,4) {(E,F)};

    \node[tcell] at (2,5) {(E,G)};

    \node[tcell] at (2,6) {(F,G)};
  \end{scope}
\end{tikzpicture}

\caption{\parbox{0.9\linewidth}{\centering
List of possible collisions occurring on a given grid cell $c_j$.}}
\end{subfigure}%
\begin{subfigure}{.3\textwidth}
\centering

\begin{tikzpicture}
  \def\cw{0.8cm}   
  \def\ch{0.4cm}  

  \begin{scope}[x=\cw, y=-\ch]
    \tikzset{
      tcell/.style={
        draw,
        line width=0.8pt,
        minimum width=\cw,
        minimum height=\ch,
        inner sep=1pt,
        align=center
      }
    }

    \node[tcell] at (0,0) {(A,B)};
    \node[tcell] at (1,0) {(A,C)};
    \node[tcell] at (2,0) {(A,E)};
    \node[tcell] at (3,0) {(A,F)};
    \node[tcell] at (4,0) {(A,G)};
    \node[tcell] at (5,0) {(B,C)};
    \node[tcell] at (6,0) {\dots};
  \end{scope}
\end{tikzpicture}

\caption{\parbox{0.9\linewidth}{\centering
A global operation sorts all collision pairs according to the first pair-member and removes duplicates.}}
\end{subfigure}%
\caption{\label{fig:collision-scheme}Three-step procedure for the global computation of all unique collision pairs, illustrated with an example (solids are represented by letters instead of numbers for better readability). Step (a) and (b) are local per-cell operations. Step (c) sorts the collision pairs and makes them unique in a global operation.}
\end{figure*}

\subsubsection{DEM forces computation}\label{subsubsec:DEMforces}
The second stage corresponds to the computation of contact forces and torques~(\ref{eq:dem-forces}) between colliding particles. These terms are first computed for all identified collision pairs before being reduced to per-particle quantities. Note that forces are stored only once for every interaction pairs, as they are equal for two interacting particles owing to the action-reaction principle, while the torque is stored separately for each of the two particles of a collision pair.

Per-collision normal and tangential forces~(\ref{eq:spring_dashpot_easy}), and their associated torque~(\ref{eq:torque}), are determined by the chosen contact model and are evaluated using per-element operations over the collision list. The tangential displacement history~(\ref{eq:trangential-history}) required by frictional models is updated in the same process.
To preserve the tangential displacement history, the displacement from the previous time step is retrieved and updated, when available. This operation is performed in parallel for each collision pair. The lookup is accelerated by exploiting the fact that the collision list from the previous time step is sorted. A sequential binary search is therefore used to efficiently identify the matching pair and copy the associated tangential displacement.
The net force and torque acting on each particle is obtained from the per-collision forces and torques through sorting and range-based accumulation as summarized below: 
\begin{center}
\begin{tabular*}{\linewidth}{@{\extracolsep{\fill}} l l l @{}}
\toprule
\textbf{Compute DEM Forces Steps} & \textbf{Primitive} & \textbf{Size} \\
\midrule
Compute force and torque & Map & $N_\text{coll}$ \\
Update tangential displacement & Map & $N_\text{coll}$ \\[3pt]
\bottomrule
\end{tabular*}
\end{center}

It is noted that the accumulation for the total force and torque acting on a given particle $i$ accounts for both pairs $(i,\cdot)$ in which $i$ appears as first member, and pairs in which $(\cdot, i)$ appears as second member. The former case is easily computed as the collision pairs are already sorted in a way in which first pair-members are memory adjacent. After computing the range of pairs for each particle, these forces are computed through a range-based accumulation (\verb+reduce_by_key+). For the latter case, the pairs are first sorted again according to their second pair-member. A renewed computation of per-particle ranges, followed by a range-based accumulation, concludes the computation of DEM forces.
All these steps are summarized below:
\begin{center}
\begin{tabular*}{\linewidth}{@{\extracolsep{\fill}} l l l @{}}
\toprule
\textbf{Reduce DEM Forces Steps} & \textbf{Primitive} & \textbf{Size} \\
\midrule
Compute particle range & Map & $N_\text{coll}$ \\
Left pair-member -- reduce force & Map & $N_\text{part}$ \\
Right pair-member -- sort & Sort & $N_\text{coll}$ \\
Right pair-member -- compute range & Map & $N_\text{coll}$ \\
Right pair-member -- reduce force & Map & $N_\text{part}$ \\
\bottomrule
\end{tabular*}
\end{center}

\begin{figure*}[t]
\centering
\scriptsize
\vspace*{0.3cm}
\renewcommand{\arraystretch}{1.5}
\setlength{\arrayrulewidth}{1pt}

\makebox[\textwidth][c]{%

\begin{tikzpicture}[remember picture]
\node[draw, rounded corners=4pt, thick, fill=blue!5, inner sep=5pt] (box1) {%
\begin{minipage}[t]{0.45\textwidth}

{\hskip 1.3cm \textbf{collision pairs} \hskip 0.60cm \textbf{collision forces}} \\
\vspace{0.2cm}
\makebox[\linewidth][c]{%
\hspace{-2.2cm}
\begin{NiceTabular}[hvlines,colortbl-like,name=CF]{c}  
\dots\\
\dots\\
\dots\\
$(i, j)$\\
$(i, k)$\\
$(i, l)$\\
\dots\\
\dots\\
\dots\\
\end{NiceTabular}
\hskip 1.2cm
\begin{NiceTabular}[hvlines,colortbl-like,name=FF]{c}  
\dots\\
\dots\\
\dots\\
$\vv{F_{ij}}$\\
$\vv{F_{ik}}$\\
$\vv{F_{il}}$\\
\dots\\
\dots\\
\dots\\
\end{NiceTabular}
}
\begin{tikzpicture}[remember picture,overlay]
\foreach \r in {4,5,6}{
  \draw[semithick, -latex, black!70, dashed] 
    ([xshift=0.30cm]CF-\r-1.east) -- ([xshift=-0.30cm]FF-\r-1.west);
}

\draw ([xshift=1.3cm]FF-5-1.west) node[scale=1.05, right, text width=2.8cm, name=sumFirst] 
  {Forces applied on $i$ are summed by one thread};
\foreach \r in {4,5,6}{
  \draw[semithick, -latex, black!70, dashed] 
    ([xshift=0.35cm]FF-\r-1.east) -- (sumFirst);
}

\draw[thick, -latex, black] ([xshift=-0.9cm]CF-4-1.west) -- ([xshift=-0.3cm]CF-4-1.west);
\draw[thick, -latex, black] ([xshift=-0.9cm]CF-6-1.west) -- ([xshift=-0.3cm]CF-6-1.west);

\draw ([xshift=-0.8cm]CF-4-1.west) node[scale=1.0,left] {Begin};
\draw ([xshift=-0.8cm]CF-6-1.west) node[scale=1.0,left] {End};
\end{tikzpicture}
\end{minipage}
};
\end{tikzpicture}
\hspace{0.5cm}
\begin{tikzpicture}[remember picture]
\node[draw, rounded corners=4pt, thick, fill=green!5, inner sep=5pt] (box2) {%
\begin{minipage}[t]{0.45\textwidth}

{\hskip 0.85cm \textbf{second pair-members} \hskip 0.3cm \textbf{collision forces}} \\
\vspace{0.2cm}
\makebox[\linewidth][c]{%
\hspace{-2.2cm}
\begin{NiceTabular}[hvlines,colortbl-like,name=CS]{c}  
\dots\\
\dots\\
\dots\\
$(l, c_1)$\\
$(l, c_2)$\\
$(l, c_3)$\\
\dots\\
\dots\\
\dots\\
\end{NiceTabular}
\hskip 1.2cm
\begin{NiceTabular}[hvlines,colortbl-like,name=FS]{c}  
\dots\\
\dots\\
$\vv{F_{il}}$\\
\dots\\
$\vv{F_{jl}}$\\
\dots\\
$\vv{F_{kl}}$\\
\dots\\
\dots\\
\end{NiceTabular}
}
\begin{tikzpicture}[remember picture,overlay]
\draw ([xshift=-0.12cm]FS-3-1.west) node[left,name=c1] {$c_1$};
\draw ([xshift=-0.12cm]FS-5-1.west) node[left,name=c2] {$c_2$};
\draw ([xshift=-0.12cm]FS-7-1.west) node[left,name=c3] {$c_3$};

\foreach \i/\c in {4/c1,5/c2,6/c3}{
  \draw[semithick, -latex, black!70, dashed] 
    ([xshift=0.30cm]CS-\i-1.east) -- ([xshift=0.10cm]\c.west);
}

\draw ([xshift=1.3cm]FS-5-1.west) node[scale=1.05, right, text width=2.8cm, name=sumSecond] 
  {Forces applied on $l$ are summed by one thread};
\foreach \r in {3,5,7}{
  \draw[semithick, -latex, black!70, dashed] 
    ([xshift=0.35cm]FS-\r-1.east) -- (sumSecond);
}

\draw[thick,-latex,black] ([xshift=-0.8cm]CS-4-1.west) -- ([xshift=-0.3cm]CS-4-1.west);
\draw[thick,-latex,black] ([xshift=-0.8cm]CS-6-1.west) -- ([xshift=-0.3cm]CS-6-1.west);

\draw ([xshift=-0.8cm]CS-4-1.west) node[scale=1.0,left] {Begin};
\draw ([xshift=-0.8cm]CS-6-1.west) node[scale=1.0,left] {End};
\end{tikzpicture}
\end{minipage}
};
\end{tikzpicture}%
} 

\caption{\label{fig:forces-sum}The particle-particle forces are computed during a parallel map operation applied to all particles. For the force acting on a given particle, a non-parallel sum over all precomputed pairwise forces in which the particle appears as the left pair-member (left panel) or the right pair-member (right panel) is carried out. The former case is easily computed, as the collision pairs and forces are already sorted according to the first pair-member, as shown in Figure~\ref{fig:collision-scheme}(c). For the latter case, the collision pairs must be sorted anew upon their second pair-member, coupled to their collision-force index, which no longer corresponds to the index of the collision pair. A similar approach applies for the torques.}
\end{figure*}

\subsubsection{Fluid-particle coupling}\label{subsubsec:coupling}
The third stage of the numerical scheme runs the LB fluid solver with partially saturated cells~(\ref{eq:lbePS}) and integrates the fluid forces acting on the particles. Given the updated particle positions, the contribution $B_k$~(\ref{eq:bk}) of individual particles on every fluid cell is computed through an element-wise operation on all fluid cells. To enhance efficiency, an upper bound $k$ is defined for the maximal number of solids that can cut a fluid cell at the same time. In the context of fully resolved particles presented in this article, a value of $k=2$ is chosen. The number of solids per cell is in practice limited to two for fully resolved LBM-DEM, as the particle diameters largely exceed the size of fluid cells. The overall solid fraction $B_{tot}$~(\ref{eq:btot}) is computed from the contributions $B_k$, and the resulting interaction forces~(\ref{eq:forceSF}) and torques~(\ref{eq:torqueSF}) are computed directly for all affected cells.
As a next step, the LBM collision-streaming kernel is executed as a single element-wise operation on all fluid cells, 
following the ``two-population'' LBM implementation described in the STLBM library~\cite{LATT_PLOSONE_16_2021}, here supplemented with the PSM [Eqs.~(\ref{eq:psm1}) and~(\ref{eq:lbePS}]. While this choice doubles the memory footprint, it greatly simplifies the implementation of the coupling steps without noticeably affecting the performance of the LBM module (see Figure 9.7 of Ref.~\cite{HOLZER_PhD_2025} for a detailed comparison).
These coupling steps are summarized as:
\begin{center}
\begin{tabular*}{\linewidth}{@{\extracolsep{\fill}} l l l @{}}
\toprule
\textbf{PSM Fluid-Particle Coupling Steps} & \textbf{Primitive} & \textbf{Size} \\
\midrule
Compute solid fractions & Map & $N_\text{fluid}$ \\
Compute forces & Map & $N_\text{fluid}$ \\
LBM collision-streaming with PSM & Map & $N_\text{fluid}$ \\[3pt]
\bottomrule
\end{tabular*}
\end{center}

After this, the computed forces are reduced to per-particle quantities through sorting and key-based reduction operations. This is achieved applying a \verb|sort| primitive over all fluid cells, and a range-based accumulation to recover per-particle forces.
\begin{center}
\begin{tabular*}{\linewidth}{@{\extracolsep{\fill}} l l l @{}}
\toprule
\textbf{Force Reduction Steps} & \textbf{Primitive} & \textbf{Size} \\
\midrule
Sort forces per particle & Sort & $N_\text{fluid}$ \\
Reduce forces & Reduce\_by\_key & $N_\text{fluid}$ \\
\bottomrule
\end{tabular*}
\end{center}

With these updated forces, the new acceleration and velocity of particles can now be computed which concludes the current time step.

\section{Validation\label{sec:validation}}

In this section, we systematically assess the accuracy and physical consistency of the LEDDS framework across a hierarchy of test cases. 
We begin by validating the DEM module in collision scenarios of increasing complexity:

\begin{enumerate}
    \item Head-on, perfectly elastic collision between two spheres, verifying energy and momentum conservation;
    \item Eccentric collision between two ellipsoids, confirming correct handling of non-spherical particle dynamics;
    \item Sphere-wall impact in the presence of friction, used to benchmark tangential force modeling; and
    \item Many-body simulation of identical spheres confined in a box, reproducing the statistical equilibrium predicted by kinetic theory for a gas of perfectly elastic particles;
    \item Many-body simulation of ellipsoids forming a sand pile, used to evaluate the angle of repose as a function of aspect ratio;
\end{enumerate}

\noindent followed by two coupled LBM-DEM benchmarks designed to further evaluate the accuracy and robustness of the fluid-particle coupling through PSM:

\begin{enumerate}
    \setcounter{enumi}{4}
    \item Settling of a single particle in a viscous fluid, used to evaluate the dynamic behavior of an isolated particle as it interacts with the surrounding flow field and approaches its terminal settling velocity;
    \item Rotational dynamics of a single ellipsoidal particle in a viscous shear flow, used to validate the computation of hydrodynamic torques and the accurate reproduction of Jeffery orbit dynamics of a non-spherical particle; and
    \item Shear-flow simulation of a densely packed particle suspension, aimed at validating the effective rheology predicted by the coupled model in many-body configurations under steady shear.
\end{enumerate}

Unless otherwise stated, quantities are either dimensionless, like to Reynold number, or expressed in physical units (shown in square brackets). Equivalent values in lattice units are reported when relevant to the numerical setup. 
All simulations are performed using single-precision arithmetic, except for the Jeffery orbit case, which uses double precision due to the very small time step required to keep the relaxation time $\tau$ below 2 --a good practice to ensure accurate LBM results in the Stokes flow regime~\cite{kruger_lattice_2017}.

\subsection{Head-on perfectly elastic collision of spheres (DEM) \label{subsec:sphere_sphere_headon_collision}}

\begin{figure}[b]
    \centering
    \begin{tikzpicture}[
        >=Stealth,
        thick,
        font=\small,
        scale=5,
        lab/.style={font=\fontsize{10}{11}\selectfont},
        partlabel/.style={lab, font=\fontsize{10}{11}\selectfont\bfseries}
    ]

    \definecolor{sphereA}{RGB}{31,119,180}
    \definecolor{sphereB}{RGB}{214,39,40}
    \definecolor{velA}{RGB}{0,0,0}
    \definecolor{velB}{RGB}{0,0,0}
    \definecolor{dimCol}{RGB}{100,100,100}

    \def\xiC{-0.42}  \def\yiC{0}
    \def\xjC{0.42}   \def\yjC{0}
    \def\Ri{0.15}    \def\Rj{0.15}
    \def\vlen{0.24}
    \def\gap{0.02}

    \begin{scope}[shift={(\xiC,\yiC)}]
        \fill[sphereA!5] (0,0) circle [radius=\Ri];
        \draw[sphereA, very thick] (0,0) circle [radius=\Ri];
        \draw[sphereA!60, thin] (0,0) -- (0,\Ri)
            node[midway, left, lab, text=sphereA!100] {$R_i$};
        \fill[sphereA] (0,0) circle [radius=0.01];
        \node[sphereA, above, partlabel] at (0,\Ri+0.06) {Sphere $i$};
        \draw[velA, ultra thick, ->] (0,0) -- ++(\vlen,0)
            node[pos=0.28, above, lab, text=velA] {$\bm{v}_i$};
    \end{scope}
    
    \begin{scope}[shift={(\xjC,\yjC)}]
        \fill[sphereB!5] (0,0) circle [radius=\Rj];
        \draw[sphereB, very thick] (0,0) circle [radius=\Rj];
        \draw[sphereB!60, thin] (0,0) -- (0,\Rj)
            node[midway, right, lab, text=sphereB!100] {$R_j$};
        \fill[sphereB] (0,0) circle [radius=0.01];
        \node[sphereB, above, partlabel] at (0,\Rj+0.06) {Sphere $j$};
        \draw[velB, ultra thick, ->] (0,0) -- ++(-\vlen,0)
            node[pos=0.28, above, lab, text=velB] {$\bm{v}_j$};
    \end{scope}

    \draw[dimCol, thin, <->] (\xiC+\Ri, -0.06) -- (\xjC-\Rj, -0.06)
        node[midway, below, lab, text=dimCol] {10~[$\mu\mathrm{m}$]};

    \end{tikzpicture}
    \caption{Initial configuration of the head-on collision between two identical spheres.
             Sphere~$i$ (blue) and sphere~$j$ (red), both of radius $R_i=R_j=150$~[mm],
             are initially separated by a gap of 10~[$\mu$m] and move toward each other
             with equal and opposite velocities $\bm{v}_i$ and $\bm{v}_j$.}
    \label{fig:sphereHeadOnScheme}
\end{figure}

To validate the integration of the LIGGGHTS DEM module into our framework, we first verify the conservation of momentum and energy during a head-on collision between two identical, frictionless spheres ($\mu = 0$) under perfectly elastic conditions (restitution coefficient $\epsilon = 1$). Energy conservation is assessed by tracking the time evolution of the total mechanical energy, defined as the sum of kinetic and elastic potential energies, throughout the collision.

As seen in Figure~\ref{fig:sphereHeadOnScheme}, the simulated system consists of two identical spheres, labeled $i$ and $j$, with radii $R_i = R_j = 150$ [mm], masses $m_i = m_j = 15.5509$ [Kg], Poisson’s ratios $\upsilon_i = \upsilon_j = 0.4999$, and Young’s moduli $E_i = E_j = 10$ [MPa]. Initially, the spheres are separated by 10 [$\mu$m] and move toward each other with equal and opposite velocities: $v_i(0)=-v_j(0)=0.1$ [m/s]. The simulation uses a time step of $\Delta t = 10^{-4}$ [s] and runs for 300 iterations. 
Single and double precision produce compatible results (see~\ref{app:single_precision}), so only single-precision results are shown.

\begin{figure}[t]
    \centering
    \includegraphics[width=0.9\linewidth]{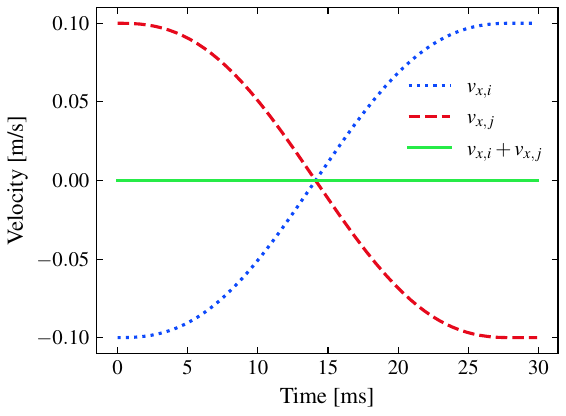} 
    \caption{Head-on elastic collision: Time evolution of the spheres' longitudinal velocities ($v_{x,i}$ and $v_{x,j}$) starting from opposite initial velocities.}
    \label{fig:demtest-velocity}
\end{figure}

Based on the above initial conditions, the total momentum of the system at $t=0$ is then $m_i v_i(0) + m_j v_j(0)=0$. As the simulation progresses, the magnitudes of the particle velocities decrease during the collision until they reach zero (see Figure~\ref{fig:demtest-velocity}). Their velocities then reverse direction and increase in magnitude as the spheres move apart in opposite directions. Clearly, the total momentum remains unaffected by the collision and retains its value of zero throughout the entire simulation.

We can also verify whether the total energy of the system, $E_t = E_{n,{el}} + E_{k,t}$,
is conserved throughout the collision. Here, $E_K(t)$ is the translational kinetic energy of the spheres, and $E_{n,el}$ is their normal elastic potential energy. The translational kinetic energy is given by
\begin{align}
    E_{k,t}(t) = \frac{1}{2} \left[ m_i v_i(t)^2 + m_j v_j(t)^2 \right].
\end{align}
while the normal elastic potential energy is computed from the normal elastic force, which is defined as $f_{n,{el}} = k_n \delta_n$ \eqref{eq:fnScal}. Explicitly, this force is
\begin{equation}
    f_{n,{el}} = k_n \delta_n = \frac{4}{3} Y^* \sqrt{R^*} \, \delta_n^{3/2}, \label{eq:fnel}
\end{equation}
where $Y^*$ and $R^*$ are defined in~\ref{app:material_properties}. Integrating this force over the overlap $\delta_n$ yields the normal elastic potential energy:
\begin{align}
    E_{n,{el}}(\delta_n) &= \int_0^{\delta_n} f_{n,{el}}(x) \, \mathrm{d}x 
    = \frac{4}{3} Y^* \sqrt{R^*} \int_0^{\delta_n} x^{3/2} \, \mathrm{d}x \\
    &= \frac{8}{15} Y^* \sqrt{R^*} \, \delta_n^{5/2}. \label{eq:enel}
\end{align}

\begin{figure}[t]
    \centering
    \includegraphics[width=.9\linewidth] {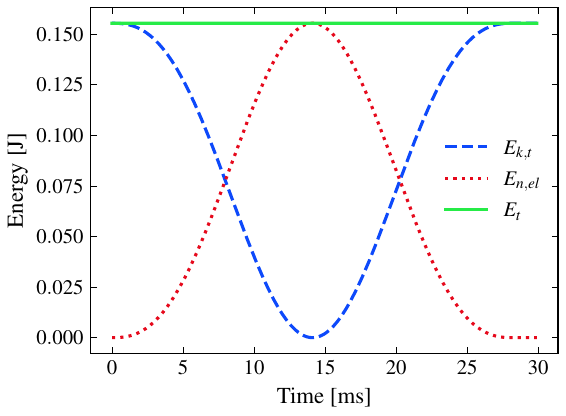}
    \caption{Head-on elastic collision: Time evolution of total $E_t$, translational kinetic $E_{k,t}$, and elastic potential energy $E_{n,el}$ of two identical spheres ($i$ and $j$) with opposite initial velocities.}
    \label{fig:demtest-energy}
\end{figure}
\noindent As illustrated in Figure~\ref{fig:demtest-energy}, the total energy of the system remains constant during the collision. Initially, all the energy in the system is of translational kinetic nature. During the initial phase of the collision, as the spheres approach each other, kinetic energy gradually transforms into elastic potential energy until particle velocities reach zero. At this point, the entire energy of the system is stored as elastic potential energy. Subsequently, the spheres start to separate under the influence of the elastic normal force, and the energy in the form of elastic potential energy transitions back into kinetic energy. This process continues until the spheres are no longer in contact, and the system's energy is solely kinetic as they move away from each other.

\subsection{Eccentric perfectly elastic collision of ellipsoids (DEM) \label{subsec:ellipsoid_ellipsoid_eccentric_collision}}

To further validate the extension of the DEM framework to non-spherical particles, we perform a second test case involving the collision of two \emph{ellipsoids}. This simulation serves two main purposes:
(1) to verify that the model correctly captures the rotational motion resulting from an \emph{eccentric} (off-center) collision, and
(2) to confirm the conservation of total mechanical energy in the absence of dissipative effects.

The physical and numerical parameters are identical to those used in the spherical benchmark, except for the particle shapes, masses, and initial positions. 
The two ellipsoids, labeled $i$ and $j$, have semi-axes
\begin{align*}
    r_i &= (0.45,\, 0.15,\, 0.15)\ \text{[mm]}, \\
    r_j &= (0.15,\, 0.45,\, 0.15)\ \text{[mm]},
\end{align*}
and identical masses $m_i = m_j = 46.6527$ [g]. 
Initially, their centers of mass are positioned at
\begin{align*}
    (x_i, y_i, z_i) &= (0,\, 0.21634,\, 0)\ \text{[mm]}, \\
    (x_j, y_j, z_j) &= (0.58232,\, 0,\, 0)\ \text{[mm]},
\end{align*}
and their initial translational velocities are $v_i(0) = -v_j(0) = 1$ [mm/s] along the $x$-axis (see Figure~\ref{fig:ellipsoidEccentricScheme}). 
This slight vertical offset introduces an \emph{eccentricity} in the impact geometry, causing the line of action of the contact force to no longer pass through the centers of mass. 
As a result, both ellipsoids experience a torque during contact, which initiates rotation. This behavior was absent from the previous simulations because it cannot occur in a perfectly frictionless sphere-sphere collision.

The simulation is carried out with a time step of $\Delta t = 10^{-5}$ [s] over 10000 iterations, which corresponds to a physical time of 100 [ms]. 

The onset of rotational motion is quantified through the rotational kinetic energy:
\begin{equation}
    E_{k,r}(t) = \frac{1}{2} \left( 
    \boldsymbol{\omega}_i^{\mathrm{T}} \mathbf{I}_i \boldsymbol{\omega}_i + 
    \boldsymbol{\omega}_j^{\mathrm{T}} \mathbf{I}_j \boldsymbol{\omega}_j 
    \right),
\end{equation}
where $\mathbf{I}_i$ and $\mathbf{I}_j$ denote the inertia tensors of the two ellipsoids, and $\boldsymbol{\omega}_i$, $\boldsymbol{\omega}_j$ are their respective angular velocity vectors. The total mechanical energy of the system then reads:
\begin{equation}
    E_t = E_{n,{el}} + E_{k,t} + E_{k,r},
\end{equation}
where $E_{n,{el}}$ is the normal elastic potential energy, $E_{k,t}$ the translational kinetic energy, and $E_{k,r}$ the rotational kinetic energy.

\begin{figure}
    \centering
    \begin{tikzpicture}[
        >=Stealth,
        thick,
        font=\small,
        scale=5,
        lab/.style={font=\fontsize{10}{11}\selectfont},
        ellabel/.style={lab, font=\fontsize{10}{11}\selectfont\bfseries}
    ]

    \definecolor{ellipseA}{RGB}{31,119,180}
    \definecolor{ellipseB}{RGB}{214,39,40}
    \definecolor{velA}{RGB}{0,0,0}
    \definecolor{velB}{RGB}{0,0,0}
    \definecolor{dimCol}{RGB}{100,100,100}

    \def\xiC{-0.7}   \def\yiC{0.21634}
    \def\xjC{0.0}    \def\yjC{0}
    \def\aiX{0.45}   \def\aiY{0.15}
    \def\ajX{0.15}   \def\ajY{0.45}
    \def\vlen{0.24}

    \begin{scope}[shift={(\xiC,\yiC)}]
        \fill[ellipseA!5] (0,0) ellipse [x radius=\aiX, y radius=\aiY];
        \draw[ellipseA, very thick] (0,0) ellipse [x radius=\aiX, y radius=\aiY];
        \draw[ellipseA!60, thin] (-\aiX,0) -- (0,0)
            node[midway, below, lab, text=ellipseA!100] {$a_i$};
        \draw[ellipseA!60, thin] (0,0) -- (0,\aiY)
            node[midway, left, lab, text=ellipseA!100] {$b_i$};
        \fill[ellipseA] (0,0) circle [radius=0.01];
        \node[ellipseA, above, ellabel] at (0,\aiY+0.06) {Ellipsoid $i$};
        \draw[velA, ultra thick, ->] (0,0) -- ++(\vlen,0)
            node[midway, above, lab, text=velA] {$\bm{v}_i$};
    \end{scope}

    \begin{scope}[shift={(\xjC,\yjC)}]
        \fill[ellipseB!5] (0,0) ellipse [x radius=\ajX, y radius=\ajY];
        \draw[ellipseB, very thick] (0,0) ellipse [x radius=\ajX, y radius=\ajY];
        \draw[ellipseB!60, thin] (0,0) -- (\ajX,0)
            node[midway, above, lab, text=ellipseB!100] {$a_j$};
        \draw[ellipseB!60, thin] (0,-\ajY) -- (0,0)
            node[midway, right, lab, text=ellipseB!100] {$b_j$};
        \fill[ellipseB] (0,0) circle [radius=0.01];
        \node[ellipseB, above, ellabel] at (0,\ajY+0.06) {Ellipsoid $j$};
        \draw[velB, ultra thick, ->] (0,0) -- ++(-\vlen,0)
            node[near start, below, lab, text=velB] {$\bm{v}_j$};
    \end{scope}

    \draw[dimCol, thin, dashed] (\xiC+\vlen,\yiC) -- (\xjC,\yiC);
    \draw[dimCol, thin, <->] (\xjC,\yjC) -- (\xjC,\yiC)
        node[midway, left, lab, text=dimCol] {$\bm{e}$};

    \end{tikzpicture}
    \caption{Initial configuration of the eccentric ellipsoid collision.
             Ellipsoid~$i$ (blue, semi-axes $a_i=0.45$~[mm], $b_i=0.15$~[mm])
             and ellipsoid~$j$ (red, semi-axes $a_j=0.15$~[mm], $b_j=0.45$~[mm])
             approach each other with equal and opposite velocities $\bm{v}_i$
             and $\bm{v}_j$. The lateral offset $e=0.216$~[mm] between centers
             induces an eccentric impact.}
    \label{fig:ellipsoidEccentricScheme}
\end{figure}

\begin{figure}[t]
    \centering
    \includegraphics[width=.9\linewidth]{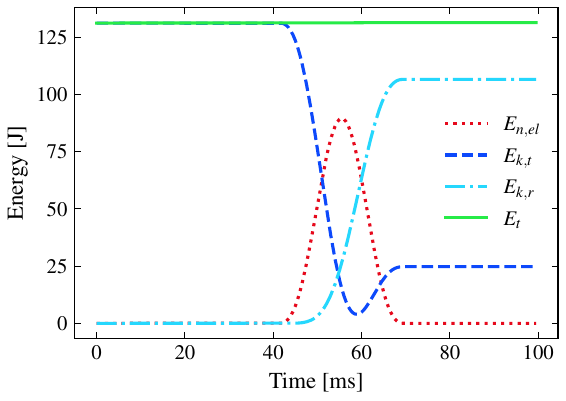}
    \caption{Eccentric collision of two identical ellipsoids: time evolution of the total ($E_t$), translational kinetic ($E_{k,t}$), rotational kinetic ($E_{k,r}$), and elastic potential ($E_{n,{el}}$) energies.}
    \label{fig:demtest-energy-ellipsoid}
\end{figure}

As shown in Figure~\ref{fig:demtest-energy-ellipsoid}, the initial translational kinetic energy of the ellipsoids is partially converted into rotational kinetic energy through the elastic potential energy developed during the collision. 
When the particles first make contact, part of the translational energy is stored elastically while a smaller portion initiates rotational motion. 
As the ellipsoids separate, the potential energy component is progressively reconverted into translational and rotational kinetic energy components. 
The total energy of the system remains nearly constant throughout the simulation, confirming good energy conservation properties of the collision model for non-spherical particles. 

Quantitatively, the relative variation between the initial and final total energies is $\left| E_{t,\mathrm{final}} - E_{t,\mathrm{initial}} \right| / E_{t,\mathrm{initial}} = 0.15\%$.
This small deviation originates from geometric and numerical approximations inherent to the discrete-element formulation for ellipsoidal particles, particularly those related to the estimation of local curvature and contact orientation (see Section~\ref{subsec:dem}).
This interpretation is further supported by repeating the same simulations using single precision (see~\ref{app:single_precision}). The resulting total-energy deviation is 
0.143\%, compared with 0.141\% obtained in double precision, indicating that numerical precision plays only a minor role relative to the geometric and algorithmic approximations discussed above.
To the best of our knowledge, the conservation of energy during collisions of non-spherical particles is rarely addressed in the DEM literature, making this work one of the few to explicitly tackle it.

Overall, the agreement with the theoretical expectations remains very strong, thereby confirming the validity and robustness of the extended collision model for ellipsoids.

\subsection{Sphere-wall collision with friction (DEM)}

\begin{figure}[b]
\centering
\begin{tikzpicture}
    [
        scale=0.8,
        every node/.style={scale=0.8}, 
        box/.style={rectangle,draw=black,thick, minimum size=1cm},
    ]

\coordinate (c1) at (2.5, 3);
\coordinate (c2) at (7.5, 3);
\coordinate (s1) at ($(c1) + (0, -1)$);
\coordinate (s2) at ($(c2) + (0, -1)$);
\coordinate (center) at (5,0);
\draw [style=dashed, color=black!40] (center) -- (5,4);
\draw [style=dashed, color=black!40] (center) -- (c1);
\draw [style=dashed, color=black!40] (center) -- (c2);

\draw[fill=black!10, draw=none]  (0,-0.5) rectangle (10,0);
\draw[pattern={Lines[angle=-45,distance={5pt}]},pattern color=black, draw=none]  (0,-0.5) rectangle (10,0);
\draw[ultra thick] (0,0) -- (10,0);

\draw[] (c1) circle (1);
\draw[] (c2) circle (1);
\node[draw,circle,inner sep=1pt,fill=black] at (c1) {};
\node[draw,circle,inner sep=1pt,fill=black] at (c2) {};
\node[draw,circle,inner sep=1pt,fill=black] at (s1) {};
\node[draw,circle,inner sep=1pt,fill=black] at (s2) {};

\coordinate (cv) at ($(center) + (0,1)$);
\pic [draw, ->, "$\theta_{in}$", angle eccentricity=1.3, angle radius=1cm] {angle = cv--center--c1};
\pic [draw, <-, "$\theta_{out}$", angle eccentricity=1.2, angle radius=1.5cm] {angle = c2--center--cv};

\draw[thick,-latex] (s1) -- +(0,-0.8) node[scale=1.2, pos=.5, below right] {$v_z^{in}$};
\draw[thick,-latex] (s1) -- +(0.8,0) node[scale=1.2, pos=1, right] {$u_x^{in}$};
\draw[thick,-latex] (s2) -- +(0,0.8) node[scale=1.2, pos=.5, above right] {$v_z^{out}$};
\draw[thick,-latex] (s2) -- +(0.8,0) node[scale=1.2, pos=1, right] {$u_x^{out}$};

\draw[-latex] ($(c2) + (0.2,1.2)$) arc (90:0:1) node[scale=1.2, pos=1, above right] {$\omega_{out}$};

\coordinate (axisCenter) at (0.6, 0.3);
\draw[thick,-latex] (axisCenter) -- +(0,0.8) node[scale=1.2, pos=1, above] {$z$};
\draw[thick,-latex] (axisCenter) -- +(0.8,0) node[scale=1.2, pos=1, right] {$x$};
\draw[thick,-latex] (axisCenter) -- +(0.5,0.3) node[scale=1.2, pos=.7, above right] {$y$};

\end{tikzpicture}
\caption{
Setup of the simulation for the test. Before the collision, the sphere is not rotating. The angles $\theta_{in}$ and $\theta_{out}$ are taken relatively to the normal direction of the wall.
}

\label{fig:collisionAngleScheme}
\end{figure}

\begin{figure*}[h!]
    \centering
    \includegraphics[width=\linewidth]{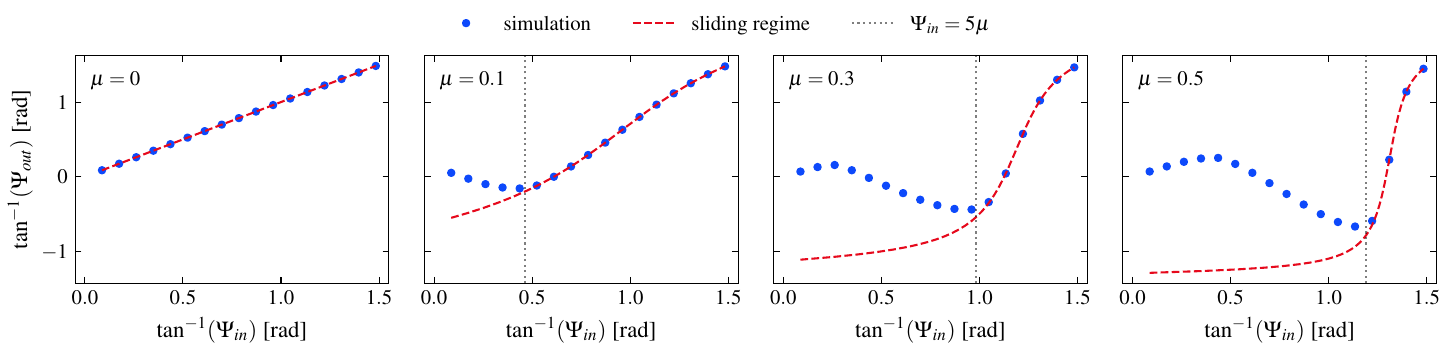}
    \caption{
    Evolution of the post-collision velocity angle ($\tan^{-1}(\Psi_{out})$) as a function of the incident angle ($\tan^{-1}(\Psi_{in})$) for different friction coefficients $\mu$. 
    DEM simulations (dots) are compared with the analytical prediction for the sliding regime (dashed lines), $\Psi_{out} = \Psi_{in} - 7\mu$. 
    The vertical dotted line marks the transition threshold $\Psi_{in} = 5\mu$ beyond which the full sliding hypothesis is correct. 
    }
    \label{fig:demtest-velocityRatio}
\end{figure*}

As a next step, we now target the validation of collisions between a single sphere and a rigid wall in the presence of friction.
To this end, we simulate a sphere approaching a flat wall with a prescribed velocity and incident angle, initially without angular velocity (see Figure~\ref{fig:collisionAngleScheme}). 
Introducing a finite friction coefficient causes part of the translational kinetic energy to be converted into rotational motion. 
Since the collision is perfectly elastic ($\epsilon = 1$), the total energy remains constant, and the translational and rotational energies are simply redistributed by the frictional interaction. 

In practice, we can quantify the effect of friction on energy transfer by tracking the sphere's angular and tangential velocity components \emph{at the point of contact} with the wall. By definition, this tangential velocity reads as
\begin{equation}
u_x = v_x - \omega R,
\end{equation}
where $\omega$ denotes the angular velocity around the $y$-axis and $R$ is the sphere radius. 
We then introduce the pre- and post-collision velocity ratios relative to the incident vertical velocity $v_z^{in}$:
\begin{equation}
\Psi_{in} = - \frac{v_x^{in}}{v_z^{in}}, \quad
\Psi_{out} = - \frac{u_x^{out}}{v_z^{in}}, \quad
u_x^{out} = v_x^{out} - \omega_{out} R,
\end{equation}
where for perfectly elastic collisions $v_z^{in} = -v_z^{out}$. These ratios are used to measure of how friction redistributes translational motion into rotational motion during the collision.

Given the complexity of tangential interactions, we focus on the regime where the sphere slides along the wall \emph{for the entire duration of contact}. In this sliding regime, the tangential force is continuously limited by Coulomb's friction law \eqref{eq:coulombCoeff}. Under this assumption, the post-collision velocity ratio can be expressed analytically as a function of $\Psi_{in}$ and the friction coefficient $\mu$ \cite{chung_benchmark_2011, rettinger_efficient_2020, domenech-carbo_analysis_2019}:
\begin{align}
\Psi_{out} = \Psi_{in} - 7 \mu.
\end{align}
A critical value of $\Psi_{in}$ exists beyond which the sliding regime is reached:
\begin{align}
\Psi_{in} = 5 \mu, \label{eq:psi-startSliding}
\end{align}
while for smaller $\Psi_{in}$, the collision exhibits a mixture of sticking and sliding, which is more difficult to analyze. The interested reader can refer to the detailed theoretical and experimental discussions of these regimes that are provided in Refs.~\cite{domenech-carbo_analysis_2019, foerster_measurements_1994}.

We conducted simulations with friction coefficients, $\mu$, ranging from $0$ to $0.5$, and incident angles from $5\degree$ to $85\degree$ (corresponding to $\arctan(\Psi_\mathrm{in}) \approx 0.087$ to $1.44$ in radians). The post-collision ratio $\Psi_\mathrm{out}$ is plotted in Figure~\ref{fig:demtest-velocityRatio} using $\arctan(\Psi_\mathrm{out})$ for direct comparison with the incident angles.
Globally speaking, the results exhibit excellent agreement with theoretical predictions in the sliding regime and accurately capture the rotational motion induced by friction. Notably, the post-collision angle matches perfectly the incident angle in case of a frictionless wall, i.e., $\Psi_{out}=\Psi_{in}$ for $\mu=0$. 
As expected for $\Psi_\mathrm{in} < 5\,\mu$ and $\mu\neq 0$, systematic deviations occur which reflects the transition to the mixed sticking/sliding regime that is also accurately captured by LEDDS.

\subsection{Many-body validation: Gas-like behavior in a box (DEM) \label{subsec:many_body}}

\begin{figure}[b]
\centering
\begin{tikzpicture}[
    scale=0.72,
    every node/.style={scale=0.8},
    >=Stealth
]

\def\W{3.2}
\def\H{3.2}
\def\dx{0.8}
\def\dy{0.55}
\def\shiftA{0.1}
\def\shiftB{6.5}
\def\r{0.06}

\def\xmin{0.4}
\def\xmax{2.8}
\def\ymin{0.4}
\def\ymax{2.8}
\def\step{0.6}

\pgfmathtruncatemacro{\Nx}{(\xmax-\xmin)/\step}
\pgfmathtruncatemacro{\Ny}{(\ymax-\ymin)/\step}


\newcommand{\drawboxback}[1]{%
    \draw[thick,dashed] ({#1+\dx},\dy) -- ({#1+\W+\dx},\dy);
    \draw[thick] ({#1+\W+\dx},\dy) -- ({#1+\W+\dx},\H+\dy);
    \draw[thick] ({#1+\W+\dx},\H+\dy) -- ({#1+\dx},\H+\dy);
    \draw[thick,dashed] ({#1+\dx},\H+\dy) -- ({#1+\dx},\dy);
    \draw[thick,dashed] (#1,0) -- ({#1+\dx},\dy);
}

\newcommand{\drawboxfront}[1]{%
    \draw[thick] (#1,0) -- ({#1+\W},0);
    \draw[thick] ({#1+\W},0) -- ({#1+\W},\H);
    \draw[thick] ({#1+\W},\H) -- (#1,\H);
    \draw[thick] (#1,\H) -- (#1,0);
    \draw[thick] ({#1+\W},0) -- ({#1+\W+\dx},\dy);
    \draw[thick] ({#1+\W},\H) -- ({#1+\W+\dx},\H+\dy);
    \draw[thick] (#1,\H) -- ({#1+\dx},\H+\dy);
}

\newcommand{\drawlayer}[5]{%
    \foreach \i in {0,...,\Nx} {
        \foreach \j in {0,...,\Ny} {
            \pgfmathsetmacro{\x}{\xmin + \i*\step}
            \pgfmathsetmacro{\y}{\ymin + \j*\step}
            \shade[ball color=#4] ({#1+\shiftA+\x+#2},{\y+#3}) circle (\r);
            \draw[#5, thin] ({#1+\shiftA+\x+#2},{\y+#3}) circle (\r);
        }
    }
}

\drawboxback{0}
\drawboxback{\shiftB}

\drawlayer{0}{0}{0}{blue!25}{blue!60!black}
\drawlayer{0}{0.25*\dx}{0.25*\dy}{red!25}{red!50!black}
\drawlayer{0}{0.50*\dx}{0.50*\dy}{green!25}{green!50!black}
\drawlayer{0}{0.75*\dx}{0.75*\dy}{black!25}{black!50!black}

\draw[thick,-latex] (\W+\dx+0.25,1.9) -- (\shiftB-0.25,1.9)
    node[midway, above] {relaxation};


\foreach \x/\y in {
0.35/0.45, 0.75/1.10, 1.20/0.70, 1.65/1.55, 2.10/0.55,
2.55/1.25, 0.55/2.05, 1.05/2.45, 1.70/2.20, 2.35/2.65,
2.75/1.85, 0.85/1.65, 1.45/0.35, 2.65/0.95, 2.85/2.35,
0.45/2.75, 1.95/2.85, 2.35/0.35, 1.35/1.95, 0.95/0.85,
2.75/1.55, 1.85/1.15, 0.65/1.45, 2.15/2.15, 1.55/2.65
}{
    \shade[ball color=blue!25] ({\shiftB+0.1+1.1*\x},{1.1*\y}) circle (\r);
    \draw[blue!60!black, thin] ({\shiftB+0.1+1.1*\x},{1.1*\y}) circle (\r);
}

\foreach \x/\y in {
0.45/0.75, 0.95/1.55, 1.35/0.45, 1.85/1.95, 2.25/1.05,
2.65/0.65, 0.65/2.35, 1.15/2.75, 1.75/2.05, 2.15/2.45,
2.55/1.55, 0.85/1.15, 1.55/1.65, 2.75/2.75, 2.85/0.45,
0.55/1.75, 1.95/2.85, 2.35/1.85, 1.25/2.25, 0.75/0.55,
2.45/2.15, 1.65/0.95, 0.95/2.05, 2.05/0.35, 1.45/2.55
}{
    \shade[ball color=red!25] ({\shiftB+0.1+1.1*\x},{1.1*\y}) circle (\r);
    \draw[red!50!black, thin] ({\shiftB+0.1+1.1*\x},{1.1*\y}) circle (\r);
}

\foreach \x/\y in {
0.40/1.15, 0.85/0.35, 1.25/1.75, 1.75/0.85, 2.10/1.45,
2.45/0.55, 0.55/2.55, 1.05/2.15, 1.55/2.75, 2.05/2.05,
2.65/2.45, 0.75/1.65, 1.35/0.95, 2.75/1.15, 2.85/2.85,
0.45/2.15, 1.85/2.35, 2.35/0.85, 1.15/2.85, 0.95/0.75,
2.55/1.85, 1.65/1.25, 0.65/1.95, 2.15/2.75, 1.45/0.45
}{
    \shade[ball color=green!25] ({\shiftB+0.1+1.1*\x},{1.1*\y}) circle (\r);
    \draw[green!50!black, thin] ({\shiftB+0.1+1.1*\x},{1.1*\y}) circle (\r);
}

\foreach \x/\y in {
0.35/0.95, 0.95/1.85, 1.45/0.65, 1.95/2.25, 2.35/1.25,
2.75/0.85, 0.55/2.65, 1.05/2.35, 1.65/2.85, 2.15/2.25,
2.55/1.65, 0.85/1.35, 1.55/1.05, 2.85/2.55, 2.65/0.45,
0.45/1.55, 1.85/2.65, 2.25/1.95, 1.25/2.05, 0.75/0.65,
2.45/2.35, 1.75/1.45, 0.95/2.15, 2.05/0.55, 1.35/2.55
}{
    \shade[ball color=black!25] ({\shiftB+0.1+\x+0.7*\dx},{\y+0.7*\dy}) circle (\r);
    \draw[black!50!black, thin] ({\shiftB+0.1+\x+0.7*\dx},{\y+0.7*\dy}) circle (\r);
}

\drawboxfront{0}
\drawboxfront{\shiftB}

\end{tikzpicture}
\caption{Pseudo-3D sketch of the many-body DEM validation. An initially ordered system of identical spheres with initial velocity $\protect\vv{v} = (1,\,1,\,1)$~[mm/s] evolves through perfectly elastic, frictionless collisions toward a Maxwell-Boltzmann equilibrium.}
\label{fig:particlesBoxSetup3D}
\end{figure}

\begin{figure*}[!t]
    \centering
    \includegraphics[width=0.8\linewidth]{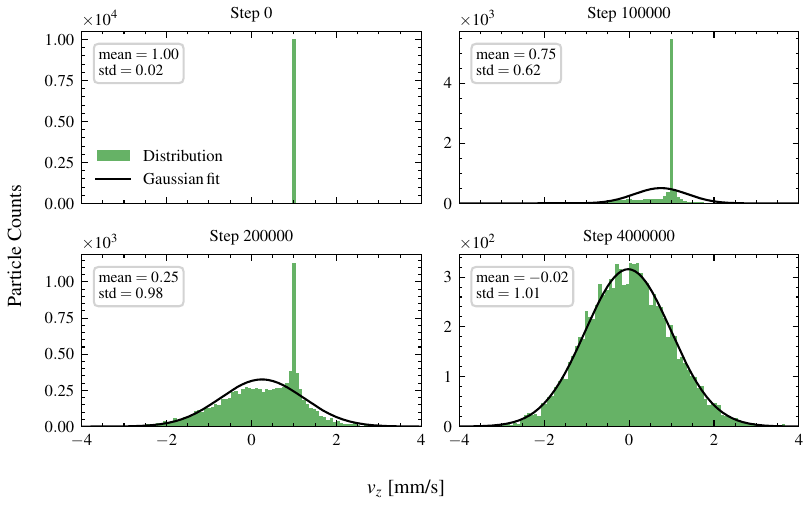}
    \caption{Many-body validation: time evolution of the particle velocity distribution ($v_z$ component) for 10000 elastic and frictionless spheres confined in a box. The distribution converges toward the theoretical Maxwell-Boltzmann form with a mean velocity of zero and a standard deviation of 1~[mm/s].}
    \label{fig:demtest-particlesBox}
\end{figure*}

As a first many-particle validation of the DEM part of LEDDS, we reproduce the collective dynamics of a large ensemble of colliding particles as predicted by the kinetic theory of gases. 
In this test, we simulate a system of $10000$ identical spheres confined within a rigid, four-wall box. 
All particles are initially assigned the same translational velocity vector, $\vv{v} = (1,\, 1,\, 1)$~[mm/s], and no rotational motion. 
Collisions are modeled as perfectly elastic ($\epsilon = 1$) and frictionless ($\mu = 0$), which ensures that the total mechanical energy of the system remains constant and predominantly resides in translational kinetic energy.
Even if there is a brief conversion of kinetic energy into elastic potential energy during contact, it is negligible in practice because the collision timescale is extremely short. As a result, energy is solely exchanged between translational degrees of freedom, much like in a monatomic gas.

When the number of particles is sufficiently large, the frequent collisions lead to momentum exchange among particles, driving the system toward a statistical equilibrium that closely resembles an ideal gas confined in a closed container. Consequently, the velocity components of the particles are expected to follow a Maxwell-Boltzmann distribution centered around zero, reflecting the fact that, in the absence of external energy input, the gas as a whole has no net motion. 

Focusing on the $z$-component of the particle velocity, the probability density function at equilibrium is given by~\cite{BOLTZMANN_1872,BHATNAGAR_PR_94_1954}:
\begin{align}
f(v_z) = \sqrt{\frac{m}{2 \pi k_b T}} \exp \!\left(-\frac{m v_z^2}{2 k_b T}\right),
\end{align}
where $m$ is the particle mass, $k_b$ the Boltzmann constant, and $T$ the absolute temperature of the system. This distribution is Gaussian, with zero mean and variance $\sigma^2 = k_b T / m$.
From the equipartition theorem, the mean kinetic energy per particle relates to the macroscopic temperature of the gas as:
\begin{align}
\frac{3}{2} k_b T = \Bar{E} = \frac{1}{2} m \langle v^2 \rangle,
\end{align}
which leads to
\begin{align}
\sigma^2 = \frac{k_b T}{m} = \frac{2}{3} \frac{\Bar{E}}{m} = \frac{1}{3} \langle v^2 \rangle.
\end{align}
Given the initial condition $\vv{v} = (1, 1, 1)$ [mm/s], the expected standard deviation of a single velocity component is therefore $\sigma = 1$ [mm/s].

The simulation was run for $400000$ iterations with particles of radius $R = 15$ [mm]. Figure~\ref{fig:demtest-particlesBox} shows the time evolution of the velocity distribution along the $z$-axis. The initially monokinetic state progressively broadens as collisions redistribute momentum among all degrees of freedom. The distribution ultimately converges to a Gaussian centered at zero with the predicted variance, confirming that the DEM formulation correctly captures the many-body statistical behavior expected for an idealized gas of perfectly elastic spheres.

\subsection{Many-body validation: Angle of repose (DEM)}

\begin{figure}[hb!]
\centering
\begin{tikzpicture}[
    x={(1cm,0cm)},
    y={(0cm,1cm)},
    z={(-0.35cm,-0.35cm)},
    scale=0.12,
    >=Stealth
]

\def\dX{55} \def\dY{25} \def\dZ{5}
\def\iX{5}  \def\iY{2}  \def\iZ{5}
\def\sH{12} \def\sL{30} 

\colorlet{pileedge}{brown!80!black}

\filldraw[gray!20, draw=gray!50] (0,0,\dZ) -- (0,\dY,\dZ) -- (0,\dY,0) -- (0,0,0) -- cycle;
\filldraw[gray!10, draw=gray!40] (0,0,0) -- (\dX,0,0) -- (\dX,0,\dZ) -- (0,0,\dZ) -- cycle;
\draw[gray!30, dashed] (\dX,0,0) -- (\dX,\dY,0) -- (\dX,\dY,\dZ) -- (\dX,0,\dZ) -- cycle;
\draw[gray!80] (0,0,\dZ) -- (0,\dY,\dZ) -- (0,\dY,0) -- (0,0,0) -- cycle;
\draw[gray!80] (0,0,0) -- (\dX,0,0) -- (\dX,0,\dZ) -- (0,0,\dZ) -- cycle;

\fill[orange!20] (0,0,\dZ) -- (0,\sH,\dZ) -- (\sL,0,\dZ) -- cycle;
\filldraw[orange!30, draw=pileedge, thick]
    (0,\sH,0) -- (0,\sH,\dZ) -- (\sL,0,\dZ) -- (\sL,0,0) -- cycle;
\filldraw[orange!45, draw=pileedge, thick]
    (0,0,\dZ) -- (0,\sH,\dZ) -- (\sL,0,\dZ) -- cycle;

\draw[thick, ->] (16,0,\dZ) arc (0:-40:-8);
\node at (14.,2,\dZ-1.5) {$\theta$};

\node[text=pileedge, font=\small] at (6,2.3,\dZ) {Sand pile};

\coordinate (A)  at (0,\sH+6,0);
\coordinate (B)  at ($(A)+(\iX,0,0)$);
\coordinate (C)  at ($(A)+(0,0,\iZ)$);
\coordinate (D)  at ($(A)+(\iX,0,\iZ)$);
\coordinate (At) at ($(A)+(0,\iY,0)$);
\coordinate (Bt) at ($(B)+(0,\iY,0)$);
\coordinate (Ct) at ($(C)+(0,\iY,0)$);
\coordinate (Dt) at ($(D)+(0,\iY,0)$);

\foreach \x/\y/\z/\a in {
    1.2/\sH+2.8/1.0/25,
    2.0/\sH+3.7/2.2/-10,
    3.0/\sH+2.5/3.4/18,
    1.8/\sH+1.9/4.0/40,
    3.8/\sH+3.2/1.4/-25,
    2.7/\sH+1.4/2.8/8,
    4.2/\sH+2.3/3.9/32,
    1.4/\sH+3.1/3.0/-35
}{
    \begin{scope}[shift={(\x,\y,\z)}, rotate=\a]
        \filldraw[fill=orange!80!black, draw=brown!90!black, line width=0.4pt]
            (0,0) ellipse [x radius=0.35, y radius=0.95];
        \fill[orange!40] (-0.08,0.12) ellipse [x radius=0.22, y radius=0.70];
    \end{scope}
}

\filldraw[fill=blue!12, draw=blue!85!black, line width=0.5pt, opacity=0.55]
    (A) -- (At) -- (Ct) -- (C) -- cycle;
\filldraw[fill=blue!18, draw=blue!85!black, line width=0.5pt, opacity=0.60]
    (A) -- (B) -- (Bt) -- (At) -- cycle;
\filldraw[fill=blue!26, draw=blue!85!black, line width=0.5pt, opacity=0.70]
    (At) -- (Bt) -- (Dt) -- (Ct) -- cycle;
\draw[draw=blue!85!black, line width=0.5pt]
    (A) -- (B) -- (D) -- (C) -- cycle
    (At) -- (Bt) -- (Dt) -- (Ct) -- cycle
    (A) -- (At) (B) -- (Bt) (C) -- (Ct) (D) -- (Dt);

\node[blue!85!black, font=\small, anchor=west]
    at ($(Bt)+(0.6,0.4,0)$) {Injection box};

\draw[gray!80] (0,\dY,0) -- (0,\dY,\dZ) -- (\dX,\dY,\dZ) -- (\dX,\dY,0) -- (0,\dY,0)-- cycle;

\draw[{Stealth[length=4.5pt,width=4.5pt]}-{Stealth[length=4.5pt,width=4.5pt]}]
    (-2,-4,0) -- (\dX-2,-4,0)
    node[midway, below] {$50 d_{eq}$};

\draw[{Stealth[length=4.5pt,width=4.5pt]}-{Stealth[length=4.5pt,width=4.5pt]}]
    (\dX+3,0,0) -- (\dX+3,\dY,0)
    node[midway, right] {$25 d_{eq}$};

\draw[line width=0.7pt,
    {Stealth[length=4pt,width=4pt]}-{Stealth[length=4pt,width=4pt]}]
    (\dX+2,-2,\dZ+1.0) -- (\dX+2,-2,-1.0)
    node[midway, below right] {$5 d_{eq}$};
    
\draw[->, thick] (45,20,0) -- ++(0,-8,0) node[midway, right] {$\mathbf{g}$};

\end{tikzpicture}
\caption{Pseudo-3D sketch of the confined injection setup used to form a sand pile. Particles are introduced at $x=0$ through the injection box and accumulate against the lateral and bottom wall, which creates the pile shown in orange. The repose angle $\theta$ is measured along the (top) free surface of the pile.}
\label{fig:sandpile_setup}
\end{figure}

As a second validation of collective particle behavior, we investigate the formation of sandpiles composed of ellipsoidal particles. The test case consists in injecting particles into a confined domain, where they accumulate against a wall to form a stable granular pile (see Figure~\ref{fig:sandpile_setup}). The angle of repose, $\theta$, defined as the maximum angle between the (top) free surface and the horizontal plane at which the pile remains stable, is used as the primary validation metric. Its accurate prediction provides a robust assessment of the contact detection algorithm and, more broadly, of the overall accuracy of LEDDS in modeling interactions between non-spherical particles.

In practice, sand piles can be generated through several procedure, including particle injection, tilting, and discharge~\cite{zhou_gm_2014,grasselli_sma_1997}. The injection method is adopted for its simplicity and its reduced dependence on rolling resistance effects~\cite{zhou_ecr_2011}. It also avoids the need for moving or partially defined boundaries, making it particularly well suited to the current capabilities of LEDDS.
Here, new particles are inserted by batches of $10$, within a box of size $5\,d_{eq} \times 2\,d_{eq} \times 5\,d_{eq}$ every $1000$ time steps.
The injection box is positioned one maximum particle radius above the highest particle in order to (1) avoid overlapping with previously injected particles, and (2) limit the falling distance and associated kinetic energy. Since the angle of repose is symmetric in practice, only half of the sand pile is simulated by injecting particles adjacent to the left wall.

In our simulations, the computational domain has dimensions $50\,d_{eq} \times 25\,d_{eq} \times 5\,d_{eq}$, where the equivalent particle diameter is fixed to $d_{eq} = 2$~[mm] for all aspect ratios ($AR$), and here, $0.5 \leq AR \leq 2$. Here, $d_{eq}$ denotes the diameter of a sphere having the same volume as the particle. For a general ellipsoid with semi-axes $a$, $b$, and $c$, its volume reads $V=\frac{4}{3}\pi abc$. Defining $d_{eq}$ as the diameter of a sphere with the same volume, $\frac{\pi}{6}d_{eq}^3=V$, which leads to $d_{eq}=2(abc)^{1/3}$.

In this study, we restrict ourselves to axisymmetric particles with $a=b$. All particles have a density $\rho_p = 2650$~[kg/m$^3$], and a small restitution coefficient $\epsilon = 0.5$ to improve robustness while speeding up convergence to a steady state. Two friction coefficients, $\mu \in \{0.5,\,1.0\}$, are considered in order to assess the sensitivity of the angle of repose to interparticle friction.
Near-rigid behavior is enforced using a Poisson ratio $\upsilon = 0.4999$ and a Young's modulus $E = 10^7$~[Pa]. The bottom and side walls are assigned identical properties to ensure consistent particle–wall interactions.
Gravity is set to $\mathbf{g} = (0, -9.81, 0)$~[m/s$^2$], and the time step is fixed to $\Delta t = 10^{-5}$~[s] (highest value for which the simulation), and simulations are performed up to $30 \times 10^6$ time steps.

\begin{figure}[t]
    \centering
    \includegraphics[width=0.9\linewidth]{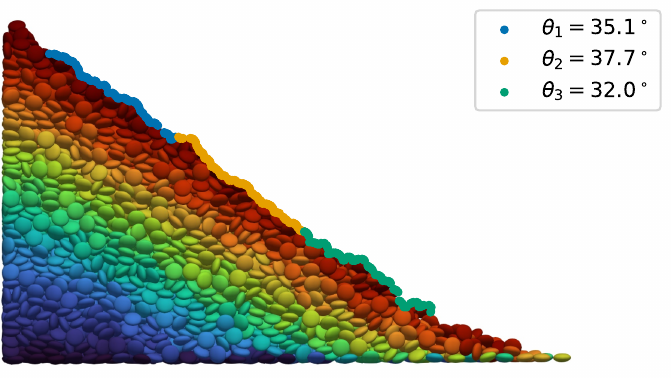}
    \caption{3D Paraview illustration of the angle of repose measurement procedure for an aspect ratio $AR = 0.5$. The extracted free surface is restricted to an intermediate region of the pile, excluding the apex and toe (lower and upper 10\% of the pile), and partitioned into segments over which local slopes are obtained by linear fitting. Particles are colored according to their deposition order, ranging from blue (earliest deposited particles) to red (most recently deposited), highlighting the successive buildup of the granular pile.}
    \label{fig:aor_method}
\end{figure}

\begin{figure}[t]
    \centering
    \includegraphics[width=0.9\linewidth]{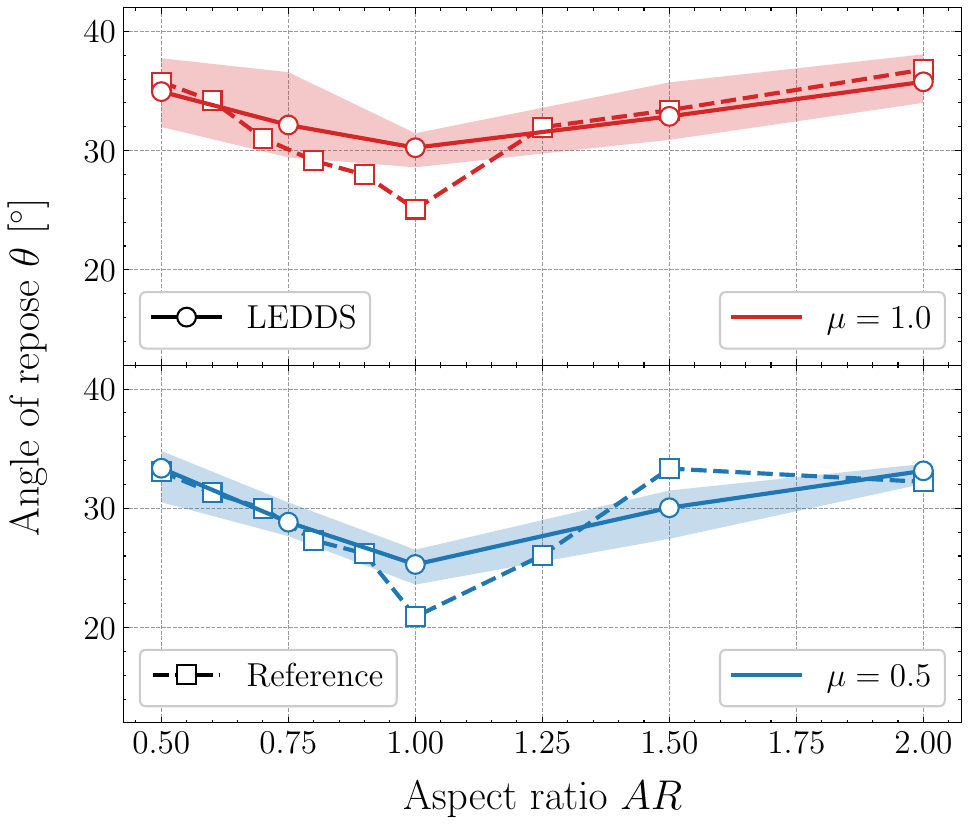}
    \caption{Sand pile validation: angle of repose $\theta$ as a function of ellipsoid aspect ratio $AR$. Comparison between LEDDS results and the reference data of Chen et al.~\cite{chen_p_2020}. Shaded regions indicate the variability range (minimum to maximum) obtained from the local slope measurements.}
    \label{fig:aor_ar}
\end{figure}

The angle of repose $\theta$ is evaluated after particle injection has ceased and the system has reached a stable equilibrium. The measurement procedure follows the methodology proposed by Chen et al.~\cite{chen_p_2020}.
Notably, the contour of the sand pile is extracted using the OpenCV library where the apex and toe regions are excluded to mitigate boundary effects (upper and lower 10\% height of the pile). The remaining surface profile is then partitioned into three equal segments, and a linear fit is applied to each segment to determine the local slope (see Figure~\ref{fig:aor_method}), from which the corresponding local angles of repose $\theta_i$ are obtained. These local values are finally averaged to define the global angle of repose $\theta$ of the sand pile.

The resulting angle of repose is reported in Figure~\ref{fig:aor_ar} over a wide range of ellipsoid aspect ratios and for two friction coefficients. Overall, very good agreement is observed with the reference data of Chen et al.~\cite{chen_p_2020}, with both datasets exhibiting a clear non-monotonic dependence on aspect ratio. In particular, the minimum angle of repose is obtained for spherical particles ($AR = 1$), while larger values are observed for both oblate ($AR < 1$) and prolate ($AR > 1$) shapes. Furthermore, increasing the friction coefficient $\mu$ from 0.5 to 1.0 leads to a systematic increase in the angle of repose across all aspect ratios, as expected from the enhanced friction between particles.
Quantitatively, deviations from the reference data remain below $5^\circ$. These discrepancies can be attributed to differences in the initial and loading conditions, which are known to influence the pile structure and angle of repose~\cite{HORABIK2017194}. Furthermore, the procedure used to estimate this angle is subject to non-negligible uncertainties (on the order of several degrees), so the level of agreement obtained here can be considered very satisfactory.

Overall, these results demonstrate that LEDDS accurately captures the influence of particle shape and friction on the angle of repose, and reliably reproduces the collective behavior of both spherical and non-spherical granular assemblies.

\subsection{Single-particle settling (LBM-DEM) \label{subsec:settling}}

As a first validation of the fluid-particle coupling, 
we simulate the settling of a single sphere in a quiescent fluid, following the experimental benchmark of a nylon particle descending through silicone oil~\cite{Cate_particle_2002}. In this experiment, a sphere of diameter $d_p = 15$~[mm] and density $\rho_p = 1120$~[kg/m$^3$] is released inside a rectangular container filled with oil, with dimensions $100 \times 100$~[mm$^2$] in cross-section and $L=160$~[mm] in height. The sphere is initially at rest at a height of $L_0=120$~[mm] (see Figure~\ref{fig:settlingSetup3D} for the full setup).

This benchmark is chosen not only to assess the particle’s terminal (maximum) settling velocity, but also to evaluate the full transient dynamics of the descent. An accurate fluid-particle coupling must capture reasonably well the initial acceleration, the subsequent approach to steady state, and the hydrodynamic interaction with solid boundaries during the final stage of motion. 
Furthermore, this benchmark allows us to identify the key parameters governing the accuracy and efficiency of the coupling strategy (PSM), so that we can propose a practical trade-off between particle resolution (i.e., the number of fluid cells per particle diameter) and the fluid time-step, which in LBM is controlled through the lattice velocity 
$u_{lb}$. This balance ensures reliable accuracy while keeping the computational cost to a minimum.

\begin{figure}[t!]
\centering
\begin{tikzpicture}[
    scale=0.95,
    every node/.style={scale=0.9},
    >=Stealth
]

\def\W{4.2}
\def\H{6.6}
\def\dx{1.15}
\def\dy{0.75}

\pgfmathsetmacro{\xc}{(\W+\dx)/2}
\pgfmathsetmacro{\yc}{(\H+\dy)*2/3}
\def\R{0.42}

\fill[black!8] (0,0) -- (\W,0) -- (\W+\dx,\dy) -- (\dx,\dy) -- cycle;

\draw[ultra thick] (0,0) -- (\W,0);
\draw[ultra thick] (\W,0) -- (\W,\H);
\draw[ultra thick] (\W,\H) -- (0,\H);
\draw[ultra thick] (0,\H) -- (0,0);

\draw[ultra thick, dashed] (\dx,\dy) -- (\W+\dx,\dy);
\draw[ultra thick] (\W+\dx,\dy) -- (\W+\dx,\H+\dy);
\draw[ultra thick] (\W+\dx,\H+\dy) -- (\dx,\H+\dy);
\draw[ultra thick, dashed] (\dx,\H+\dy) -- (\dx,\dy);

\draw[ultra thick, dashed] (0,0) -- (\dx,\dy);
\draw[ultra thick] (\W,0) -- (\W+\dx,\dy);
\draw[ultra thick] (\W,\H) -- (\W+\dx,\H+\dy);
\draw[ultra thick] (0,\H) -- (\dx,\H+\dy);

\shade[ball color=blue!25] (\xc,\yc) circle (\R);
\fill[blue!70] (\xc,\yc) circle (1.1pt);

\draw[thick,-latex] (\xc,\yc) -- ++(0,-1.2)
    node[pos=1, below] {$\mathbf{g}$};

\draw[thin,<->] (-0.55,0) -- (-0.55,\H)
    node[midway,left] {$L$};

\draw[thin,<->] (\W+\dx+0.55,0) -- (\W+\dx+0.55,\yc)
    node[midway,right] {$L_0$};

\draw[thin,<->] (\xc+\R+0.18,\yc-\R) -- (\xc+\R+0.18,\yc+\R)
    node[midway,right] {$d_p$};

\end{tikzpicture}
\caption{Pseudo-3D sketch of the single-particle settling benchmark. A spherical particle is released at a height $L_0$ in a quiescent fluid contained in a rectangular domain and settles under gravity $\mathbf{g}$. The transient settling velocity and the interaction with the bottom wall are used to assess the accuracy of the fluid-particle coupling.}
\label{fig:settlingSetup3D}
\end{figure}

In our numerical setup, gravity is applied only to the particle, not to the fluid. Consequently, the buoyancy force is not generated through the fluid’s hydrostatic pressure field but is instead imposed analytically as
\begin{equation}
\mathbf{F}_{\text{buoyancy}} = (1 - \rho_f/\rho_p) \, V_p \, \mathbf{g},
\end{equation}
where $\rho_f$ is the fluid density, and $V_p$ is the particle volume. This buoyancy force is then subtracted from the gravitational force acting on the particle. This approach notably reduces the amplitude of spurious pressure waves generated at initialization, and which are inherent to any compressible fluid solver such as LBM.

We will evaluate the settling behaviour under four different fluid conditions, varying both density and viscosity as listed in Table~\ref{tab:particle_settling}. For each case, the Reynolds number is computed using the particle’s settling velocity $u_s$.
\begin{table}[h!]
    \centering
    \renewcommand{\arraystretch}{1.4}
    \begin{tabular}{c|c|c|c|c}
        \textbf{Case} & $\rho_f$ [kg/m$^3$] & $\mu_f$ [Pa$\cdot$s] & $u_s$ [m/s] & $\mathrm{Reynolds}$ \\
        \hline
        1 & 970 & 0.373 & 0.036 & \phantom{1}1.5 \\
        2 & 965 & 0.212 & 0.057 & \phantom{1}4.1 \\
        3 & 962 & 0.113 & 0.087 & 11.6 \\
        4 & 960 & 0.058 & 0.122 & 31.9 \\
    \end{tabular}
    \caption{Fluid parameters considered for the particle settling setups.}
    \label{tab:particle_settling}
\end{table}

\begin{figure}[t]
    \centering
    \includegraphics[width=.90\linewidth]{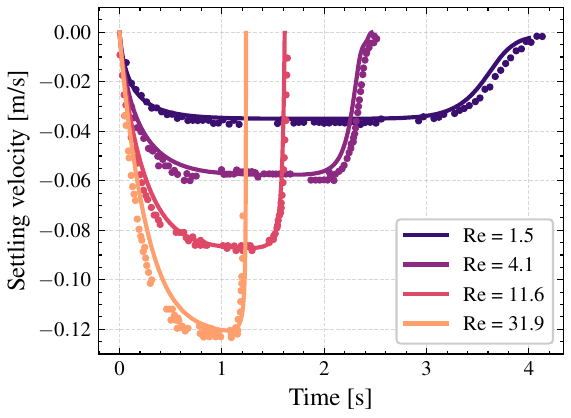}
    \caption{Single particle settling simulation: Time evolution of the settling velocity for a particle resolved with $D/\Delta x = 20$ fluid cells and an LB velocity of $u_{lb} = 0.01$. Simulation results (solid lines) are compared against the reference data (dots) of ten Cate et al.~\cite{Cate_particle_2002} for several Reynolds numbers.}
    \label{fig:demtest-particleSettling}
\end{figure}
Figure~\ref{fig:demtest-particleSettling} shows the time evolution of the settling velocity for all four cases. The particle is resolved with $D/\Delta x = 20$ fluid cells and simulated using a characteristic LB velocity of $u_{lb} = 0.01$. The numerical results (solid lines) are compared directly with the experimental measurements of ten Cate et al.~\cite{Cate_particle_2002} (symbols).

Globally speaking, the simulation results show very good agreement with the experimental measurements across 
all Reynolds numbers for all the different stages. 
More quantitatively, a slight delay is observed for initial stage of the motion with the highest Reynolds number. This behavior is likely due to the initial pressure disturbance generated when the particle begins to move: at higher Reynolds numbers, viscous damping is weaker, allowing this pressure wave to persist longer and momentarily affect the early acceleration. Conversely, for the lowest Reynolds numbers, the particle tends to decelerate slightly earlier as it approaches the bottom wall. This is expected in high-viscosity regimes, where lubrication effects become more pronounced; since no explicit lubrication model is used here, the near-wall resistance is slightly over-predicted, causing an earlier reduction in velocity (see Section 3.2 of Ref.~\cite{RETTINGER_JCP_453_2022} for more details). In all cases, the particle terminal velocity is very well predicted by LEDDS.

Interestingly, these trends are consistent across a wide range of numerical parameters, 
including particle resolutions from $D/\Delta x = 5$ to $30$ and lattice velocities $u_{lb}$ from $0.002$ to $0.08$, 
as detailed in~\ref{app:param_study}. For most Reynolds numbers, the simulation runtime can 
be reduced by more than a factor of 10 by lowering the particle resolution to $D/\Delta x = 10$ and 
increasing the lattice velocity to $u_{lb} = 0.02$, while still maintaining good accuracy.

\subsection{Ellipsoid evolution in a shear flow (LBM-DEM)}

\begin{figure}[b]
\centering
\begin{tikzpicture}[
    scale=0.8,
    every node/.style={scale=0.8},
    >=Stealth
]

\def\Lx{10}
\def\Ly{5}
\def\xc{5}
\def\yc{2.5}
\def\a{1.2}
\def\b{0.55}
\def\ang{35}

\draw[fill=black!10, draw=none] (0,\Ly) rectangle (\Lx,\Ly+0.35);
\draw[pattern={Lines[angle=-45,distance={5pt}]}, pattern color=black, draw=none]
    (0,\Ly) rectangle (\Lx,\Ly+0.35);
\draw[ultra thick] (0,\Ly) -- (\Lx,\Ly);

\draw[fill=black!10, draw=none] (0,-0.35) rectangle (\Lx,0);
\draw[pattern={Lines[angle=-45,distance={5pt}]}, pattern color=black, draw=none]
    (0,-0.35) rectangle (\Lx,0);
\draw[ultra thick] (0,0) -- (\Lx,0);

\draw[dashed, color=black!40] (0,0) -- (0,\Ly);
\draw[dashed, color=black!40] (\Lx,0) -- (\Lx,\Ly);

\foreach \y/\len in {0.5/0.3, 1.3/0.8, 2.1/1.3, 2.9/1.8, 3.7/2.3, 4.5/2.8}
    \draw[thick,-latex,blue!70] (1.0,\y) -- ++(\len,0);

\draw[thick,-latex] (7.8,\Ly-0.22) -- ++(1.2,0)
    node[pos=0.5, below=2pt] {$\mathbf{u}=(\dot{\gamma}L,0,0)$};

\node at (8.4,0.3) {$\mathbf{u}=\mathbf{0}$};

\begin{scope}[shift={(\xc,\yc)}, rotate=\ang]
    \fill[red!5] (0,0) ellipse [x radius=\a, y radius=\b];
    \draw[red, thick] (0,0) ellipse [x radius=\a, y radius=\b];
    \draw[red!70, thin] (0,0) -- (\a,0)
        node[midway, above, black] {$a$};
    \draw[red!70, thin] (0,0) -- (0,\b)
        node[midway, left, black] {$b$};
\end{scope}

\node[circle, inner sep=1.2pt, fill=red!70, draw=none] at (\xc,\yc) {};

\draw[dashed, color=black!40] (\xc,\yc) -- ({\xc+2.0},{\yc});
\draw[dashed, color=black!40] (\xc,\yc) -- ({\xc+2*cos(\ang)},{\yc+2*sin(\ang)});

\coordinate (C)    at (\xc,\yc);
\coordinate (xref) at ({\xc+2.6},{\yc});
\coordinate (xell) at ({\xc+2.6*cos(\ang)},{\yc+2.6*sin(\ang)});

\draw[dashed, color=black!40] (C) -- (xref);
\draw[dashed, color=black!40] (C) -- (xell);

\pic [
    draw,
    <-,
    "$\theta_z$",
    angle eccentricity=1.3,
    angle radius=1.4cm
] {angle = xref--C--xell};

\draw[latex-]
    ({\xc-1.2},{\yc-0.55})
    arc[start angle=180,end angle=270,radius=0.5]
    node[pos=0.55, below left] {$\omega_z$};

\end{tikzpicture}
\caption{Schematic representation of the Jeffery orbit simulation. A prolate ellipsoid is placed at the center of a plane channel of height $L$, bounded by two horizontal walls. A steady Couette flow is imposed by moving the upper wall with velocity $\mathbf{u}=(\dot{\gamma}L,0,0)$ while the lower wall remains at rest, generating a linear shear profile that drives the periodic rotation of the ellipsoid.}
\label{fig:jefferySetup}
\end{figure}

\begin{figure}[t]
    \centering
    \includegraphics[width=0.95\linewidth]{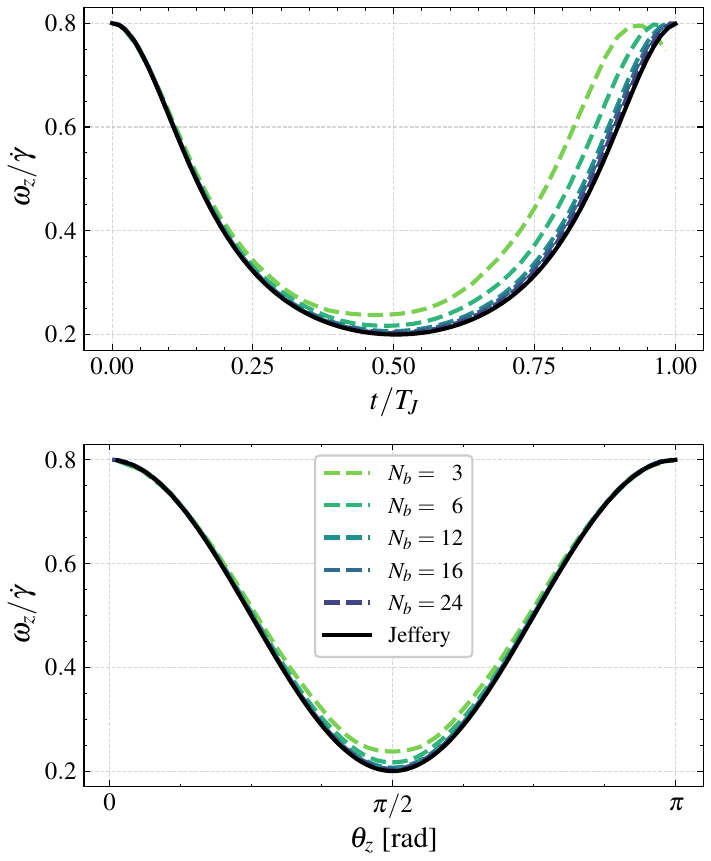}
    \caption{Simulation of Jeffery orbit at Reynolds $\mathrm{Re}=0.08$ for various fluid cell resolutions of the minor axis $N_b\in\{3,6,12,24\}$. Top: Time evolution of the angular velocity $\omega_z$. Bottom: Angular velocity $\omega_z$ as a function of its orientation $\theta_z$.}
    \label{fig:jeffery-orbits}
\end{figure}

As an additional validation of the fluid-solid coupling, we consider the rotational dynamics of an ellipsoid immersed in a viscous shear flow. This configuration is a classical benchmark for non-spherical particle dynamics, as the motion of an ellipsoid at vanishing Reynolds number is governed by the analytical solutions derived by Jeffery~\cite{JEFFERY_PRSL_1922}. These so-called Jeffery orbits describe the periodic rotational motion of the particle and provide an excellent test of a numerical method’s ability to accurately capture hydrodynamic torques on anisotropic solids.

As illustrated in Figure~\ref{fig:jefferySetup}, the setup consists of a prolate ellipsoid with semi-axes $a = 0.5$~[m] and $b = c = 0.25$~[m] (aspect ratio of 2) placed at the center of a rectangular domain of size $\tfrac{L}{2} \times L \times \tfrac{L}{2}$, with $L = 20\,a = 10$~[m]. Such a large domain notably prevents the influence of the no-slip (bounce-back) boundaries on the ellipsoid dynamics. The domain is periodic in the $x$-direction, allowing the ellipsoid to rotate freely. The rotation of the prolate is controlled by imposing a steady Couette flow through a constant velocity difference between two parallel walls: the upper wall moves at $\mathbf{u} = (\dot{\gamma}\, L, 0, 0)$, with a shear rate $\dot{\gamma} = 0.002$~[s$^{-1}$], while the bottom wall remains stationary.  The fluid velocity is initialized with a linear profile, while the ellipsoid is left in a rest state.

In this Stokes regime, the rotation of a prolate ellipsoid induced by a shear flow is governed by Jeffery’s theory, which provides analytical expressions for the orientation angle $\theta_z$ about the $z$-axis and its angular velocity $omega_z$~\cite{JEFFERY_PRSL_1922}:
\begin{align}
    theta_z(t) &= \tan^{-1}\!\left[\frac{a}{b}\,\tan\!\left(\frac{ab\,\dot{\gamma}\,t}{a^2+b^2}\right)\right],\\
    omega_z &= \frac{\dot{\gamma}}{a^2+b^2}\left(a^2\cos^2theta_z + b^2\sin^2theta_z\right).
\end{align}
The characteristic time scale of the rotational motion is the Jeffery period, $T_J$, which corresponds here to the time required for the ellipsoid to complete one $\pi$-orientation cycle (``Jeffery orbit''). 
It is derived by defining the inner phase $\phi(t) = \frac{ab\,\dot{\gamma}\,t}{a^2+b^2}$, and noting that the tangent function in the solution is periodic. Requiring that the ellipsoid completes a rotation corresponding to a $\pi$ change in $\phi$, i.e.,
\begin{equation}
    theta_z(t+T_J)-theta_z(t) = \pi \:\: \longrightarrow \:\: T_J = \frac{\pi\,(a^2+b^2)}{ab\,\dot{\gamma}}\approx 3927\,\text{[s]}.
\end{equation}

To remain in the Stokes regime, the fluid and particle densities are set equal to eliminate buoyancy effects, and the simulation parameters are chosen such that the Reynolds number is fixed at $\mathrm{Re} = 0.08$, thereby minimizing inertial contributions. As with the settling benchmark, such low Reynolds numbers require small time steps to maintain accuracy. We therefore set $u_{lb} = 8\times 10^{-4}$, ensuring $\tau < 2$ for all tested resolutions of the ellipsoid. Each simulation is run for a non-dimensional time of $8$, corresponding to $5000$~[s] of physical time (about $1.27\,T_J$ for the chosen parameters).

Figure~\ref{fig:jeffery-orbits} presents the angular velocity of the prolate ellipsoid as a function of its orientation angle, together with the corresponding time evolution of the particle rotation, for several spatial resolutions (defined as the number of fluid cells across the particle’s minor axis $b$). At low resolutions, the rotation rate is slightly overestimated, which results in a Jeffery period shorter than the theoretical prediction. This behavior stems from insufficiently resolved hydrodynamic torques acting on the particle. Once the resolution reaches 12 cells per minor axis (or higher) the agreement with the analytical Jeffery solution becomes significantly more accurate: both the angular-velocity profile and the oscillation period closely match the theoretical curves. Further refinement provides only minor improvements, indicating that the essential hydrodynamic torques governing the particle motion are already well captured at these moderate resolutions.

These results are fully consistent with the trends observed for spherical particles in the settling benchmark (Section~\ref{subsec:settling} and~\ref{app:param_study}), where resolutions of approximately 10 fluid cells per particle diameter provided a good balance between accuracy and computational cost. Here, despite the added geometric complexity of an anisotropic particle, similar resolution levels for the minor axis remain sufficient to reproduce the correct rotational dynamics. This benchmark therefore confirms the accuracy of the solid-fluid coupling (PSM) for non-spherical particles and demonstrates that the framework accurately captures torque-driven motion of single particles in viscous flows.

\subsection{Effective viscosity in a shear flow (LBM-DEM)\label{subsec:effective_visco}}

\begin{figure}[b!]
    \centering
    \begin{subfigure}[b]{0.48\columnwidth} 
        \centering
        \includegraphics[width=\linewidth]{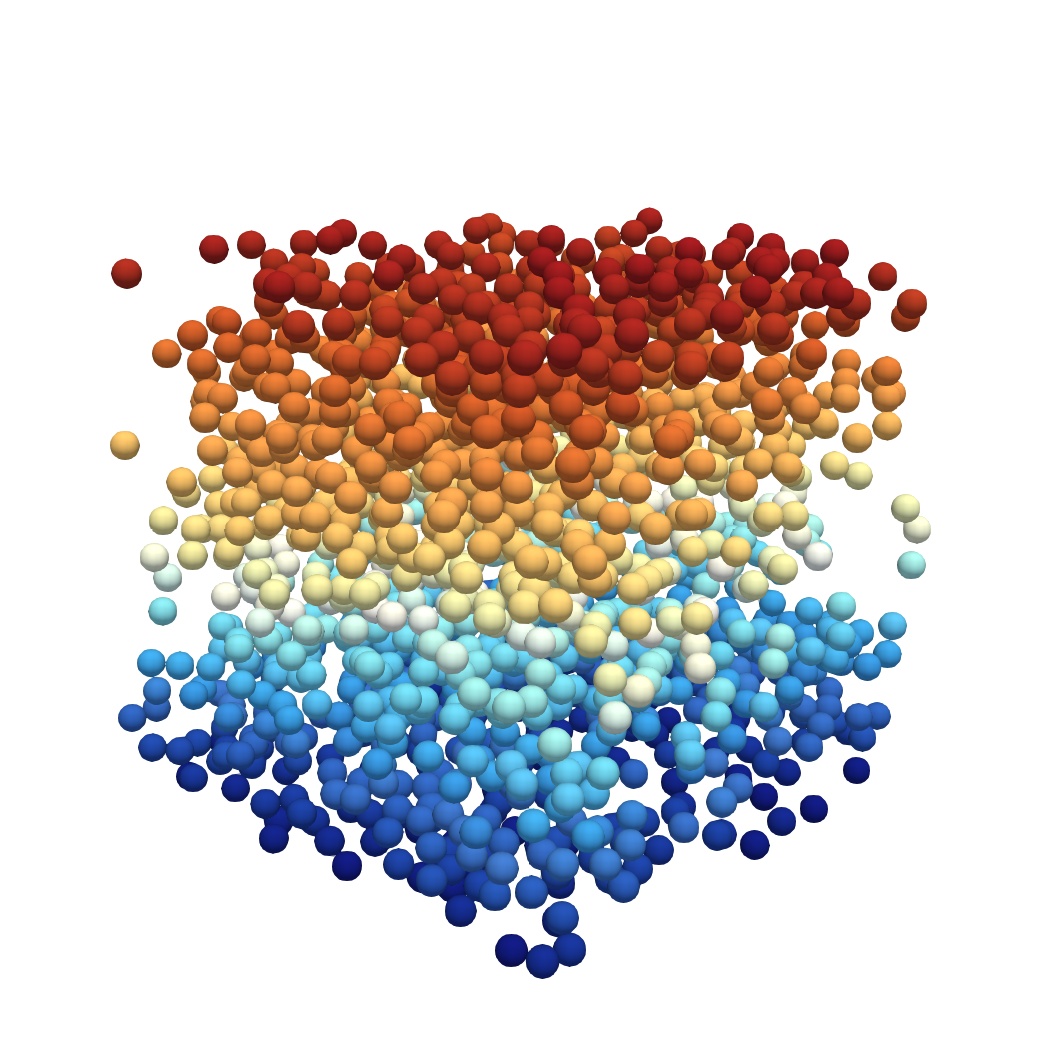}
        \caption{Initial condition with $\phi = 0.1$}
    \end{subfigure}
    \hfill 
    \begin{subfigure}[b]{0.48\columnwidth}
        \centering
        \includegraphics[width=\linewidth]{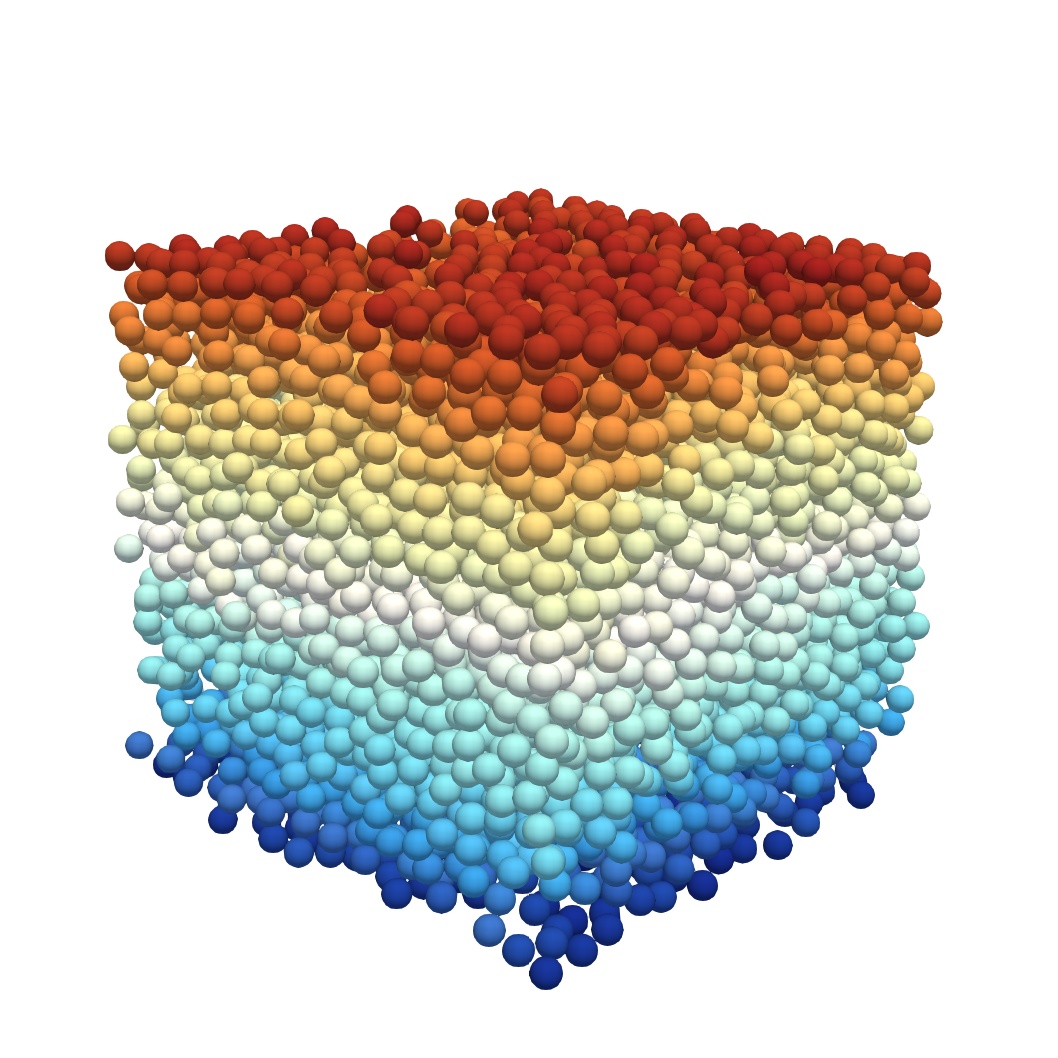}
        \caption{Initial condition with $\phi = 0.4$}
    \end{subfigure}
    
    \caption{
        3D Paraview shear flow illustrations: Instantaneous particle configurations for two volume fractions $\phi = 0.1$ and $\phi = 0.4$. 
        Only the solid phase is shown; particles are colored by their streamwise velocity component $u_x$, with red indicating faster motion and blue indicating nearly quiescent regions.
    }
    
    \label{fig:shear}
\end{figure}

The final validation test considered in this work combines the many-body contact dynamics of Section~\ref{subsec:many_body} with the fluid-particle interaction encountered in Section~\ref{subsec:settling}. More precisely, we examine the effective rheology of a suspension of rigid spheres subjected to a steady shear flow~\cite{EINSTEIN_AP_324_1906,KRIEGER_TSR_3_1959}. This configuration is widely used to assess the ability of a coupled LBM-DEM solver to reproduce bulk suspension properties, such as the increase in apparent viscosity with particle volume fraction~\cite{LISHCHUK_PRE_74_2006, LORENZ_PRE_79_2009,THORIMBERT_CF_166_2018,GUO_PoF_33_2021}.

The suspension consists of neutrally buoyant spheres of radius $R=0.1$~[m] placed in a Newtonian fluid of dynamic viscosity $\eta_0 = 1.0$~[Pa$\cdot$s]. The particles then have density equal to that of the fluid, and gravity is omitted. The solids are modeled with Poisson ratio $\upsilon = 0.4$, Young’s modulus $E = 100$~[MPa], friction coefficient $\mu = 0.45$, and restitution coefficient $\epsilon = 0.95$. While these parameters are chosen to mimick non-elastic aspects of real particle collisions, the result shows little sensitivity to specific collision parameters in practice, as the interaction with the viscous fluids dominates energy dissipation. The fluid domain is discretized by $200^3$ lattice nodes, ensuring that each particle diameter is resolved by 10 fluid cells. The LBM time step is $\Delta t = 5\times 10^{-4}$~[s], and the DEM solver performs 10 substeps per fluid iteration to warrant the stability of the DEM solver.

To characterize the macroscopic rheology of the suspension, we compute the effective viscosity from the mean wall shear stress, which is directly related to the time-averaged streamwise velocity profile $u_x(y)$.
In a Newtonian fluid, the wall shear stress is defined as
\begin{equation}
\tau_w = \eta_0 \left.\frac{\partial u_x}{\partial y}\right|_{y=0}.
\end{equation}
However, directly approximating the gradient from the first off-wall node introduces large discretization errors because of the moderate resolution.
To obtain a more accurate estimate of the wall shear rate, we then fit a second-order polynomial through the wall node and the first two interior fluid nodes and evaluate its derivative analytically at the wall location,
\begin{equation}
u_x'(0) = a_1,
\end{equation}
where $a_1$ is the linear coefficient of the fitted polynomial. 
The effective suspension viscosity is finally obtained from
\begin{equation}
\frac{\eta}{\eta_0} = \frac{u_x'(y_0)}{s},
\end{equation}
where $s$ is the measured wall shear rate.
For validation, this quantity is compared to the classical Krieger-Dougherty model~\cite{krieger_mechanism_1959},
\begin{equation}
\frac{\eta}{\eta_0}
= \left(1 - \frac{\phi}{\phi_m}\right)^{-[\eta],\phi_m},
\end{equation}
where $\phi$ is the particle volume fraction, $\phi_m = 0.64$ being the maximum packing fraction for monodisperse spheres, and $[\eta]=2.5$ the intrinsic dynamic viscosity.

\begin{figure}[t!]
    \centering
    \includegraphics[width=.9\linewidth]{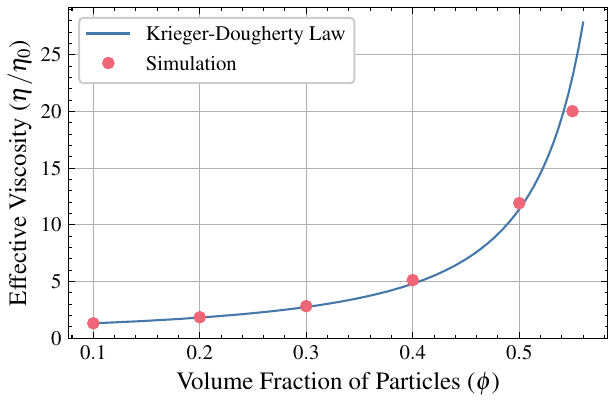}
    \caption{Evolution of the apparent viscosity ($\eta/\eta_0$) as a function of the volume fraction ($\phi$). Simulation data (symbols) are compared to the Krieger-Dougherty semi-analytical law~\cite{KRIEGER_TSR_3_1959}.}
    \label{fig:app_viscosity}
\end{figure}

While generating initial particle configurations by random placement (with velocities matched to the Couette profile) is straightforward in the dilute regime, this approach becomes increasingly inefficient for volume fractions above $\phi=0.25$ due to geometric crowding.
To robustly initialize dense suspensions, we successively rely on two pre-processing steps. First, a sedimentation procedure ensures all particles are accommodated within the domain without overlap.
Second, a brief relaxation stage with small random velocities is applied to remove any residual spatial bias in the particle arrangement before the shear flow is imposed.
Once both particles and fluid have been properly initialized, all simulations are performed over approximately $t^* = 30$ convective times, which corresponds to 3 million LBM time iterations.
Although this duration is insufficient to obtain statistically converged data, the statistical homogeneity of the flow in the $x$- and $z$-directions allows us to average $u_x(y)$ over these directions to get sufficient statistical convergence for analysis.

The resulting numerical measurements are compared with the theoretical prediction in Figure~\ref{fig:app_viscosity} for volume fractions ranging from $\phi=0.1$ to $0.55$. Perfect match is obtained for $\phi\leq 0.3$ (dilute regime), and very good agreement with the Krieger-Dougherty model is obtained for higher volume fractions.

\section{Performance analysis\label{sec:perfo}}

To assess the computational efficiency and scalability of LEDDS, we perform a series of performance benchmarks using a fluidized-bed configuration across a range of solid volume fractions~\cite{KEMMLER_IJHPCA_39_2025}. The analysis is conducted on both a 128-core AMD EPYC~7742 CPU and an NVIDIA A100-SXM4~(80GB) GPU.
The CPU benchmarks provide a reference baseline for speedup evaluation and highlight the effects of floating-point precision on performance.  
The GPU benchmarks, in turn, quantify the benefits of hardware acceleration and Thrust’s GPU-optimized primitives, evaluate precision-dependent performance, and provide a direct comparison with waLBerla to gauge the efficiency of LEDDS’s high-level, portable parallelization strategy relative to a highly optimized CUDA implementation.
Eventually, the A100 GPU card was chosen notably because it provides native double precision (FP64) support with a peak throughput approximately half that of single precision (FP32), which allows for meaningful comparisons between FP64 and FP32 performance.
Additional results on the GPU memory footprint of LEDDS are provided in~\ref{app:gpu_memory_usage}.

\subsection{Numerical setup for performance comparisons}

The performance benchmarks are based on a fluidized-bed configuration at various solid volume fractions, following closely the setup of the performance study conducted with waLBerla~\cite{KEMMLER_IJHPCA_39_2025}. This reference is particularly well suited for comparison, since both LEDDS and waLBerla rely on (1) the PSM-based fluid-particle coupling strategy, and (2) the distance-based solid-fraction computation~\eqref{eq:naiveSolidFrac}. These methodological similarities then enable us to propose a meaningful GPU performance comparison against a state-of-the-art CUDA implementation.

The simulation parameters are chosen to match the reference study, i.e., the particle diameter is resolved by 20 fluid cells, and the DEM solver performs 10 sub-steps per LBM iteration. The computational domain spans 
$[500 \times 200 \times 800]$ fluid cells (about 80 millions) and is periodic in the $x$- and $z$-directions. No-slip bounce-back walls are imposed in the \(y\)-direction. The Reynolds number is set to $\mathrm{Re}=1$, gravity acts in the negative $y$-direction with magnitude $g = 9.81$ [m/s$^2$], and the particle-to-fluid density ratio is fixed at 1.1.

The benchmark covers particle volume fractions from $\phi = 0$ up to $\phi = 0.45$, corresponding to a maximum of 8494 spheres. This range provides insight into the computational scaling of the PSM coupling strategy from dilute to highly packed regimes. 
For volume fractions above $\phi \approx 0.25$, we use the same sedimentation-relaxation strategy to get the initial positioning of particles as for the shear flow simulations conducted in Section~\ref{subsec:effective_visco}.

\subsection{Performance metric and compilation flags}

To ensure consistent and reproducible performance measurements, all benchmarks are executed after an initial warm-up phase designed to remove startup overheads such as memory allocation, or data transfers. 
On the GPU, $500$ warm-up iterations are performed before measurements begin, followed by $2500$ iterations used to compute the reported execution times.  
On the CPU, $100$ warm-up iterations are used, and performance is evaluated over $500$ iterations.  
These iteration counts are sufficiently large to average out short-term fluctuations while keeping the benchmarking stage computationally affordable.

Performance results are reported in lattice updates per second (LUPS), the standard metric in the LBM community that measures how many fluid cells are updated per second. Because this number is typically very large, especially on modern GPUs, it is common to express it in millions (MLUPS).
Although the MLUPS performance metric does not directly quantify the cost of the DEM component, it reflects the overall performance of the coupled LBM-DEM simulation because it is computed from the total wall-clock time per iteration.
As a result, all operations (LBM, DEM, and PSM-based coupling) are implicitly included in the metric.  
This makes MLUPS suitable for comparing different configurations and for assessing the overhead introduced by particle-fluid interactions as the solid volume fraction increases.

All simulations are compiled with the \texttt{nvc++} compiler (version 24.9). It should be noted that the current version of LEDDS is incompatible with older versions of \texttt{nvc++}, while upward compatibility is currently guaranteed.
LEDDS is built for the GPU using the \texttt{stdpar=gpu} backend and for multicore CPU execution with the \texttt{stdpar=multicore} backend.  
Both builds use the following optimization flags \texttt{fast} and \texttt{O3} which enables a collection of aggressive optimizations.
The only hardware-specific flag is \texttt{-march=native}, which is applied on CPUs to allow generation of machine code optimized for the host architecture.
These settings provide a consistent and performance-oriented compilation strategy across platforms while preserving C++ standard compliance.  
CPU benchmarks are performed on a single AMD EPYC~7742 node at 2.25~[GHz], using 128 cores. 
This choice of CPU hardware provides a node-level comparison with the server-class A100-SXM4 (80GB) GPU.

\subsection{CPU performance}

The first analysis investigates how the solid volume fraction affects overall CPU performance and assesses the potential benefits of using reduced floating-point precision. 

\begin{figure}[t!]
    \centering
    \includegraphics[width=.95\linewidth]{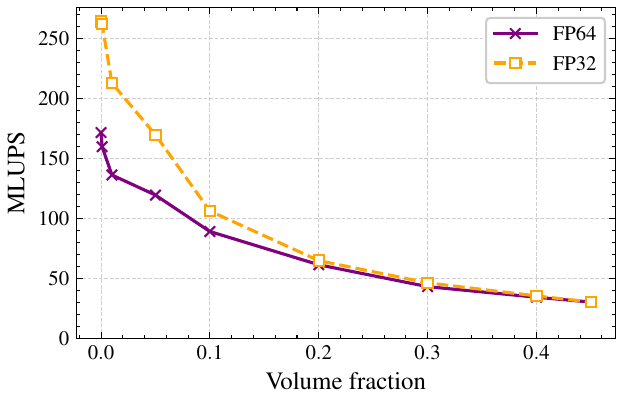}
    \caption{CPU performance comparison of single (FP32) and double (FP64) precision for STL implementation. MLUPS are shown as a function of the solid volume fraction $0 \leq \phi \leq 0.45$.}
    \label{fig:mlupsCPU}
\end{figure}

Figure~\ref{fig:mlupsCPU} shows that increasing $\phi$ has a substantial impact on performance whatever the precision: compared to pure-fluid simulations ($\phi = 0$), the MLUPS rate decreases by a factor of 5 to 10 at $\phi = 0.45$. This degradation reflects the growing cost of DEM updates and particle–fluid coupling, which progressively overshadow the memory-bound LBM module as the suspension becomes more packed.

In dilute regimes, where the LBM dominates and performance is mainly limited by memory bandwidth, switching from FP64 to FP32 leads to a speedup of about 1.5. As the solid volume fraction $\phi$ increases, however, this advantage gradually diminishes. In dense configurations, the runtime becomes increasingly dominated by DEM-related operations such as collision-list construction, contact handling, and force reduction. Since these operations rely heavily on integer keys, sorting, binary searches, and irregular memory accesses, their cost does not scale as favorably with floating-point precision as the LBM update does. As a result, FP32 brings a clear performance benefit in dilute flows, while in densely packed suspensions it is mainly advantageous because it reduces memory consumption and thus increases the number of particles and fluid cells that can be handled on a given architecture.

\subsection{GPU performance}

Continuing with the GPU performance analysis of LEDDS, we first compare the use of STL algorithms and Thrust primitives for the DEM and PSM-based coupling modules. 

\begin{figure}[t]
    \centering
    \includegraphics[width=.95\linewidth]{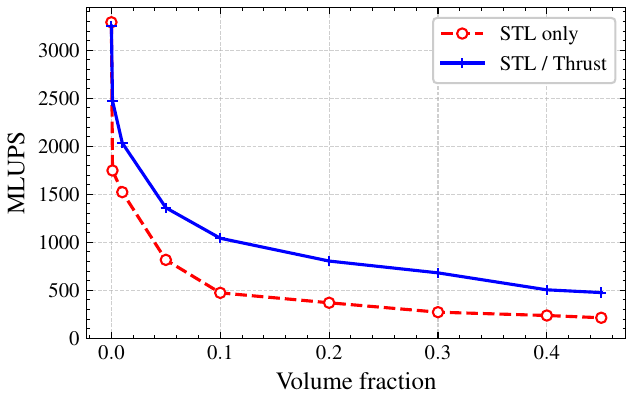}
    \caption{GPU performance comparison of a pure C++ STL implementation against an accelerated version, using the \texttt{reduce\_by\_key} algorithm and a radix-sort based version of \texttt{sort} from the Thrust library. Both implementations use double precision (FP64).}
    \label{fig:mlups_gpu_std_thrust}
\end{figure}

Figure~\ref{fig:mlups_gpu_std_thrust} compares both approaches using double precision (FP64). While both implementations achieve very high and nearly identical performance for pure fluid simulations ($\phi = 0$), performance rapidly decreases as the solid volume fraction increases, and significant differences emerge. The results demonstrate that the STL/Thrust version, relying on the \texttt{reduce\_by\_key} algorithm and a radix-sort based version of \texttt{sort} from the Thrust library, consistently outperforms the pure STL version on GPU, achieving an almost constant speedup of about 2. Based on this observation, Thrust algorithms are used for particle/force sorting and reduction operations in all subsequent GPU benchmarks, while the remaining parts of LEDDS continue to rely on STL algorithms, as summarized in Table~\ref{tab:workflow}.

\begin{figure}[h]
    \centering
    \includegraphics[width=.95\linewidth]{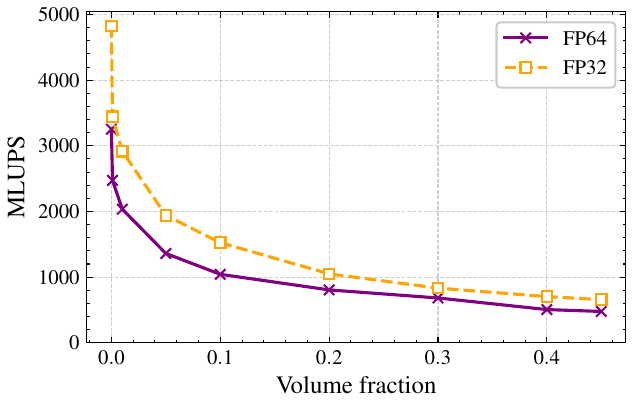}
    \caption{GPU performance of the STL/Thrust version of LEDDS in single (FP32) and double precision (FP64) across varying solid volume fractions. Both set of data were generated with the STL/Thrust version of LEDDS.}
    \label{fig:mlups_gpu}
\end{figure}

Building on the previous analysis of algorithm selection, we next investigate the impact of floating-point precision on GPU performance of the STL/Thrust version of LEDDS. 
Figure~\ref{fig:mlups_gpu} presents MLUPS results for LEDDS using single (FP32) and double precision (FP64). 
In dilute configurations ($\phi = 0$), where the solver is predominantly memory-bound due to the LBM updates, switching from FP64 to FP32 yields a significant performance gain similar to the one obtained on CPU. 
As the solid volume fraction increases, however, the relative advantage of FP32 gradually decreases because the computational cost becomes dominated by particle-fluid coupling and DEM updates. For high volume fractions ($\phi \ge 0.3$), the performance improvement decreases to approximately 30\%. Consequently, as observed for CPU simulations, FP32 offers significant speedups in dilute regimes, while in dense suspensions it primarily reduces memory usage without substantially impacting overall performance.

\subsection{GPU speedup}

\begin{figure}[t]
\centering
\includegraphics[width=.95\linewidth]{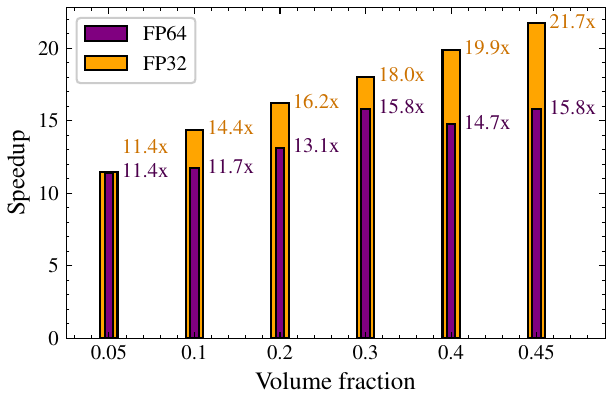}
\caption{GPU speedup relative to the CPU baseline. Results are shown for FP32 and FP64 precision and for dilute ($\phi = 0.05$) and dense ($\phi = 0.45$) suspensions.}
\label{fig:speedup}
\end{figure}

The primary motivation for employing algorithmic primitives (from STL and Thrust) is to develop a simulation framework capable of achieving substantial, nearly out-of-the-box GPU acceleration. Figure~\ref{fig:speedup} reports the speedup obtained for the optimal CPU (STL-only) and GPU (STL/Thrust) implementations of LEDDS. Results are presented for both FP32 and FP64 precision, across solid volume fractions ranging from dilute suspensions ($\phi = 0.05$) to densely packed ones ($\phi = 0.45$).

These benchmarks indicate that running LEDDS on an A100 GPU provides performance comparable to 11-18 CPUs with 128 cores for FP64, with speedups reaching up to 23 for FP32 in the dense case ($\phi = 0.45$). 

In practical terms, the measured speedups correspond to node-level performance gains of approximately 10-20$\times$ when comparing the A100-SXM4 (80GB) to a 128-core CPU node. 
While the CPU version of LEDDS could likely benefit from additional optimization, it nevertheless provides a consistent reference for comparison. The overall conclusion remains that the use of algorithmic primitives enables substantial runtime acceleration on GPUs, without sacrificing portability or ease of implementation.

\subsection{Comparison with CUDA-based solver}

To have a more global view on the GPU performance of LEDDS, we compared it against the state-of-the-art implementation in waLBerla~\cite{KEMMLER_IJHPCA_39_2025}, which employs a hybrid GPU-CPU approach: the LBM and coupling modules are highly optimized on the GPU using CUDA, while an optimized DEM module runs on the CPU, with communication overhead effectively hidden (see their Figure 7).

To ensure correct usage of waLBerla, we first reproduced the performance results reported in Ref.~\cite{KEMMLER_IJHPCA_39_2025} on the same hardware: A100-SXM4 (40GB) GPU. Following this verification, we measured waLBerla’s performance on an A100-SXM4 (80GB) GPU paired with an AMD EPYC-7742 CPU (64 cores), with LEDDS performance solely evaluated on the GPU.

For a broader performance comparison, simulations were conducted at solid volume fractions ranging from $\phi \approx 0.03$ to $\phi \approx 0.42$ by adjusting the particle generation spacing, while remaining within the range studied in previous studies. It should be noted that in waLBerla, only the particle spacing can be specified. The number of particles, and hence the actual solid volume fraction, is automatically determined by the framework. This makes a one-to-one comparison at exactly the same $\phi$ challenging.

\begin{figure}[t]
    \centering
    \includegraphics[width=.95\linewidth]{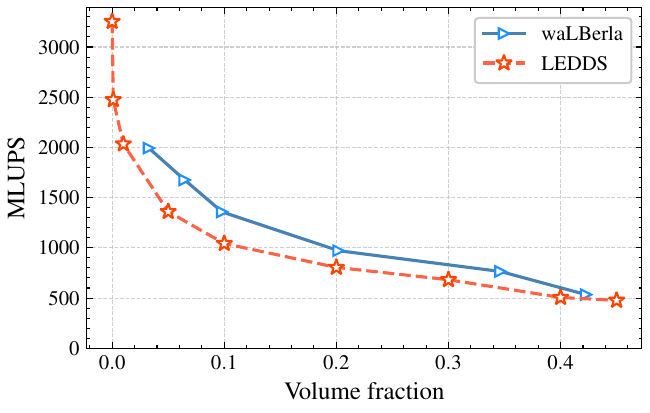}
\caption{GPU performance comparison between LEDDS and waLBerla framework~\cite{KEMMLER_IJHPCA_39_2025}. MLUPS are reported as a function of the solid volume fraction $\phi$ for FP64 precision.}
    \label{fig:walberla_comparison_gpu}
\end{figure}

The results, shown in Figure~\ref{fig:walberla_comparison_gpu}, indicate that LEDDS generally achieves performance within a factor of two of waLBerla using FP64. Importantly, the gap between the two solvers systematically decreases as the suspension becomes denser, and at the upper end of the tested regime their performance becomes comparable. This demonstrates that a design based on high-level algorithmic primitives can achieve GPU performance in the same order of magnitude as a hand-optimized CUDA-CPU hybrid framework, while offering good levels of portability and maintainability.

\section{Summary and outlook\label{sec:summary_and_outlook}}

\begin{figure*}[t]
    \centering
        \includegraphics[width=0.49\textwidth, trim={0 0 6cm 0}, clip]{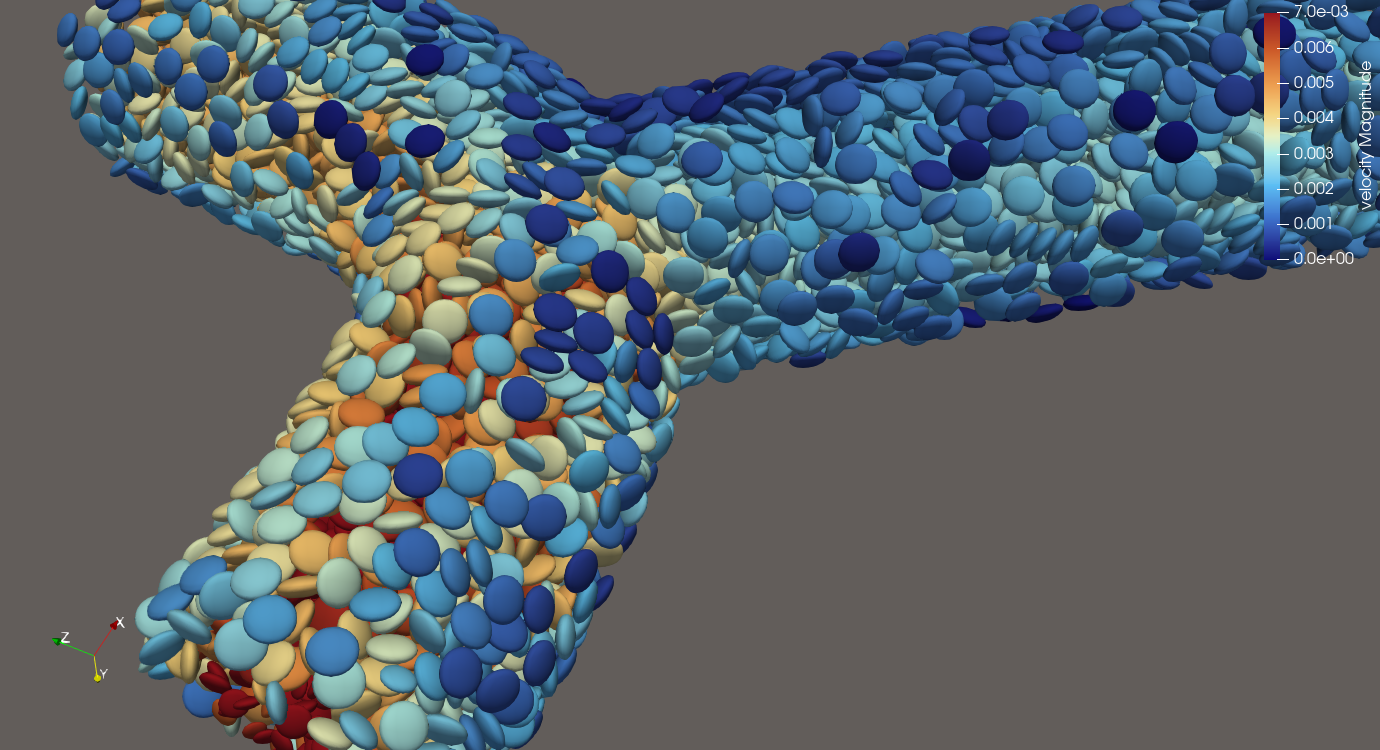}
    \hfill
        \includegraphics[width=0.49\textwidth, trim={0 0 4.6cm 0}, clip]{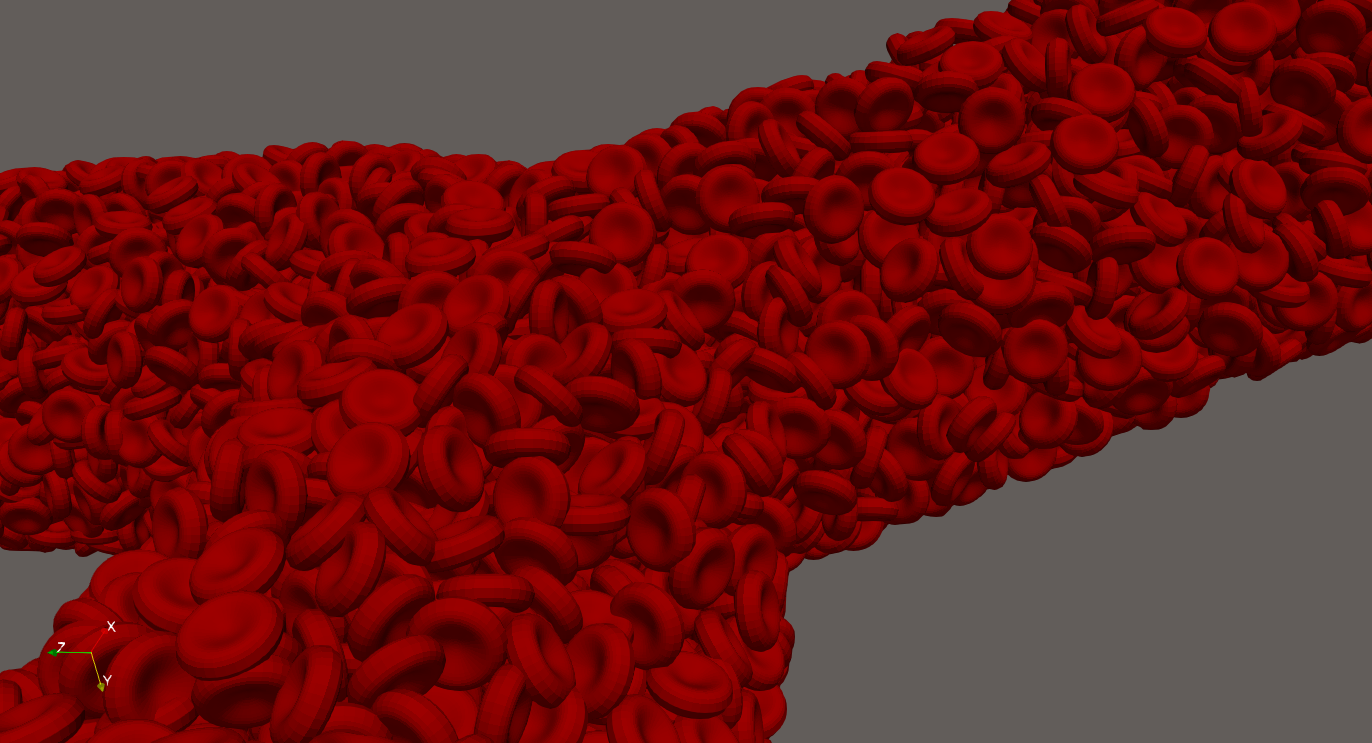}
    \caption{3D Paraview visualizations from demonstrator biofluidic flow simulations in a simplified arterial geometry using LEDDS. Left: Flow with axisymmetric ellipsoids colored by their velocity magnitude, ranging from low values near the vessel walls (blue) to higher values in the central region (red). Right: Preliminary simulation of more realistic red blood cell flows using biconcave particle shapes.}
    \label{fig:bioflow_demonstrators}
\end{figure*}

This work presented LEDDS, a portable and fully algorithmic LBM-DEM framework for GPU-accelerated simulations of fluid-particle systems. The central philosophy underlying LEDDS is the systematic decomposition of all computational stages into high-level algorithmic primitives. By expressing neighbor search, collision detection, DEM force evaluation, fluid-solid coupling and lattice Boltzmann updates solely in terms of map, sort, scan and reduction operations, the entire codebase becomes free of device-specific kernels and relies only on compiler-supported parallelism. This design demonstrates that complex multiphysics solvers can be implemented in a manner that is simultaneously computationally efficient, readable and portable, and thus establishes a practical methodology that can be adopted by other scientific computing codes targeting heterogeneous platforms. The performance analysis confirms that the primitive-based implementation reaches throughput comparable to optimized CUDA solvers for both D3Q19-LBM and DEM components, even in demanding fluid-particle benchmarks. These results suggest that high-level algorithmic programming, when supported by efficient backend runtimes, is sufficient to deliver very good performance without resorting to handcrafted GPU kernels. 

Regarding future HPC-related developments, we will focus on extending the framework toward distributed-memory, multi-GPU execution using MPI-enabled domain decomposition, which is essential for simulations involving large particle counts and high-resolution fluid grids. The algorithmic character of the implementation is expected to ease this transition, as most data-dependent operations are already structured as bulk-synchronous primitives. The memory footprint of the fluid module could be further reduced by adopting one of the ``single-population'' LBM implementations~\cite{HOLZER_PhD_2025}, so that even larger systems can be simulated on a single GPU.
Another important direction is to extend the investigation of performance portability across GPU architectures from multiple vendors. Although the current implementation targets NVIDIA GPUs, the use of high-level algorithmic primitives makes the code largely platform-agnostic. Portability on AMD and Intel GPUs will first be assessed for pure LBM, before extending the analysis to coupled LBM-DEM simulations.

From a physical modeling standpoint, the representation of complex geometries through sphere packings and composite solids is of particular interest, as the current PSM formulation and uniform-grid DEM structures naturally accommodate such extensions. This would broaden the applicability of LEDDS to more realistic particulate systems commonly encountered in geophysics, process engineering, and biomechanics.
Preliminary demonstrator simulations in a simplified arterial geometry, using both axisymmetric ellipsoids and a first approximation of biconcave particles mimicking red blood cells, illustrate the potential of the framework notably for biofluidic applications (see Figure~\ref{fig:bioflow_demonstrators}).
A final line of research aims at exploring the behavior of the solver for high density ratios, as it was previously done with an IBM-based approach~\cite{BHOWMICK2025106696}.

Overall, LEDDS provides evidence that algorithmic primitives constitute a robust foundation for portable, high-performance multiphysics computing. The approach outlined in this work may therefore serve as a blueprint for future scientific software targeting heterogeneous architectures and emerging accelerator technologies.

\section*{Acknowledgements}
The authors would like to thank the reviewers for their careful reading of the manuscript and their constructive comments, which helped improve the quality and clarity of this work.
This research was supported by the SNSF grant 200020{\_}197223, ``Numerical simulation of melt extraction in magmatic crystal mush''.
C.C. further acknowledges financial support from the BNBU Research Fund $\mathrm{n^o UICR0700094}$-$24$ entitled ``GPU-accelerated multidisciplinary methods for industrial computational fluid dynamics''.

\appendix

\section{Material properties}
\label{app:material_properties}

The elastic and damping coefficients ($k_n, k_t, \gamma_n, \gamma_t$) can be obtained from the material properties of the solids which are the the Young's modulus and the Poisson's ratio.
Heinrich Hertz was the first to derive formulas that link the deformation to the force exerted between two spheres~\cite{Hertz1882}. The following formulas allow obtaining the damping and spring coefficients from the physical properties of the materials of the spheres $i$ and $j$~\cite{poschel_computational_2005, norouzi_coupled_2016}
\footnote{We have chosen to keep the same notation as in the LIGGGHTS documentation: \url{https://www.cfdem.com/media/DEM/docu/gran_model_hertz.html}.}:
\begin{align}
&k_n = \frac{4}{3} Y^* \sqrt{R^* \vb \delta_n},
&&k_t = 8 G^* \sqrt{R^* \vb \delta_n},\\
&\gamma_n = -2 \sqrt{\frac{5}{6}} \beta \sqrt{S_n m^*} \geq 0,
&&\gamma_t = -2 \sqrt{\frac{5}{6}} \beta \sqrt{S_t m^*} \geq 0,
\end{align}
where~\cite{brilliantov_1996}:
\begin{align}
&S_n = 2 E^* \sqrt{R^* \delta_n}\\
&S_t = 8 G^* \sqrt{R^* \delta_n}\\
&\beta = \frac{\ln(\epsilon)}{\sqrt{\ln^2(\epsilon) + \pi^2}}\\
&Y^* = \left( \frac{1 - \upsilon_i^2}{E_i} + \frac{1 - \upsilon_j^2}{E_j} \right)^{-1}\\
&G^* = \left( \frac{2(2 - \upsilon_i)(1 + \upsilon_i)}{E_i} + \frac{2(2 - \upsilon_j)(1 + \upsilon_j)}
{E_j} \right)^{-1}\\
&R^* = \left( \frac{1}{R_i} +  \frac{1}{R_j} \right)^{-1} \label{eq:relative_radius}\\
&m^* = \left( \frac{1}{m_i} +  \frac{1}{m_j} \right)^{-1} \label{eq:relative_mass}
\end{align}
where $\delta_n$ is the distance between the surfaces of the particles given by equation \eqref{eq:distanceSpheres}, $E_i$ and $E_j$ are the Young's modulus of the particles, $\upsilon_i$ and $\upsilon_j$ are the Poisson's ratio, $m_i$ and $m_j$ are the masses, and $R_i$ and $R_j$ are the radii. As specified in Eq.~(\ref{eq:local-curvature}), for ellipsoid the radius is locally approximated by the radius of an equivalent sphere. The parameter $\epsilon$ is the coefficient of restitution of the collision. Formally, it corresponds to the ratio of relative velocity of the spheres after the collision to the relative velocity of the particles before the collision. The coefficient of restitution is within the interval $[0, 1]$. When it equals 1, the collision is perfectly elastic, and the damping terms $\gamma_n$ and $\gamma_t$ are zero. When epsilon is smaller than 1, energy loss occurs during the collision, and it is no longer perfectly elastic.
If the particles $i$ and $j$ have different coefficient of restitution, we simply use the arithmetic mean of the two coefficients \cite{poschel_computational_2005}:
\begin{align}
    \epsilon = \frac{\epsilon_i + \epsilon_j}{2}
\end{align}

\section{Single precision}
\label{app:single_precision}

\begin{figure}[t]
\centering
\includegraphics[width=.9\linewidth]{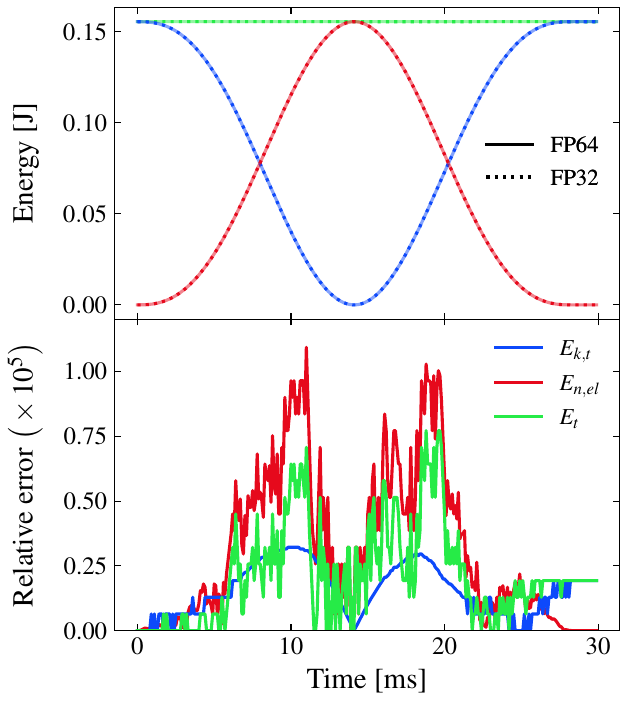}
\caption{FP64 vs FP32 comparison for a head-on sphere collision. Top: Time evolution of energy components ($E_{k,t}$, $E_{n,el}$, $E_t$). Bottom: Relative error normalized by the total energy, $\eta = |E_{\text{FP64}} - E_{\text{FP32}}| / |E_{t,\text{FP64}}|$.}
\label{fig:dem_totalEnergy_32_64_sphere}
\end{figure}

\begin{figure}[h]
\centering
\includegraphics[width=.9\linewidth]{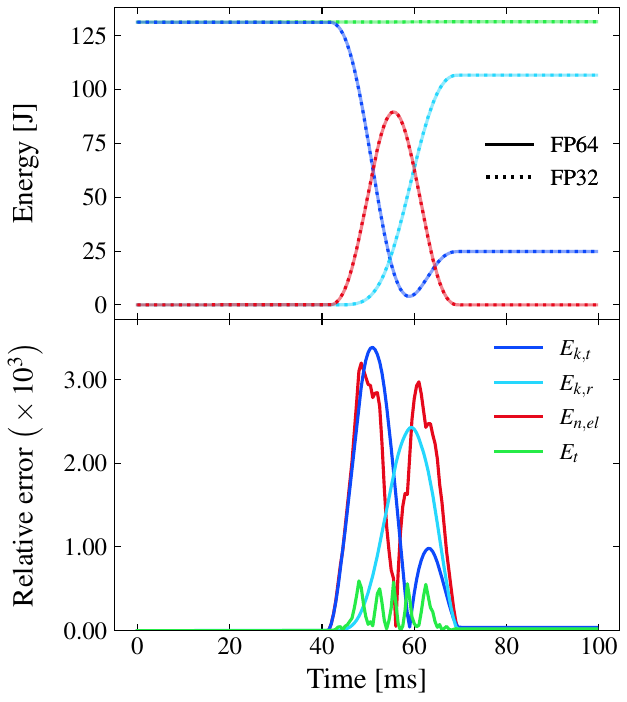}
\caption{FP64 vs FP32 comparison for an eccentric ellipsoid collision. Top: Time evolution of energy components ($E_{k,t}$, $E_{k,r}$, $E_{n,el}$, $E_t$). Bottom: Relative error normalized by the total energy, $\eta = |E_{\text{FP64}} - E_{\text{FP32}}| / |E_{t,\text{FP64}}|$.}
\label{fig:dem_totalEnergy_32_64_ellipsoid}
\end{figure}

As the present version of LEDDS relies on single-GPU executions, it is essential to exploit GPU capabilities as efficiently as possible. In this context, the use of single precision is highly desirable, as it not only accelerates simulations but also significantly reduces memory consumption, thereby allowing larger numbers of particles and fluid cells to be simulated on the same hardware. However, its impact on the accuracy of DEM collision modeling must be carefully assessed, since contact resolution directly affects energy conservation and overall physical fidelity.

To this end, we compare the evolution of kinetic and elastic energies during representative collision events for both single precision (FP32) and double precision (FP64). Two benchmark cases are considered: a head-on collision between spheres (Section~\ref{subsec:sphere_sphere_headon_collision}) and an eccentric collision between ellipsoids (Section~\ref{subsec:ellipsoid_ellipsoid_eccentric_collision}). For each case, we report both the energy evolution and a normalized error measure defined as
$\eta = |E_{\text{FP64}} - E_{\text{FP32}}| / |E_{t,\text{FP64}}|$,
which avoids spurious amplification when individual energy components approach zero. The spherical and ellipsoidal cases are shown in Figures~\ref{fig:dem_totalEnergy_32_64_sphere} and~\ref{fig:dem_totalEnergy_32_64_ellipsoid}, respectively.

For both benchmark cases, the energy evolutions obtained with FP32 and FP64 are nearly indistinguishable, indicating a very close agreement between the two precision levels. The normalized error remains small throughout the simulations. In the ellipsoidal case, deviations are observed during the collision because the transition between contact regimes is more sensitive to numerical precision, in particular due to orientation matrix manipulations and non-spherical contact mechanics. A similar, though less pronounced, behavior is observed for spherical particles, with deviations two orders of magnitude smaller than for the non-spherical case. This is primarily attributed to the simpler and more robust numerical treatment of spherical contact mechanics.

Importantly, the error remains bounded in time and vanishes after contact, indicating that no long-term drift or energy bias is introduced by single precision. These results demonstrate that FP32 provides sufficient accuracy for DEM collision modeling in the present framework, while enabling improved performance and larger simulation sizes on GPU architectures. For pure fluid (LBM) simulations, the use of FP32 is also standard practice, and we have shown in previous works that LBM solvers based on modern C++ can accurately simulate a wide range of flow regimes with FP32, from aerodynamics to aeroacoustics and even supersonic flows~\cite{THYAGARAJAN_PoF_35_2023,COREIXAS_CPC_317_2025,LATT_CPC_323_2026,COREIXAS_3AF_2026,COREIXAS_AIAA_2026}. Additional studies conducted on single-particle settling (Section~\ref{subsec:settling}) further extend these observations to coupled LBM-DEM simulations, although the corresponding results are not shown here for brevity.

\section{Parametric study for particle settling}
\label{app:param_study}

Here, we present parametric studies to identify an optimal balance 
between accuracy, robustness, and computational efficiency for the particle settling benchmark, but which can more generally be used for simulations with fluid-particle interactions. Such studies are particularly 
critical in LBM-DEM coupling, where the LBM velocity ($u_{lb}$) controls the time step, which must be sufficiently small to accurately capture the Stokes regime, yet excessively small 
steps also reduce computational efficiency. Similarly, the ratio of particle diameter to fluid cell size 
$(D/\Delta x)$ strongly affects the accuracy of the fluid--particle coupling, while higher resolution 
increases computational cost and memory usage. Thus, practical simulations require a careful trade-off 
between accuracy and efficiency.

\begin{figure}[t]
    \centering
    \includegraphics[width=\linewidth]{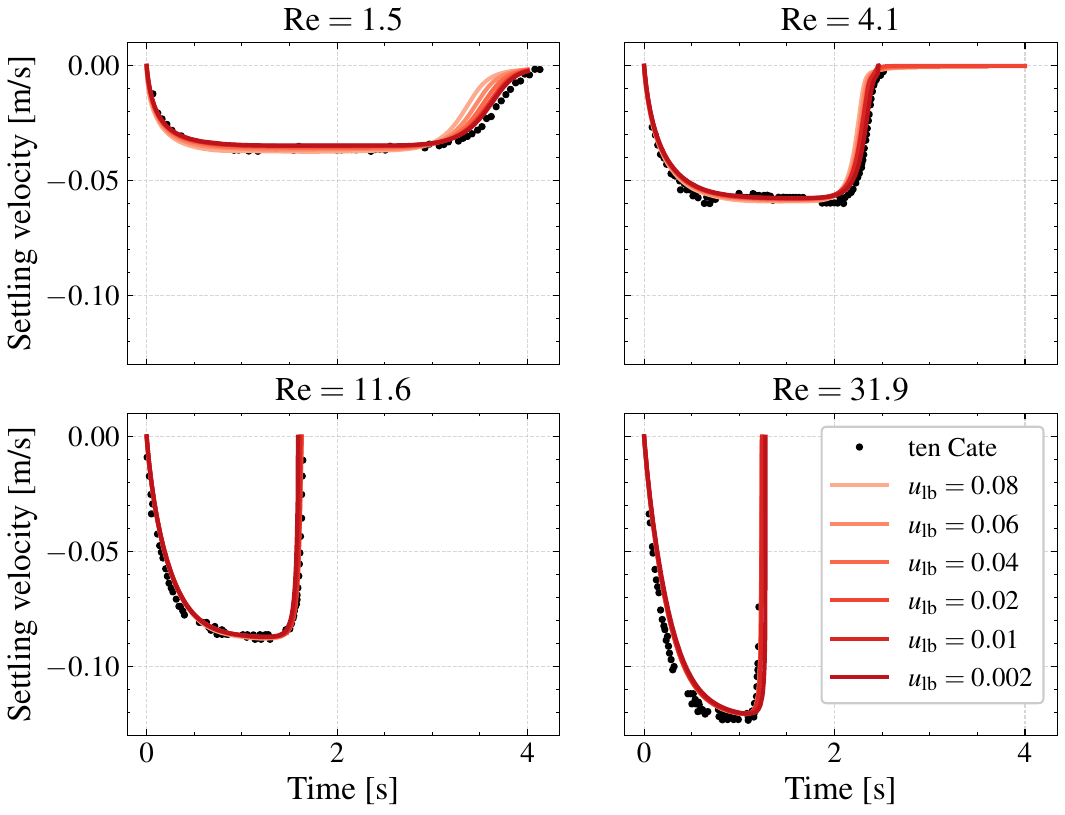}
    \caption{Single particle settling simulations: Impact of the LB velocity (time step) on the accuracy of LEDDS simulations for a fixed resolution of the particle ($D/\Delta x = 20$).}
    \label{fig:param_study_particleSettling_ulb_impact}
\end{figure}

We first quantify the impact of the LB velocity on the particle settling simulations for a fixed particle 
resolution of \(D/\Delta x = 20\) (Figure~\ref{fig:param_study_particleSettling_ulb_impact}). Reducing 
\(u_{lb}\) improves accuracy in the low Reynolds number regime (Re \(\lesssim 5\)), where viscous forces 
dominate. In higher Reynolds number regimes, where inertial effects are more significant, the results are 
relatively insensitive to the choice of \(u_{lb}\). Overall, a value of \(u_{lb} = 0.01\) to \(0.02\) 
represents a good compromise between accuracy and computational efficiency, consistent with the main text 
observations regarding transient dynamics and the early-stage pressure-wave effects.

\begin{figure}[t]
    \centering
    \includegraphics[width=\linewidth]{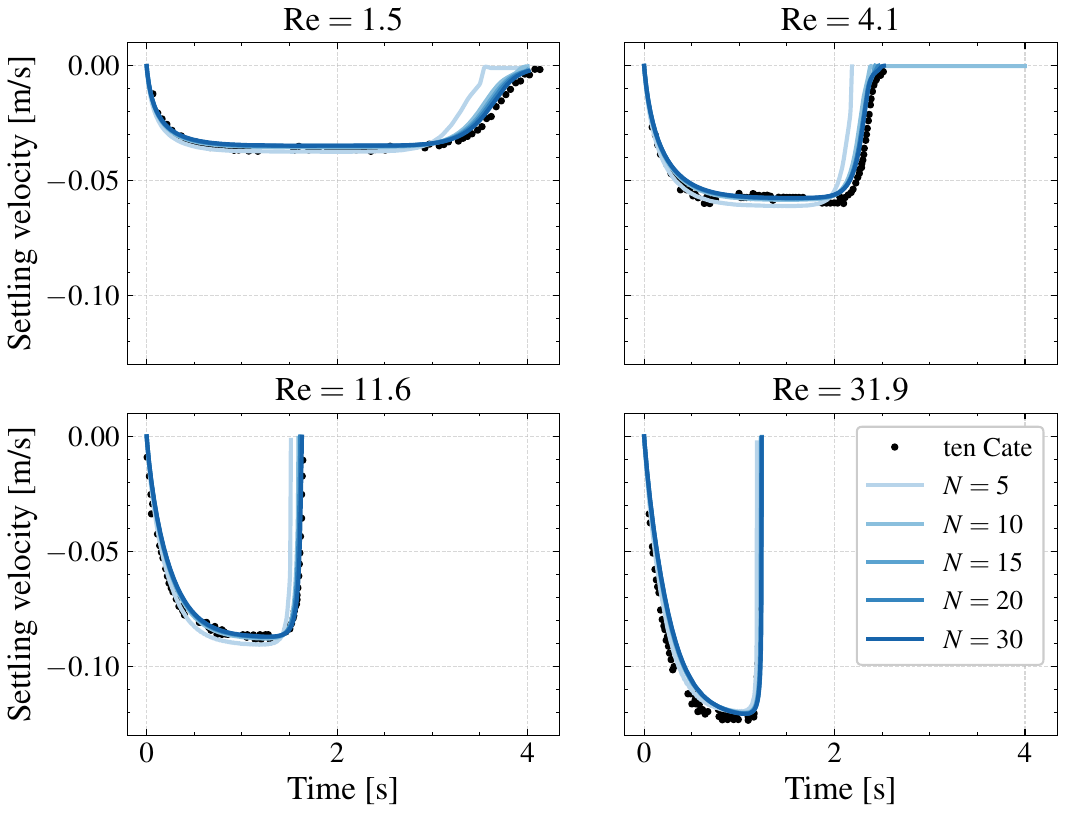}
    \caption{Single particle settling simulations: Impact of the particle resolution, $N=D/\Delta x$, on the accuracy of LEDDS simulations for a fixed LB velocity ($u_{lb} = 0.01$).}
    \label{fig:param_study_particleSettling_cpd_impact}
\end{figure}
Next, we investigate the effect of particle resolution at a fixed lattice velocity \(u_{lb} = 0.01\) 
(Figure~\ref{fig:param_study_particleSettling_cpd_impact}). For resolutions of at least 10 fluid cells 
per particle diameter, the terminal settling velocity is essentially unchanged, while small deviations 
appear near the no-slip boundary at the bottom wall. Resolutions between \(D/\Delta x = 10\) and 20 
provide a good compromise for capturing both transient dynamics and near-wall effects in the Stokes 
regime, while keeping the memory footprint reasonable. Indeed, reducing the resolution from 20 to 10 
reduces the fluid memory requirement by a factor of approximately 8, with minimal loss of accuracy, as 
also highlighted in the main text.

In summary, if the primary objective is to solely recover the correct terminal settling velocity, simulations can be accelerated by several orders of magnitude when moving from the most demanding parameter set \((D/\Delta x, u_{lb}) = (30,\,0.002)\) to a more relaxed configuration such as \((D/\Delta x, u_{lb}) = (10,\,0.04)\). As a general best practice for LBM-DEM simulations with PSM coupling, we recommend using \(D/\Delta x = 10\) together with \(u_{lb} = 0.01\text{--}0.02\), which provides a good balance between accuracy and efficiency. Lower resolutions such as \(D/\Delta x = 5\) should be restricted for large scale particle laden flows --typically involving hundreds/thousands of particles-- where a resolution of \(D/\Delta x = 10\) would exceed the memory capacity of a single GPU, and the corresponding runtime would be prohibitively long.

\section{GPU memory usage}
\label{app:gpu_memory_usage}

Memory usage is a key constraint for GPU-based simulations. Since LEDDS currently targets single-GPU executions, this metric directly determines the class of GPUs required to run a given simulation and, consequently, the range of feasible configurations. In this work, we rely on high-end GPUs with large memory capacity (e.g., NVIDIA A100-SXM4 with 80~GB), which allows us to prioritize computational performance and simplified data structures, at the cost of a higher memory footprint.
In the following, we first derive simple memory estimates for the LBM and DEM modules separately, and then combine them into a global model that provides order-of-magnitude predictions for the total memory requirement of a given simulation.

\subsection{DEM module: Theoretical models \label{subsec:gpu_memory_usage_dem}}

The memory requirements of the DEM module are summarized in Table~\ref{tab:dem_memory}. The minimal memory footprint per particle reads
\begin{equation}\label{eq:mem_model_dem}
m_p = 47s + 1,
\end{equation}
where $s$ denotes the size (in Bytes) of a floating-point value ($4$ for FP32 and $8$ for FP64). This corresponds to $189$~[B] in FP32 and $377$~[B] in FP64.
The memory associated with fluid--particle coupling is
\[
m_{p,\mathrm{cpl}} = 6s,
\]
which represents a fraction
\[
\frac{m_{p,\mathrm{cpl}}}{m_p} = \frac{6s}{47s+1}
\]
of the total memory per particle. This corresponds to approximately $13\%$, independently of the precision, which indicates that the coupling overhead remains relatively small within the DEM module.
Based on these values, one gigabyte of memory can store on the order of $5 \times 10^6$ particles in FP32 and $2.5 \times 10^6$ in FP64, providing a useful rule of thumb for estimating the memory requirements of the DEM module. 

\begin{table}
    \centering
    \small
    \caption{Memory footprint per particle (DEM and coupling variables). Comp.: number of components; Mem.: memory in Bytes; $s$: size of a floating-point value (4~Bytes for FP32, 8~Bytes for FP64).}
    \begin{tabular}{l c c c}
        \toprule
        Variable & Comp. & Mem. [B] & Usage \\
        \midrule
        Type & 1 & $1$ & DEM \\
        Position & 3 & $3s$ & DEM \\
        Orientation & 9 & $9s$ & DEM \\
        Velocity & 3 & $3s$ & DEM \\
        Angular vel. & 3 & $3s$ & DEM \\
        Acceleration & 3 & $3s$ & DEM \\
        Angular accel. & 3 & $3s$ & DEM \\
        Semi-axes & 3 & $3s$ & DEM \\
        Inertia & 3 & $3s$ & DEM \\
        Mass & 1 & $s$ & DEM \\
        Young mod. & 1 & $s$ & DEM \\
        Poisson ratio & 1 & $s$ & DEM \\
        Restitution coeff. & 1 & $s$ & DEM \\
        Friction coeff. & 1 & $s$ & DEM \\
        Force & 3 & $3s$ & DEM \\
        Torque & 3 & $3s$ & DEM \\
        \midrule
        Fluid force & 3 & $3s$ & Coupling \\
        Fluid torque & 3 & $3s$ & Coupling \\
        \midrule
        \textbf{DEM} & & $\bm{41s + 1}$ & \\
        \textbf{Coupling} & & $\bm{6s}$ & \\
        \textbf{Total} & & $\bm{47s + 1}$ & \\
        \bottomrule
    \end{tabular}
    \label{tab:dem_memory}
\end{table}

\subsection{DEM module: Measurements}
\label{subsec:gpu_memory_usage_meas_dem}

Building on this baseline, we now assess the effective memory usage in practical DEM simulations. The following measurements quantify the impact of dynamic data structures on the overall GPU memory footprint.

In this setup, spheres are initially axis-aligned and distributed on a uniform grid used for collision detection. A one-to-one mapping between particles and grid cells is enforced by selecting a DEM cell size (see Section~\ref{subsubsec:grid}) slightly larger than the particle diameter ($l_{DEM} = 1.1\,d_{p}$). This configuration limits the number of non-physical candidate pairs generated during collision detection and controls the number of particles per cell, while preserving numerical stability. 
In contrast to the benchmark in Section~\ref{subsec:many_body}, the computational domain is fully periodic and does not include walls, thereby avoiding additional dynamic memory allocations associated with particle-wall interactions. The total number of spheres is varied from approximately $8000$ to a bit more than $4$ millions by increasing the domain size.

\begin{figure}
\centering
\includegraphics[width=.95\linewidth]{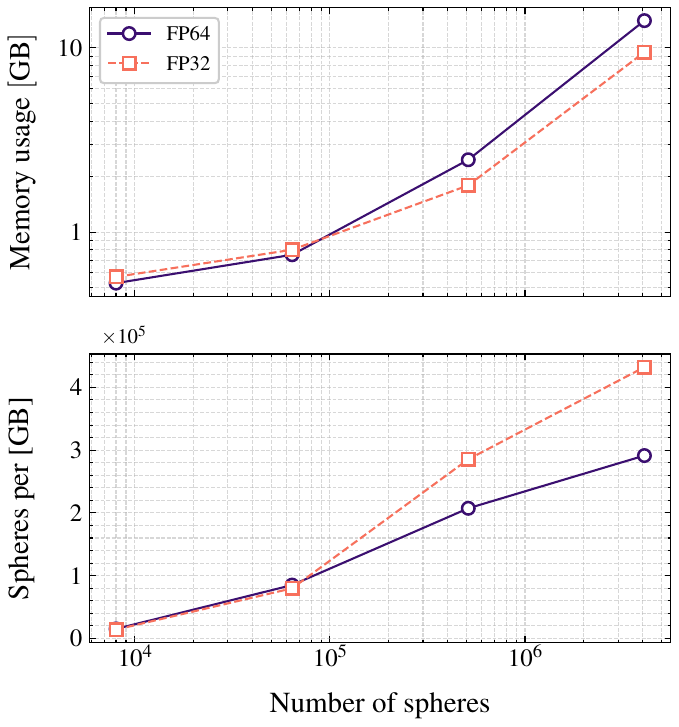}
\caption{GPU memory usage (top) and corresponding memory efficiency, expressed as the number of spheres per [GB] (bottom), as functions of system size.}
\label{fig:memory_usage_dem}
\end{figure}

Figure~\ref{fig:memory_usage_dem} reports both the total GPU memory usage and the corresponding memory efficiency (number of spheres per [GB]), as functions of the number of simulated spheres. In both precisions, the memory footprint increases markedly with system size, from about $0.54$ to $14.4$~[GB] in FP64 and from about $0.59$ to $9.70$~[GB] in FP32 as the number of spheres increases. The difference between FP32 and FP64 becomes more pronounced for large systems, while for small cases both curves remain relatively close because the fixed memory overhead of the application is still significant.
Focusing on the memory efficiency (number of spheres per [GB]), it increases with system size which reflects the progressive amortization of fixed overheads. However, even for the largest configurations, the achieved efficiency remains about one order of magnitude below the prediction of the particle-based memory model.

A more quantitative view is obtained by examining the effective memory cost per sphere. For the largest simulation, the measured memory usage corresponds to about $2.7\times10^5$ spheres per [GB] in FP64 versus $4\times10^5$ spheres per [GB] in FP32, i.e., about $3.5$~[kB] and $2.4$~[kB] per sphere, respectively. This is roughly one order of magnitude larger than the ideal per-particle estimate derived from the static memory model~(\ref{eq:mem_model_dem}). The discrepancy is mainly explained by the storage of collision pairs: even in the present simplified configuration, up to $5.3\times10^{7}$ candidate pairs are generated and stored, accounting for approximately $80\%$ of the total memory usage.

Overall, these results indicate that omitting dynamic memory contributions, notably those arising from collision-pair storage and contact handling, results in a substantial underestimation of the actual memory footprint. In practice, this discrepancy can be compensated for by applying a simple correction factor of about $10$ to the particle-based model~(\ref{eq:mem_model_dem}), leading to more realistic estimates of DEM memory requirements.

\subsection{LBM and LBM-DEM: Theoretical models \label{subsec:gpu_memory_usage_lbm}}

As detailed in Table~\ref{tab:lbm_memory}, the memory footprint per fluid cell is given by the general expression
\begin{equation}
m_f = (2Q + 7n_s + 1)s + 12n_s + 6,
\end{equation}
where $Q$ is the number of lattice populations, $n_s$ is the number of solids that overlap the fluid cell, and $s$ denotes the size (in Bytes) of a floating-point value ($4$ for FP32 and $8$ for FP64).

\begin{table}
    \centering
    \small
    \caption{Memory per fluid cell. Comp.: number of components; Mem.: memory in Bytes; Use: usage (LBM: fluid solver, Coupling: fluid-particle coupling). $s$: size of a floating-point value (4 for FP32, 8 for FP64); $Q$: number of lattice populations; $n_s$: maximal number of solids per fluid cell; $\phi_s$: solid fraction per solid; $\phi$: total solid fraction.}
    \begin{tabular}{l c c c}
        \toprule
        Var. & Comp. & Mem. [B] & Use \\
        \midrule
        Type & 1 & $1$ & LBM \\
        Pop. & $Q$ & $2Qs$ & LBM \\
        \midrule
        $\phi_s$ & 1 & $n_s s$ & Coupling \\
        $\phi_s$ ids & 1 & $4n_s$ & Coupling \\
        $\phi$ & 1 & $s$ & Coupling \\
        $n_s$ & 1 & $1$ & Coupling \\
        Force & 3 & $3n_s s$ & Coupling \\
        Torque & 3 & $3n_s s$ & Coupling \\
        Force ids & 1 & $4n_s$ & Coupling \\
        \#Force & 1 & $4$ & Coupling \\
        Buffer & 1 & $4n_s$ & Coupling \\
        \midrule
        \textbf{LBM} & & $\bm{2Qs + 1}$ & \\
        \textbf{Coupling} & & $\bm{(7n_s + 1)s + 12n_s + 5}$ & \\
        \textbf{Total} & & $\bm{(2Q + 7n_s + 1)s + 12n_s + 6}$ & \\
        \bottomrule
    \end{tabular}
    \label{tab:lbm_memory}
\end{table}

Introducing $\phi_0$, $\phi_1$, and $\phi_2$ as the fractions of fluid cells containing zero, one, and two solids (the latter being the hard implementation limit in LEDDS), the average memory per fluid cell can be expressed as
$$
\langle m_f \rangle = \phi_0(2Qs+1) + \phi_1\big[(2Q+8)s+18\big] + \phi_2\big[(2Q+15)s+30\big].
$$
In fully resolved LBM-DEM simulations, particles are typically much larger than fluid cells, so most cells contain at most one solid ($\phi_2 \ll \phi_1$), and $\phi_1 \approx \phi$, where $\phi$ is the solid volume fraction. This leads to the practical approximation
\begin{equation}
\langle m_f \rangle \approx (1-\phi)(2Qs+1) + \phi\big[(2Q+8)s+18\big],
\end{equation}
and, for the D3Q19-LBM used in LEDDS, this gives:
\begin{itemize}[noitemsep, topsep=0pt]
    \item $n_s=0$ (fluid cell): $162/318$~[B] for FP32/FP64,
    \item $n_s=1$ (typical case): $202/386$~[B] for FP32/FP64,
    \item $n_s=2$ (upper bound): $242/454$~[B] for FP32/FP64.
\end{itemize}

Focusing only on the memory associated with fluid-particle coupling per fluid cell, it reads
\begin{equation}
m_{f,\mathrm{cpl}} = (7n_s+1)s + 12n_s + 5,
\end{equation}
which represents about $35\%$ of the total memory per cell in the worst case ($n_s=2$), and about $20$-$25\%$ for the more realistic case $n_s=1$. 
Since most cells contain at most one or no solid, the effective coupling overhead is expected to remain moderate in practice, typically below $30\%$ of the pure fluid memory footprint when accounting for a safety margin.

Based on the above estimates and assuming an average coupling overhead of about $30\%$, the effective memory per fluid cell is on the order of $200$~[B] in FP32 and $400$~[B] in FP64 for a D3Q19-LBM.
This corresponds to a storage capacity of approximately $5 \times 10^6$ fluid cells per [GB] in FP32 and $2.5 \times 10^6$ in FP64. These values provide a useful rule of thumb for estimating the memory requirements of LBM-DEM simulations performed with LEDDS.

Combining both contributions, the total memory usage of the LBM-DEM model can be approximated as
\begin{equation}
m_{\mathrm{total}}\approx m_f N_{\mathrm{fluid}} + m_p N_{\mathrm{part}},    
\end{equation}
where $N_{\mathrm{fluid}}$ and $N_{\mathrm{part}}$ denote the number of fluid cells and particles, respectively. Since $m_f$ and $m_p$ are of the same order of magnitude, the relative contribution of each module is primarily governed by the ratio $N_{\mathrm{fluid}}/N_{\mathrm{part}}$. In practice, \emph{fully resolved} LBM-DEM simulations typically require at least $\mathcal{O}(10)$ fluid cells per particle diameter, so that $N_{\mathrm{fluid}} \gg N_{\mathrm{part}}$. As a result, the total memory usage is dominated by the LBM module in most practical configurations.

\subsection{LBM-DEM: Measurements}
\label{subsec:gpu_memory_usage_meas_dem}

Using the fluidized-bed configuration introduced in Section~\ref{sec:perfo}, which involves up to $80$ million fluid cells and approximately $8600$ particles, we now evaluate the total memory usage of LEDDS under realistic conditions. In contrast to the idealized estimates above, these measurements also include the dynamic data structures required in practice for collision handling and coupling, such as the cell-linked list, collision pair lists, tangential displacement history, and temporary buffers used during sorting and reduction operations.

\begin{figure}[h]
\centering
\includegraphics[width=.95\linewidth]{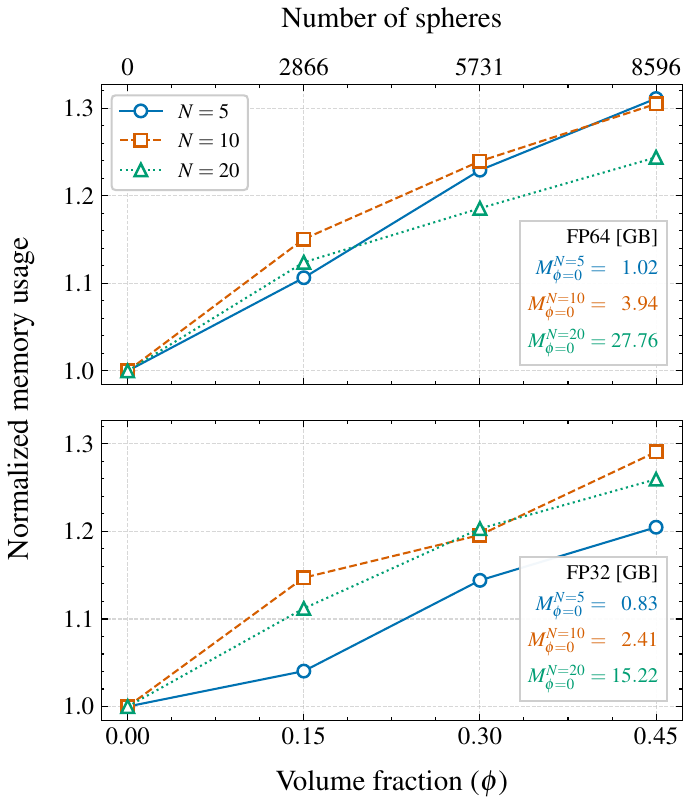}
\caption{Normalized GPU memory usage as a function of the solid volume fraction $\phi$ for different resolutions ($N$). Top: Double precision. Bottom: Single precision}
\label{fig:memory_usage_gpu_single_double}
\end{figure}

Figure~\ref{fig:memory_usage_gpu_single_double} shows the normalized GPU memory usage as a function of the solid volume fraction for different resolutions. The presence of particles leads to an increase in memory usage compared with the pure-fluid case. More precisely, increasing the solid volume fraction up to $\phi=0.45$ raises the total memory footprint by roughly $20$-$30\%$, depending on the resolution and precision. The dependence on the resolution is not strictly monotonic, which indicates that this increase is not governed by particle storage itself, but rather by the additional dynamic memory required for collision handling and fluid-particle coupling.

The reduction in memory footprint when switching from FP64 to FP32 remains significant, although it is smaller than a factor of two. For the pure-fluid reference case ($\phi=0$), the memory usage decreases from $1.02$, $3.94$, and $27.76$~GB in FP64 to $0.83$, $2.41$, and $15.22$~GB in FP32 for $N=5$, $10$, and $20$, respectively (see Figure~\ref{fig:memory_usage_gpu_single_double}). This behavior is expected, as part of the memory footprint is associated with data structures (e.g., indices and buffers) that are independent of floating-point precision.

A more detailed interpretation of this increase can be obtained by separating the respective contributions of the fluid and particle data. Using the actual particle counts of the fluidized-bed configuration, namely $N_{\mathrm{part}}=0$, $2866$, $5731$, and $8596$ for $\phi=0$, $0.15$, $0.30$, and $0.45$, respectively, the minimal DEM storage remains negligible compared with the fluid contribution. In FP32, the DEM memory associated with $8596$ particles is only about $1.62$~[MB] (and $3.24$~[MB] in FP64), which does not significantly affect the total memory estimate. 
This confirms that the observed $20$-$30\%$ increase with $\phi$ cannot be attributed to particle storage itself, but is primarily due to additional dynamic data structures associated with collision detection and fluid-particle coupling.

A more detailed comparison can then be made against the theoretical estimate of the fluid contribution alone. Using an effective memory footprint of about $200$-$260$~[B] per fluid cell in FP32 (including coupling and a safety margin), the expected memory usage for $N=5$, $10$, and $20$ is approximately $0.23$, $1.81$, and $14.5$~[GB], to be compared with the measured $0.83$, $2.41$, and $15.22$~[GB]. 
The agreement improves with increasing resolution, confirming that the model captures the dominant fluid-storage contribution, while the remaining discrepancy reflects the influence of auxiliary GPU data structures.

In summary, these results confirm that the present minimal memory model provides a reliable lower bound for fully resolved LBM-DEM simulations. In practice, applying a safety factor of approximately $30$-$40\%$ to the pure-fluid prediction allows for a robust prediction of the actual memory requirements for LBM-DEM simulations.
This offers a simple and practical guideline to anticipate GPU memory needs and to select appropriate hardware for a given simulation.




\end{document}